\documentclass[11pt, authoryear]{article}
\usepackage{graphicx}
\usepackage{lscape}
\usepackage{amsmath}
\usepackage{amsthm}
\usepackage{graphicx}
\usepackage{amsfonts}
\usepackage{amssymb}
\usepackage{subfigure}
\usepackage{booktabs}
\usepackage{float}
\usepackage{natbib}
\usepackage{enumitem}
\usepackage[hidelinks,colorlinks=true,linkcolor=blue,citecolor=blue]{hyperref}

\newcommand{\EE}{\mathbb{E} }
\newcommand{\Var}{\operatorname{Var} }

\usepackage{color}

\newcommand{\be}{\begin{equation}}
\newcommand{\en}{\end{equation}}

\newcommand{\ea}{\end{eqnarray}}
\newcommand{\ba}{\begin{eqnarray}}
\newcommand{\ean}{\end{eqnarray*}}
\newcommand{\ban}{\begin{eqnarray*}}

\graphicspath{{hetergroup/}}

\usepackage[margin=1.25in]{geometry}

\begin{document}
\title{Change‑point estimation for \text{Weibull} time series with copula‑based Markov models}
\author{Li-Hsien Sun \thanks{\textit{Corresponding author}. Graduate Institute of Statistics, National Central University, Chung-Li, Taiwan, 32001 {\em lihsiensun@ncu.edu.tw}. Work supported by NSTC-112-2628-M-008-002-MY3}, Zong-Yuan Huang\thanks{Graduate Institute of Statistics, National Central University, Chung-Li, Taiwan, 32001},
Yi-Ling Huang\thanks{Graduate Institute of Statistics, National Central University, Chung-Li, Taiwan, 32001},
Chi-Yang Chiu\thanks{Division of Biostatistics, Department of Preventive Medicine, University of Tennessee Health Science Center, Memphis, Tennessee, USA},
Ning Ning\thanks{\textit{Corresponding author}. Department of Statistics,
	Texas A\&M University, College Statione, USA. {\em patning@tamu.edu}}
}

\date{\today}
\maketitle

\begin{abstract}
We study offline change-point estimation for time series data exhibiting nonlinear serial dependence. To address this problem, we propose a copula-based Markov chain model with Weibull marginal distributions, which is suitable for modeling nonnegative data such as event times and volatility measures. Nonlinear dependence is incorporated through the Clayton and Joe copulas, allowing the model to capture asymmetric lower-tail and upper-tail dependence structures, respectively. We derive the corresponding likelihood function and estimate the change point and model parameters using maximum likelihood estimation implemented through the Newton--Raphson algorithm. Confidence intervals are constructed via a parametric bootstrap Monte Carlo procedure. Extensive numerical studies are conducted to evaluate the finite-sample performance and robustness of the proposed method under different dependence structures and copula misspecification scenarios. The results demonstrate that the proposed estimators perform well in terms of RMSE and relative error, particularly for the estimation of the change point. An empirical application to the VIX index during the COVID-19 pandemic further illustrates the practical usefulness of the proposed approach in detecting structural changes in both the marginal distributions and serial dependence structure.
\end{abstract}


\renewcommand{\thepage}{\arabic{page}}
\setcounter{page}{1}
\section{Introduction}

Detecting change-points in sequential data is an important problem in statistical applications because structural changes often reflect significant shifts in the underlying data-generating mechanism. Change-point analysis has been widely applied in many areas, including finance \cite{lavielle2007adaptive}, climatology \cite{reeves2007review}, reliability studies, and medical research \cite{parpoula2022distribution}. In general, change-point methodology can be divided into two categories: offline estimation and online monitoring. Offline change-point analysis aims to estimate structural changes retrospectively using a complete dataset observed over a fixed time horizon, whereas online monitoring procedures are designed to detect changes sequentially in real time as new observations become available.

Most classical change-point methods are developed under assumptions of independent observations or linear dependence structures, such as autoregressive models \cite{timmer2003change}. Under these settings, change-points are commonly estimated through likelihood-based methods, particularly maximum likelihood estimation (MLE). However, linear dependence assumptions are often inadequate for describing the nonlinear and asymmetric dependence structures frequently observed in practical time series data \cite{ning2023iterated,ionides2024iterated}. To overcome this limitation, copula-based Markov chain models provide a flexible framework for modeling nonlinear serial dependence \cite{chen2006estimation,joe1997multivariate}. By separating the marginal distributions from the dependence structure, copula models allow greater modeling flexibility and can capture asymmetric lower- and upper-tail dependence that conventional linear models fail to describe. Moreover, copula-based Markov models satisfy strict stationarity under mild regularity conditions and have been shown to perform well in nonlinear time series settings; see, for example, \cite{beare2010copulas} and \cite{huang2021model}.

In this paper, we propose a copula-based Markov chain model with Weibull marginal distributions for offline change-point estimation in nonnegative time series exhibiting nonlinear dependence. The Weibull distribution is widely used in survival analysis, reliability modeling, and duration analysis because of its flexibility in accommodating different hazard rate behaviors. To capture asymmetric tail dependence structures, we incorporate the Clayton and Joe copulas into the Markov framework, where the Clayton copula is capable of modeling stronger lower-tail dependence and the Joe copula captures upper-tail dependence. This framework allows us to simultaneously model nonlinear serial dependence and structural changes in the marginal distributions and dependence structures. Related copula-based approaches have been considered in the literature. For example, Emura and Murotani \cite{emura2015algorithm} proposed estimation procedures for copula-based dependent truncation models, while Sun et al. \cite{sun2020bayesian} developed Bayesian inference methods for copula-based Markov time series models with Student's $t$ marginals.

For statistical inference, we develop a likelihood-based estimation procedure using maximum likelihood estimation. Since the likelihood function is not differentiable with respect to the change point, we employ a profile likelihood approach combined with the Newton--Raphson algorithm to obtain the parameter estimates numerically \cite{ELS2021,SWLEC2025}. To ensure numerical stability under parameter constraints, a suitable re-parameterization technique is adopted. In addition, confidence intervals for both the model parameters and the change point are constructed using a parametric bootstrap Monte Carlo procedure. The finite-sample performance of the proposed method is investigated through extensive simulation studies under different sample sizes, dependence strengths, change-point locations, and copula misspecification scenarios. The simulation results demonstrate that the proposed estimators perform well in terms of RMSE and relative error (RE), particularly for the estimation of the change point. The results also show that the proposed procedure remains relatively robust under moderate copula misspecification, although the estimation of tail dependence parameters becomes more sensitive when the copula family is misspecified.

To illustrate the practical usefulness of the proposed framework, we apply the model to the VIX index during the COVID-19 pandemic period. The VIX index, commonly referred to as the ``fear index,'' measures market expectations of future volatility over a 30-day horizon and is widely used as an indicator of financial market uncertainty. The empirical results identify structural changes associated with the onset of the COVID-19 pandemic and show that the Clayton copula-based Markov model provides the best fit according to the Akaike Information Criterion (AIC), suggesting the presence of stronger downside dependence during periods of market stress.

The remainder of this paper is organized as follows. Section~\ref{Sec: Model} introduces the proposed copula-based Markov chain model with Weibull marginal distributions and a single change point. Section~\ref{Sec: PE} presents the likelihood-based estimation procedure, the Newton--Raphson algorithm, and the bootstrap Monte Carlo interval estimation method. Section~\ref{Sec: Sim} reports the simulation studies and investigates the effects of dependence strength and copula misspecification. Section~\ref{Sec: Em} presents the empirical analysis of the VIX data during the COVID-19 pandemic. Finally, Section~\ref{Sec: Con} concludes the paper and discusses possible directions for future research.

\section{The Proposed Model}\label{Sec: Model}
We describe the copula-based Markov \text{Weibull} time series with one structural change. We first introduce the \text{Weibull} distribution and then the copula functions. Finally, we propose the conditional density for the likelihood inference based on copulas and marginals. 

\subsection{\text{Weibull} distribution Marginals} 
 Let \( X \) be a random variable driven by the \text{Weibull} distribution with the scale parameter $\lambda> 0$ and the shape parameter $k> 0$, respectively. The cumulative distribution function (CDF) is
\[
F(x) = 1 - e^{-(\frac{x}{\lambda})^{k}}, \quad x \geq 0,
\]
and the probability density function (PDF) is
\[
f(x) = \frac{k}{\lambda} \left(\frac{x}{\lambda}\right)^{k-1} e^{-(\frac{x}{\lambda})^{k}}, \quad x \geq 0. 
\]
The mean and variance are  
$
\EE(X) = \lambda \Gamma \left( 1 + \frac{1}{k} \right)
$
and 
$
\Var(X) =\lambda^2 \left[ \Gamma \left( 1 + \frac{2}{k} \right) - \Gamma \left( 1 + \frac{1}{k} \right)^2 \right]
$,
respectively. 


The shape parameter \( k \) has a more significant impact on the distribution because it directly determines the fundamental shape and characteristics of the distribution, including whether a peak exists, the direction of skewness, and the presence of long tails. When \( k > 1 \), the distribution becomes more symmetric and exhibits a distinct peak, meaning that the data is more concentrated, resembling a normal distribution. When \( k < 1 \), the distribution is right-skewed, with a pronounced long-tail effect. On the other hand, the scale parameter \( \lambda \) only affects the translation and stretching of the distribution. As \( \lambda \) increases, the distribution spreads out more, but its shape remains unchanged.

When the scale parameter is fixed, the estimation of the shape parameter is relatively stable, and the estimation error remains small. This is because the scale parameter primarily affects the spread of the distribution without altering its overall shape characteristics. In contrast, when the shape parameter is fixed, increasing the scale parameter leads to a wider data distribution and a broader range of sample observations. As a result, the estimation process becomes more sensitive to random fluctuations, which in turn increases the estimation error.


\subsection{Copulas and Serial Dependence} 
A copula is a multivariate distribution function whose marginal distributions are uniform \cite{sklar1959fonctions}. Copula functions are widely used to characterize the dependence structure among multiple random variables. In this study, we focus on bivariate copulas within the Markov model framework, as they effectively capture tail dependence. In particular, the Clayton copula \citep{clayton1978model} emphasizes lower-tail dependence, whereas the Joe copula \citep{Joe1993} is well suited for modeling upper-tail dependence.

The Clayton copula \citep{clayton1978model} is defined by
\[
C_{\alpha}(u,v)
=
\left[
\max\left(u^{-\alpha}+v^{-\alpha}-1,\,0\right)
\right]^{-1/\alpha},
\qquad 0 \leq u,v \leq 1,
\]
where the dependence parameter satisfies
$
\alpha \in (-1,\infty)\setminus\{0\}.
$
A positive value of $\alpha$ corresponds to positive dependence, whereas a negative value indicates negative dependence. The density function of the Clayton copula is given by
\[
c_{\alpha}(u,v)
=
(1+\alpha)
u^{-(1+\alpha)}
v^{-(1+\alpha)}
\left(
u^{-\alpha}+v^{-\alpha}-1
\right)^{-\left(\frac{1}{\alpha}+2\right)}
\mathbf{I}\!\left(
u^{-\alpha}+v^{-\alpha}-1>0
\right),
\]
where $\mathbf{I}(\cdot)$ denotes the indicator function.
Moreover, the Joe copula \citep{Joe1993} is defined as
\[
C_{\alpha}(u,v)
=
1-
\left[
(1-u)^{\alpha}
+
(1-v)^{\alpha}
-
(1-u)^{\alpha}(1-v)^{\alpha}
\right]^{1/\alpha},
\]
where the dependence parameter satisfies $\alpha \geq 1$. Its corresponding density function is
\[
c_{\alpha}(u,v)
=
\mathcal{J}(u,v,\alpha)^{\frac{1}{\alpha}-2}
(1-u)^{\alpha-1}
(1-v)^{\alpha-1}
\left[
(\alpha-1)+\mathcal{J}(u,v,\alpha)
\right],
\]
with
\[
\mathcal{J}(u,v,\alpha)
=
(1-u)^{\alpha}
+
(1-v)^{\alpha}
-
(1-u)^{\alpha}(1-v)^{\alpha}.
\]
Note that the Joe copula is only capable of modeling positive dependence. In both copula families, larger values of $|\alpha|$ correspond to stronger dependence.

\subsection{Copula-based Markov Models} 
We focus on the \text{Weibull} distribution as the marginal distribution, which is characterized by two parameters, \( \gamma = (\lambda, k) \). Consider a time series \( \{X_t : t = 1,2,\ldots,T\} \), where \(T\)  represents the length of data. We propose a copula-based Markov chain model for a \text{Weibull} time series with a two-stage structure:
\begin{eqnarray}
	X_1, X_2, \ldots, X_{\tau} &\sim& F_{\gamma_0}
	\quad \rightarrow \quad \text{Weibull}(\lambda_0, k_0), \nonumber \\
	X_{\tau+1}, X_{\tau+2}, \ldots, X_T &\sim& F_{\gamma_1}
	\quad \rightarrow \quad \text{Weibull}(\lambda_1, k_1), \nonumber
\end{eqnarray}
where \(F_{\gamma_0}\) and \(F_{\gamma_1}\) denote \text{Weibull} distributions with parameters \(\gamma_0=(\lambda_0,k_0)\) and \(\gamma_1=(\lambda_1,k_1)\), respectively, and \(\tau\) represents a change point.

The Markov property is defined as
\[
P(X_t = x_t \mid X_{t-1} = x_{t-1}, X_{t-2} = x_{t-2}, \ldots)
=
P(X_t = x_t \mid X_{t-1} = x_{t-1}),
\quad \forall t.
\]
That is, \(X_t\) depends only on the previous observation \(X_{t-1}\), and the dependence structure between consecutive observations is modeled using copulas. To accommodate sequential data with a single change point, we specify the joint distributions of two consecutive observations as follows:
\begin{eqnarray}
	P(X_{t-1} \leq x_{t-1}, X_t \leq x_t)
	&=&
	C_{\alpha_0}
	\left(
	F_{\gamma_0}(x_{t-1}),
	F_{\gamma_0}(x_t)
	\right),
	\quad t=2,3,\ldots,\tau,
	\nonumber \\
	P(X_{\tau} \leq x_{\tau}, X_{\tau+1} \leq x_{\tau+1})
	&=&
	C_{\alpha_{01}}
	\left(
	F_{\gamma_0}(x_{\tau}),
	F_{\gamma_1}(x_{\tau+1})
	\right),
	\nonumber \\
	P(X_{t-1} \leq x_{t-1}, X_t \leq x_t)
	&=&
	C_{\alpha_1}
	\left(
	F_{\gamma_1}(x_{t-1}),
	F_{\gamma_1}(x_t)
	\right),
	\quad t=\tau+2,\ldots,T,
	\nonumber
\end{eqnarray}
where \(F_{\gamma}(\cdot)\) denotes the CDF with parameter \(\gamma\), and \(C_{\alpha}(\cdot,\cdot)\) denotes the copula function with dependence parameter \(\alpha\). Under the proposed model, observations from \(X_1\) to \(X_{\tau}\) correspond to the period before the change point, where the marginal distribution follows \(\text{Weibull}(\lambda_0,k_0)\) with dependence parameter \(\alpha_0\). Observations from \(X_{\tau+1}\) to \(X_T\) correspond to the period after the change point, where the marginal distribution follows \(\text{Weibull}(\lambda_1,k_1)\) with dependence parameter \(\alpha_1\). The dependence between \(X_{\tau}\) and \(X_{\tau+1}\) is characterized by the parameter \(\alpha_{01}\).
Accordingly, the conditional PDFs are given by
\begin{eqnarray}
	f(x_t \mid x_{t-1})
	&=&
	c_{\alpha_0}
	\left(
	F_{\gamma_0}(x_{t-1}),
	F_{\gamma_0}(x_t)
	\right)
	f_{\gamma_0}(x_t),
	\quad t=2,3,\ldots,\tau,
	\nonumber \\
	f(x_{\tau+1} \mid x_{\tau})
	&=&
	c_{\alpha_{01}}
	\left(
	F_{\gamma_0}(x_{\tau}),
	F_{\gamma_1}(x_{\tau+1})
	\right)
	f_{\gamma_1}(x_{\tau+1}),
	\nonumber \\
	f(x_t \mid x_{t-1})
	&=&
	c_{\alpha_1}
	\left(
	F_{\gamma_1}(x_{t-1}),
	F_{\gamma_1}(x_t)
	\right)
	f_{\gamma_1}(x_t),
	\quad t=\tau+2,\ldots,T,
	\nonumber
\end{eqnarray}
where \(f_{\gamma}(\cdot)\) denotes the PDF with parameter \(\gamma\), and \(c_{\alpha}(\cdot,\cdot)\) is the corresponding copula density function.

 \section{Parameter Estimation} \label{Sec: PE}
In this section, we develop the likelihood-based estimation procedure for the proposed copula-based change-point model and describe the corresponding Monte Carlo bootstrap method for interval estimation. Specifically, maximum likelihood estimation is carried out using the Newton--Raphson algorithm, while confidence intervals for the model parameters and the change point are constructed through a parametric bootstrap Monte Carlo procedure.

\subsection{Parameter Space}
The parameter space is defined as 
\begin{equation*}
   \begin{gathered}
\Theta=\Big\{(\lambda_{0},\lambda_{1},k_{0},k_{1},\alpha_{0},\alpha_{1},\tau)\mid\lambda_{0},\lambda_{1}\in(0,\infty), k_{0},k_{1}\in(0,\infty),\alpha_0, \alpha_1 \in (\alpha_{\min}, \alpha_{\max}),\\\hspace{11cm}\tau\in\{3,4,...,T-3\}\Big\}.
\end{gathered}
\end{equation*}
Here, $\alpha_{\min}$ and $\alpha_{\max}$ denote the admissible bounds for the copula dependence parameters. For example, one may choose $\alpha_{\min}=-1$ and $\alpha_{\max}=1000$, which corresponds approximately to Kendall’s $\tau$ approaching one for both the Clayton and Joe copulas.
Note that estimation of the dependence parameter $\alpha$ requires at least three time points to identify a structural change. Consequently, for the pair $(X_{\tau},X_{\tau+1})$ at the change point, the dependence parameter $\alpha_{01}$ cannot be reliably estimated. Therefore, we assume that $\alpha_{01}$ is fixed and known. In practice, setting $\alpha_{01}=2$ is a reasonable choice, as suggested by \cite{ELS2021}; see also Remark 1 in \cite{SWLEC2025}.

\subsection{Maximum Likelihood Estimation} 
The corresponding log-likelihood function is 
\begin{align}
\nonumber &\ell(\boldsymbol{\theta}) = \sum_{t=1}^{\tau} \mbox{log}(f_{\gamma_0}(x_t)) + \sum_{t=\tau+1}^T \mbox{log}(f_{\gamma_1}(x_t)) + \sum_{t=1}^{\tau-1} \mbox{log}[c_{\alpha_0}(F_{\gamma_0}(x_{t}),F_{\gamma_0}(x_{t+1}))] \\
&\qquad \quad + \mbox{log}[c_{\alpha_{01}}(F_{\gamma_0}(x_{\tau}),F_{\gamma_1}(x_{\tau+1}))] + \sum_{t=\tau+1}^{T-1}\mbox{log}[c_{\alpha_1}(F_{\gamma_1}(x_{t}),F_{\gamma_1}(x_{t+1}))]\label{loglikelihood}
\end{align}
where $\boldsymbol{\theta} = \{\lambda_{0},\lambda_{1},k_{0},k_{1},\alpha_{0},\alpha_{1},\tau \}$. Here, $\ F_{\gamma_0}(\cdot)$, $F_{\gamma_1}(\cdot)$, $f_{\gamma_0}(\cdot)$, and $f_{\gamma_1}(\cdot)$  are the distribution functions and the corresponding density functions of the \text{Weibull} distribution, respectively.

We employ the Newton-Raphson algorithm to compute the maximum likelihood estimates using \eqref{loglikelihood}. However, since the log-likelihood function is not differentiable at the change point \(\tau\), we adopt a two-stage procedure following \cite{ELS2021,SWLEC2025}. In the first stage, we fix the change-point value \(\tau\) and then apply the Newton-Raphson algorithm to estimate the remaining parameters as follows:
\[
(\hat{k}_0, \hat{k}_1, \hat{\lambda}_0, \hat{\lambda}_1, \hat{\alpha}_0, \hat{\alpha}_1) \big|_{\tau} = \mathop{\mbox{argmax}}\limits_{(k_0,k_1,\lambda_0,\lambda_1,\alpha_0,\alpha_1)}\ell(k_0, k_1, \lambda_0, \lambda_1, \alpha_0, \alpha_1 \mid \tau).
\]
Next, through the obtained $(\hat{k}_0, \hat{k}_1, \hat{\lambda}_0, \hat{\lambda}_1, \hat{\alpha}_0, \hat{\alpha}_1) \big|_{\tau} $, the maximum likelihood estimates for $\tau$ is obtained by 
\[
\hat{\tau} = \mathop{\mbox{argmax}}\limits_{\tau \in \{3,4,\ldots,T-3\}} \ell(\tau | \hat{k}_0, \hat{k}_1, \hat{\lambda}_0, \hat{\lambda}_1, \hat{\alpha}_0, \hat{\alpha}_1).
\]
To address convergence issues arising from the constrained parameters in the Newton-Raphson algorithm, we adopt the re-parameterization method suggested in \cite{macdonald2014does}.
The log-likelihood function is re-parameterized as
\[
\tilde{\ell}(K_0, K_1, \Lambda_0, \Lambda_1, A_0, A_1) = \ell(k_0, k_1, \lambda_0, \lambda_1, \alpha_0, \alpha_1),
\]
where \( K_0 = \log (k_0) \), \( K_1 = \log (k_1) \), \( \Lambda_0 = \log (\lambda_0) \), \( \Lambda_1 = \log (\lambda_1) \), \( A_0 = \log (\alpha_0 + 1) \), and \( A_1 = \log (\alpha_1 + 1) \). This transformation ensures that \( K_0, K_1, \Lambda_0, \Lambda_1, A_0, A_1 \in (-\infty, \infty) \).

 The partial derivatives are obtained via the chain rule (see the supplementary material \cite{SUN2026} for details). The Newton-Raphson algorithm proceeds as follows:
 
 \begin{enumerate}[label=Step (\arabic*):, leftmargin=*]
 	\item Set the convergence tolerance \(\varepsilon > 0\).
 	
 	\item Choose initial values \((k_0^{(0)}, k_1^{(0)}, \lambda_0^{(0)}, \lambda_1^{(0)}, \alpha_0^{(0)}, \alpha_1^{(0)})\).
 	
 	\item Transform the initial values to the unconstrained parameters:
 	\[
 	K_0^{(0)} = \log k_0^{(0)}, \quad 
 	K_1^{(0)} = \log k_1^{(0)}, \quad 
 	\Lambda_0^{(0)} = \log \lambda_0^{(0)}, \quad 
 	\Lambda_1^{(0)} = \log \lambda_1^{(0)},
 	\]
 	\[
 	A_0^{(0)} = \log(\alpha_0^{(0)} + 1), \quad 
 	A_1^{(0)} = \log(\alpha_1^{(0)} + 1).
 	\]
 	
 	\item At iteration \(k = 0,1,2,\dots\), compute the gradient vector \(G^{(k)}\) and the Hessian matrix \(H^{(k)}\) of \(\tilde{\ell}\) evaluated at \((K_0^{(k)}, K_1^{(k)}, \Lambda_0^{(k)}, \Lambda_1^{(k)}, A_0^{(k)}, A_1^{(k)})\), where
{\renewcommand{\arraystretch}{1.8}   
	\[
	H =
	\begin{bmatrix}
		\frac{\partial^2 \tilde{\ell}}{\partial K_0^2} & \frac{\partial^2 \tilde{\ell}}{\partial K_0 \partial K_1} & \frac{\partial^2 \tilde{\ell}}{\partial K_0 \partial \Lambda_0} & \frac{\partial^2 \tilde{\ell}}{\partial K_0 \partial \Lambda_1} & \frac{\partial^2 \tilde{\ell}}{\partial K_0 \partial A_0} & \frac{\partial^2 \tilde{\ell}}{\partial K_0 \partial A_1} \\
		\frac{\partial^2 \tilde{\ell}}{\partial K_1 \partial K_0} & \frac{\partial^2 \tilde{\ell}}{\partial K_1^2} & \frac{\partial^2 \tilde{\ell}}{\partial K_1 \partial \Lambda_0} & \frac{\partial^2 \tilde{\ell}}{\partial K_1 \partial \Lambda_1} & \frac{\partial^2 \tilde{\ell}}{\partial K_1 \partial A_0} & \frac{\partial^2 \tilde{\ell}}{\partial K_1 \partial A_1} \\
		\frac{\partial^2 \tilde{\ell}}{\partial \Lambda_0 \partial K_0} & \frac{\partial^2 \tilde{\ell}}{\partial \Lambda_0 \partial K_1} & \frac{\partial^2 \tilde{\ell}}{\partial \Lambda_0^2} & \frac{\partial^2 \tilde{\ell}}{\partial \Lambda_0 \partial \Lambda_1} & \frac{\partial^2 \tilde{\ell}}{\partial \Lambda_0 \partial A_0} & \frac{\partial^2 \tilde{\ell}}{\partial \Lambda_0 \partial A_1} \\
		\frac{\partial^2 \tilde{\ell}}{\partial \Lambda_1 \partial K_0} & \frac{\partial^2 \tilde{\ell}}{\partial \Lambda_1 \partial K_1} & \frac{\partial^2 \tilde{\ell}}{\partial \Lambda_1 \partial \Lambda_0} & \frac{\partial^2 \tilde{\ell}}{\partial \Lambda_1^2} & \frac{\partial^2 \tilde{\ell}}{\partial \Lambda_1 \partial A_0} & \frac{\partial^2 \tilde{\ell}}{\partial \Lambda_1 \partial A_1} \\
		\frac{\partial^2 \tilde{\ell}}{\partial A_0 \partial K_0} & \frac{\partial^2 \tilde{\ell}}{\partial A_0 \partial K_1} & \frac{\partial^2 \tilde{\ell}}{\partial A_0 \partial \Lambda_0} & \frac{\partial^2 \tilde{\ell}}{\partial A_0 \partial \Lambda_1} & \frac{\partial^2 \tilde{\ell}}{\partial A_0^2} & \frac{\partial^2 \tilde{\ell}}{\partial A_0 \partial A_1} \\
		\frac{\partial^2 \tilde{\ell}}{\partial A_1 \partial K_0} & \frac{\partial^2 \tilde{\ell}}{\partial A_1 \partial K_1} & \frac{\partial^2 \tilde{\ell}}{\partial A_1 \partial \Lambda_0} & \frac{\partial^2 \tilde{\ell}}{\partial A_1 \partial \Lambda_1} & \frac{\partial^2 \tilde{\ell}}{\partial A_1 \partial A_0} & \frac{\partial^2 \tilde{\ell}}{\partial A_1^2}
	\end{bmatrix},
	\]
}
\[
G = 
\begin{bmatrix}
	\frac{\partial \tilde{\ell}}{\partial K_0} &
	\frac{\partial \tilde{\ell}}{\partial K_1} &
	\frac{\partial \tilde{\ell}}{\partial \Lambda_0} &
	\frac{\partial \tilde{\ell}}{\partial \Lambda_1} &
	\frac{\partial \tilde{\ell}}{\partial A_0} &
	\frac{\partial \tilde{\ell}}{\partial A_1}
\end{bmatrix}^T.
\]
 Update the parameter vector:
 	\[
 	I^{(k+1)} = I^{(k)} - \left( H^{(k)} \right)^{-1} G^{(k)},
 	\]
 	where \(I^{(k)} = (K_0^{(k)}, K_1^{(k)}, \Lambda_0^{(k)}, \Lambda_1^{(k)}, A_0^{(k)}, A_1^{(k)})^\top\).
 	
 	\item Repeat until the convergence criterion is satisfied:
 	\[
 	\max \Bigl\{ |K_0^{(k+1)} - K_0^{(k)}|, \, |K_1^{(k+1)} - K_1^{(k)}|, \, |\Lambda_0^{(k+1)} - \Lambda_0^{(k)}|, 
 	\]
 	\[\hspace{4cm}
 	|\Lambda_1^{(k+1)} - \Lambda_1^{(k)}|, \, |A_0^{(k+1)} - A_0^{(k)}|, \, |A_1^{(k+1)} - A_1^{(k)}| \Bigr\} < \varepsilon.
 	\]
 	
 	\item Upon convergence, set \((\hat{K}_0, \hat{K}_1, \hat{\Lambda}_0, \hat{\Lambda}_1, \hat{A}_0, \hat{A}_1) = I^{(k+1)}\) and transform back to the original parameters if needed.
 \end{enumerate}

\subsection{Interval Estimation} 
We then construct confidence intervals (CIs) using the parametric bootstrap method \citep{tibshirani1993introduction}, which is a Monte Carlo procedure. The detailed steps are as follows:

\begin{enumerate}[label=Step (\roman*):, leftmargin=*]
	\item Fit the proposed model to the observed data $\{X_1, \dots, X_T\}$ and obtain the maximum likelihood estimators (MLEs). Denote the resulting estimates by 
	\[
	\hat{\theta} = (\hat{K}_0, \hat{K}_1, \hat{\Lambda}_0, \hat{\Lambda}_1, \hat{A}_0, \hat{A}_1).
	\]
	
	\item For each replication $b = 1, \dots, B$:
	\begin{itemize}
		\item Generate a bootstrap sample $\{X_1^{(b)}, \dots, X_T^{(b)}\}$ by Monte Carlo simulation from the fitted model using the parameter estimates $\hat{\theta}$ obtained in Step (i).
		
		\item Refit the model to the bootstrap sample $\{X_1^{(b)}, \dots, X_T^{(b)}\}$ and obtain the bootstrap estimates
		\[
		\hat{\theta}^{(b)} = \bigl(\hat{K}_0^{(b)}, \hat{K}_1^{(b)}, \hat{\Lambda}_0^{(b)}, \hat{\Lambda}_1^{(b)}, \hat{A}_0^{(b)}, \hat{A}_1^{(b)}\bigr).
		\]
	\end{itemize}
	
	\item Repeat Step (ii) $B$ times (where $B$ is a large number, e.g., $1000$ or more) to obtain $B$ independent bootstrap replications $\hat{\theta}^{(b)}$, $b=1,\dots,B$.
	
	\item For each continuous model parameter $\theta \in \{K_0, K_1, \Lambda_0, \Lambda_1, A_0, A_1\}$, construct the $100(1-\alpha)\%$ percentile bootstrap confidence interval by ordering the $B$ bootstrap estimates and taking the $\left\lceil \frac{\alpha}{2}B \right\rceil$-th and $\left\lfloor (1-\frac{\alpha}{2})B \right\rfloor$-th ordered values.

	\item Identify the set of distinct bootstrap estimates $\{a_1, \dots, a_{\tilde{N}_B}\}$, where $\tilde{N}_B \leq B$ is the number of unique values among $\{\hat{\tau}^{(b)} : b=1,\dots,B\}$.
	
	\item Arrange these distinct values in ascending order: $\{a_{(1)} < a_{(2)} < \dots < a_{(\tilde{N}_B)}\}$.
	
	\item Select the smallest contiguous subset $\{a_{(1)}, \dots, a_{(\tilde{N}_\tau)}\}$ such that at least $(1-\alpha)$ proportion of all bootstrap estimates $\hat{\tau}^{(b)}$ (i.e., at least $(1-\alpha)B$ replications) fall within $[a_{(1)}, a_{(\tilde{N}_\tau)}]$. This interval is the $100(1-\alpha)\%$ bootstrap confidence interval for $\tau$.
\end{enumerate}
 
\section{Simulation Studies}\label{Sec: Sim}
We conduct an extensive simulation study to evaluate the performance and robustness of the proposed estimation procedure. The experiments consider various sample sizes, different locations of the change point, multiple dependence structures, different strengths of serial dependence, and scenarios involving copula model misspecification.

\subsection{Simulation Setting}
We consider sample sizes $T = 100$ and $T = 250$, with change-points $\tau$ occurring at various positions within the series. Each scenario is independently replicated $R = 500$ times. In each replication, the observations $\{X_1, X_2, \dots, X_T\}$ are generated from a copula-based Markov model with a single change-point in both the marginal distributions and the dependence structure. The data before the change-point, $\{X_1, \dots, X_\tau\}$, are generated from the Weibull distribution with parameters $(k_0, \lambda_0) = (1.8, 1.2)$ and copula parameter $\alpha_0$. The data after the change-point, $\{X_{\tau+1}, \dots, X_T\}$, follow the Weibull distribution with parameters $(k_1, \lambda_1) = (2.1, 1.5)$ and copula parameter $\alpha_1$. The transition copula parameter between the two regimes is denoted by $\alpha_{01}$.

We examine two levels of dependence strength under both the Clayton and Joe copulas:
\begin{itemize}
	\item Weak to moderate dependence: $(\alpha_0, \alpha_{01}, \alpha_1) = (2, 2, 2)$, corresponding to Kendall's $\tau \approx 0.50$ (Clayton) and $\tau \approx 0.426$ (Joe).
	\item Strong dependence: $(\alpha_0, \alpha_{01}, \alpha_1) = (8, 8, 8)$, corresponding to Kendall's $\tau \approx 0.80$ (Clayton) and $\tau \approx 0.883$ (Joe).
\end{itemize}
Additionally, we investigate the effect of varying the transition parameter $\alpha_{01}$ while keeping the within-regime dependence fixed, to better understand its impact on the estimation of the change-point $\tau$ and the model parameters.

The change-point $\tau$ and all model parameters are estimated via the MLE method. To assess the performance of the estimators, we compute the following metrics, where $\theta$ denotes the true parameter value and $\hat{\theta}_i$ is the estimate from the $i$-th replication:
\[
\mathrm{RMSE}(\hat{\theta}) = \sqrt{\frac{1}{R} \sum_{i=1}^{R} (\hat{\theta}_i - \theta)^2}, \qquad
\mathrm{RE}(\hat{\theta}) = \frac{\mathrm{RMSE}(\hat{\theta})}{\theta},
\]
and the empirical average
\[
\hat{\mathrm{E}}(\hat{\theta}) = \frac{1}{R} \sum_{i=1}^{R} \hat{\theta}_i.
\]

\begin{figure}[htbp!]
    \centering
    \includegraphics[width=1\linewidth]{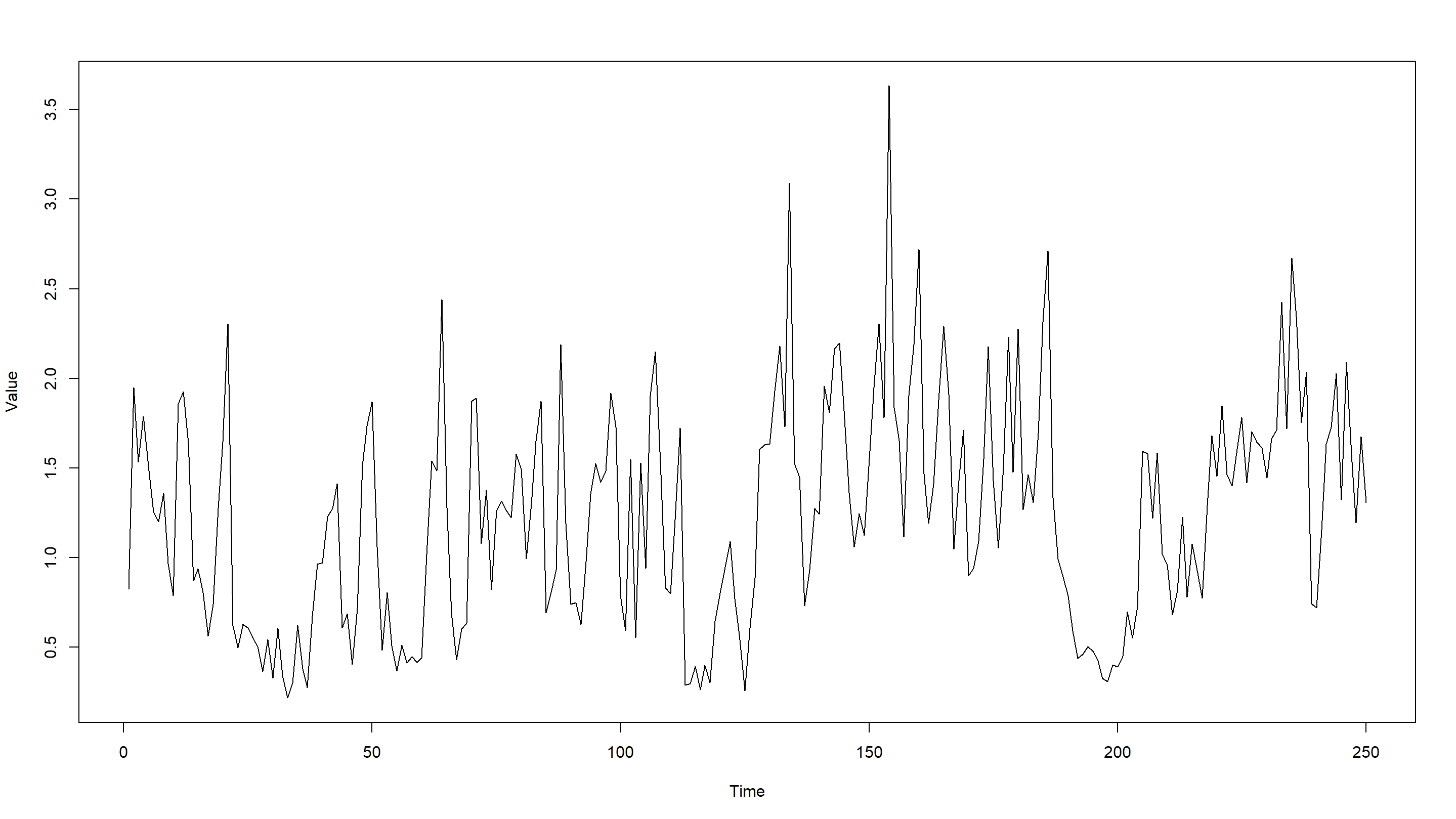}
    \caption{The time series plot of the simulated dataset.}
    \label{fig:timeseries}
\end{figure}

Figure \ref{fig:timeseries} shows a representative time series plot from the simulated data, exhibiting noticeable fluctuations, multiple oscillations, and several spikes, which indicate the presence of structural changes in both the marginal distributions and the serial dependence structure.

\subsection{Simulation Results}

From Table~\ref{tab:copulaestimation1}, we observe that the estimation performance for all parameters is generally better under weak dependence, as reflected by the smaller RMSE values. For example, in the case of the Clayton copula, the RMSE of $\hat{\tau}$ is 1.5588 (with RE equal to 0.0125) when $(\alpha_0,\alpha_{01},\alpha_1)=(2,2,2)$. In contrast, the RMSE increases slightly to 1.6125 (with RE equal to 0.0129) when $(\alpha_0,\alpha_{01},\alpha_1)=(8,8,8)$. A similar pattern is observed for the Joe copula. Specifically, when $(\alpha_0,\alpha_{01},\alpha_1)=(2,2,2)$, the RMSE of $\hat{\tau}$ is 1.1958 (with RE equal to 0.0096), whereas the RMSE increases to 1.4526 (with RE equal to 0.0116) when $(\alpha_0,\alpha_{01},\alpha_1)=(8,8,8)$.

We next consider scenarios with different change points $\tau$. Tables~\ref{tab:copulaestimation2}--\ref{tab:copulaestimation5} show that when the change occurs at an early stage, such as $\tau=25$ or $\tau=50$, the RMSE of $\hat{\tau}$ tends to be larger. In addition, the estimation of the distribution before the change point deteriorates, as reflected by the increased RMSE values. For instance, under the Clayton copula with $(\alpha_0,\alpha_{01},\alpha_1)=(2,2,2)$, the RMSE of $\hat{\tau}$ is 1.9131 (RE = 0.0765) when $\tau=25$, compared with 1.5588 (RE = 0.0125) when $\tau=125$. A similar phenomenon is observed under the Joe copula with the same parameter setting, where the RMSE decreases from 1.5232 (RE = 0.0609) for $\tau=25$ to 1.1958 (RE = 0.0096) for $\tau=125$.

The same tendency is evident when the dependence parameters are set to $(8,8,8)$. In the Clayton copula case, the RMSE of $\hat{\tau}$ is 1.8735 (RE = 0.0749) when $\tau=25$, while the RMSE values for $\tau=83$ and $\tau=125$ are 1.4491 (RE = 0.0175) and 1.6125 (RE = 0.0129), respectively. Likewise, under the Joe copula, the RMSE values are 1.4730 (RE = 0.0589), 1.3892 (RE = 0.0167), and 1.4526 (RE = 0.0116) for $\tau=25$, $\tau=83$, and $\tau=125$, respectively. Although the RMSE values do not decrease monotonically, the relative error becomes substantially smaller as $\tau$ increases. Moreover, when $\tau=25$, the parameters $k_0$, $\lambda_0$, and $\alpha_0$ exhibit poorer estimation performance, as indicated by their larger RMSE values. These results suggest that the sample sizes before and after the change point, which depend on $\tau$, play an important role in estimation accuracy. In particular, the RMSE and RE generally improve as the effective sample size increases.

Next, we investigate the case of different values of $\alpha_{01}$. Since only one pair of observations, namely $(X_{\tau},X_{\tau+1})$, is available for estimating $\alpha_{01}$, this parameter is not identifiable. Therefore, in practical applications, $\alpha_{01}$ is treated as a fixed and known quantity in the modeling framework. However, because the true value of $\alpha_{01}$ is generally unknown in practice, we conduct a sensitivity analysis to evaluate the robustness of the proposed model under possible misspecification. In this study, the assumed value used for estimation is denoted by $\alpha_{01}^{as}$.

Tables~\ref{tab:copulaestimation6}--\ref{tab:copulaestimation9} report the results of the sensitivity analysis, where the assumed dependence parameter $\alpha_{01}^{as}$ takes values 1, 2, 4, and 8, while the true parameter setting is $(\alpha_0,\alpha_{01},\alpha_1)=(2,2,2)$ or $(8,8,8)$. Overall, the estimation of the change point $\tau$ remains relatively stable and close to the true value in most scenarios. For example, under the Clayton copula with $(\alpha_0,\alpha_{01},\alpha_1)=(2,2,2)$, the RMSE of $\hat{\tau}$ is 1.5673 (RE = 0.0125) when $\alpha_{01}^{as}=1$, 1.5588 (RE = 0.0125) when $\alpha_{01}^{as}=2$, and 1.4595 (RE = 0.0117) when $\alpha_{01}^{as}=4$. However, the RMSE increases to 1.7176 (RE = 0.0137) when $\alpha_{01}^{as}=8$. Similarly, under the Joe copula with $(\alpha_0,\alpha_{01},\alpha_1)=(2,2,2)$, the RMSE of $\hat{\tau}$ equals 1.2767 (RE = 0.0102), 1.1958 (RE = 0.0096), 1.3453 (RE = 0.0108), and 1.7000 (RE = 0.0136) for $\alpha_{01}^{as}=1,2,4$, and $8$, respectively.

A similar pattern is observed when the true dependence parameters are $(8,8,8)$. Under the Clayton copula, the RMSE of $\hat{\tau}$ is 1.6032 (RE = 0.0128), 1.6101 (RE = 0.0129), 1.6425 (RE = 0.0131), and 1.6125 (RE = 0.0129) for $\alpha_{01}^{as}=1,2,4$, and $8$, respectively. Under the Joe copula, the corresponding RMSE values are 1.3341 (RE = 0.0107), 1.2922 (RE = 0.0103), 1.4248 (RE = 0.0114), and 1.4526 (RE = 0.0116). These results indicate that the estimation procedure is relatively robust to moderate misspecification of $\alpha_{01}$, although stronger assumed dependence may lead to some deterioration in estimation accuracy. Therefore, selecting $\alpha_{01}^{as}$ between 1 and 4 appears to be a reasonable practical strategy. 

Tables~\ref{tab:copulaestimation10} and \ref{tab:copulaestimation11} summarize the estimation results under different sample sizes. In general, increasing the sample size leads to smaller RMSE and RE values, indicating improved estimation accuracy and consistency. For instance, under the Clayton copula with $(\alpha_0,\alpha_{01},\alpha_1)=(2,2,2)$, increasing the sample size from 100 to 250 reduces the RMSE of $\hat{\tau}$ from 1.6912 to 1.5588, while the RE decreases from 0.0338 to 0.0125. Similarly, under the Joe copula with the same parameter setting, the RMSE of $\hat{\tau}$ decreases from 1.4142 to 1.1958, and the RE decreases from 0.0283 to 0.0096. For the stronger dependence setting $(8,8,8)$, the Clayton copula shows a reduction in RMSE from 1.9950 to 1.6125 (RE decreases from 0.0399 to 0.0129), whereas under the Joe copula the RMSE changes from 1.2961 to 1.4526 while the RE decreases from 0.0259 to 0.0116. Overall, these results support the consistency of the proposed estimators as the sample size increases.

\begin{table}[htbp!]
\centering
\caption{The results of parameter estimation under different copulas.}
\resizebox{0.8\textwidth}{!}{%
\begin{tabular}{*{7}{c}}
   \toprule
        \textbf{Copula} & \textbf{($\boldsymbol{\alpha_0}, \boldsymbol{\alpha_{01}}, \boldsymbol{\alpha_1}$)} & \textbf{Parameter} & \textbf{True value} & \textbf{$\boldsymbol{\hat{\mathrm{E}}(\cdot)}$} & \textbf{RMSE}$\boldsymbol{(\cdot)}$ & \textbf{RE}$\boldsymbol{(\cdot)}$ \\
        \hline
        Clayton & (2,2,2) & $\tau$ & 125 & 124.75 & 1.5588 & 0.0125\\
        &  & $k_0$ & 1.8 & 1.7843 & 0.2571 & 0.1428\\
        &  & $k_1$ & 2.1 & 2.2184 & 0.3616 & 0.1722\\
        &  & $\lambda_0$ & 1.2 & 1.1493 & 0.1595 & 0.1329\\
        &  & $\lambda_1$ & 1.5 & 1.5276 & 0.1635 & 0.1090\\
        &  & $\alpha_0$ & 2 & 2.1556 & 0.6248 & 0.3124\\
        &  & $\alpha_1$ & 2 & 1.9370 & 0.6095 & 0.3048\\
        \hline
        Clayton & (8,8,8) & $\tau$ & 125 & 124.66 & 1.6125 & 0.0129\\
        &  & $k_0$ & 1.8 & 1.9041 & 0.5741 & 0.3189\\
        &  & $k_1$ & 2.1 & 2.4262 & 0.7456 & 0.3550\\
        &  & $\lambda_0$ & 1.2 & 1.2047 & 0.3937 & 0.3281\\
        &  & $\lambda_1$ & 1.5 & 1.5481 & 0.2956 & 0.1971\\
        &  & $\alpha_0$ & 8 & 8.8051 & 4.4219 & 0.5527\\
        &  & $\alpha_1$ & 8 & 7.6146 & 5.3206 & 0.6651\\
        \hline
        Joe & (2,2,2) & $\tau$ & 125 & 125.11 & 1.1958 & 0.0096\\
        &  & $k_0$ & 1.8 & 1.8318 & 0.1577 & 0.0876\\
        &  & $k_1$ & 2.1 & 2.1258 & 0.1835 & 0.0874\\
        &  & $\lambda_0$ & 1.2 & 1.1928 & 0.1469 & 0.1224\\
        &  & $\lambda_1$ & 1.5 & 1.5133 & 0.1690 & 0.1127\\
        &  & $\alpha_0$ & 2 & 2.0079 & 0.4824 & 0.2412\\
        &  & $\alpha_1$ & 2 & 2.0316 & 0.4646 & 0.2323\\
        \hline
        Joe & (8,8,8) & $\tau$ & 125 & 125.31 & 1.4526 & 0.0116\\
        &  & $k_0$ & 1.8 & 1.8361 & 0.2878 & 0.1599\\
        &  & $k_1$ & 2.1 & 2.1421 & 0.2699 & 0.1285\\
        &  & $\lambda_0$ & 1.2 & 1.1656 & 0.4649 & 0.3874\\
        &  & $\lambda_1$ & 1.5 & 1.4118 & 0.3601 & 0.2401\\
        &  & $\alpha_0$ & 8 & 8.8307 & 8.2639 & 1.0330\\
        &  & $\alpha_1$ & 8 & 7.6063 & 4.1442 & 0.5180\\
        \bottomrule
\end{tabular}%
}
\label{tab:copulaestimation1}
\end{table}

\begin{table}[htbp!]
\centering
\caption{The simulation results of the Clayton copula at $\alpha=2$  for different settings of $\tau$ over various time periods.}
\resizebox{1\textwidth}{0.185\textheight}{%
\begin{tabular}{*{12}{c}}
   \toprule
        \textbf{($\boldsymbol{\alpha_0}, \boldsymbol{\alpha_{01}}, \boldsymbol{\alpha_1}$)} & \textbf{Parameter} & \textbf{True value} & \textbf{$\boldsymbol{\hat{\mathrm{E}}(\cdot)}$} & \textbf{RMSE}$\boldsymbol{(\cdot)}$ &
        \textbf{RE}$\boldsymbol{(\cdot)}$ &
        \textbf{($\boldsymbol{\alpha_0}, \boldsymbol{\alpha_{01}}, \boldsymbol{\alpha_1}$)} & \textbf{Parameter} & \textbf{True value} & \textbf{$\boldsymbol{\hat{\mathrm{E}}(\cdot)}$} & \textbf{RMSE}$\boldsymbol{(\cdot)}$ &
        \textbf{RE}$\boldsymbol{(\cdot)}$ 
        \\
        \hline
        (2,2,2) & $\tau$ & 25 & 24.18 & 1.9131& 0.0765& (2,2,2) & $\tau$ & 50 & 49.71 &1.7349 &0.0347\\
        & $k_0$ & 1.8 & 2.201 & 1.0260& 0.5700 && $k_0$ & 1.8 & 1.8826 & 0.4499 &0.2499 \\
        & $k_1$ & 2.1 & 2.1932 & 0.2925&0.1393 && $k_1$ & 2.1 & 2.1541 & 0.2877&0.1370 \\
        & $\lambda_0$ & 1.2 & 1.3400 & 1.6586 &1.3822 && $\lambda_0$ & 1.2 & 1.1553 & 0.2350 &0.1958\\
        & $\lambda_1$ & 1.5 & 1.5126 & 0.1286&0.0857 && $\lambda_1$ & 1.5 & 1.5096 & 0.1375&0.0917 \\
        & $\alpha_0$ & 2 & 2.2685 & 1.5984 &0.7992 & & $\alpha_0$ & 2 & 2.1859 & 0.9138 &0.4569\\
        & $\alpha_1$ & 2 & 1.9249 & 0.6700 &0.3350 & & $\alpha_1$ &  2 & 1.9152 & 0.6287 &0.3144 \\
        \hline
        (2,2,2) & $\tau$ & 83 & 82.53 & 1.6643 & 0.0201 &(2,2,2) & $\tau$ & 125 & 124.75 & 1.5588 &0.0125 \\
        & $k_0$ & 1.8 & 1.8052 & 0.3484 &0.1936 && $k_0$ & 1.8 & 1.7843 & 0.2571 &0.1428 \\
        & $k_1$ & 2.1 & 2.1784 & 0.2996 &0.1427 && $k_1$ & 2.1 & 2.2184 & 0.3616 &0.1722\\
        & $\lambda_0$ & 1.2 & 1.1431 & 0.1934 &0.1612 && $\lambda_0$ & 1.2 & 1.1493 & 0.1595 & 0.1329\\
        & $\lambda_1$ & 1.5 & 1.5169 & 0.1383 &0.0922 && $\lambda_1$ & 1.5 & 1.5276 & 0.1635 &0.1090\\
        & $\alpha_0$ & 2 & 2.1727 & 0.8129 &0.4065 && $\alpha_0$ & 2 & 2.1556 & 0.6248 &0.3124 \\
        & $\alpha_1$ & 2 & 1.9698 & 0.5332 &0.2666 && $\alpha_1$ & 2 & 1.9370 & 0.6095 &0.3048 \\
        \bottomrule
\end{tabular}%
}
\label{tab:copulaestimation2}
\end{table}

\begin{table}[htbp!]
\centering
\caption{The simulation results of the Joe copula at $\alpha=2$  for different settings of $\tau$ over various time periods.}
\resizebox{1.01\textwidth}{0.185\textheight}{%
\begin{tabular}{*{12}{c}}
   \toprule
        \textbf{($\boldsymbol{\alpha_0}, \boldsymbol{\alpha_{01}}, \boldsymbol{\alpha_1}$)} & \textbf{Parameter} & \textbf{True value} & \textbf{$\boldsymbol{\hat{\mathrm{E}}(\cdot)}$} & \textbf{RMSE}$\boldsymbol{(\cdot)}$ &
        \textbf{RE}$\boldsymbol{(\cdot)}$ &
        \textbf{($\boldsymbol{\alpha_0}, \boldsymbol{\alpha_{01}}, \boldsymbol{\alpha_1}$)} & \textbf{Parameter} & \textbf{True value} & \textbf{$\boldsymbol{\hat{\mathrm{E}}(\cdot)}$} & \textbf{RMSE}$\boldsymbol{(\cdot)}$ &
        \textbf{RE}$\boldsymbol{(\cdot)}$ 
        \\
        \hline
        (2,2,2) & $\tau$ & 25 & 25.32 & 1.5232& 0.0609 &(2,2,2) & $\tau$ & 50 & 50.10 & 1.5360 &0.0307\\
        & $k_0$ & 1.8 & 2.0525 & 0.5978 & 0.3321&& $k_0$ & 1.8 & 1.9062 & 0.3168 &0.1760 \\
        & $k_1$ & 2.1 & 2.1083 & 0.1235& 0.0588 && $k_1$ & 2.1 & 2.1088 & 0.1453 & 0.0692\\
        & $\lambda_0$ & 1.2 & 1.2339 & 0.4567 &0.3806 && $\lambda_0$ & 1.2 & 1.2555 & 0.3066 & 0.2555\\
        & $\lambda_1$ & 1.5 & 1.4938 & 0.1041 &0.0694 && $\lambda_1$ & 1.5 & 1.4935 & 0.1194 & 0.0796\\
        & $\alpha_0$ & 2 & 2.2730 & 2.4456 &1.2228 && $\alpha_0$ & 2 & 2.2329 & 1.2064 & 0.6032\\
        & $\alpha_1$ & 2 & 1.9907 & 0.2634 &0.1317&& $\alpha_1$ & 2 & 1.9924 & 0.2944 & 0.1472\\
        \hline
        (2,2,2) & $\tau$ & 83 & 82.98 & 1.4212& 0.0171 &(2,2,2) & $\tau$ & 125 & 125.11 & 1.1958 & 0.0096\\
        & $k_0$ & 1.8 & 1.8501 & 0.1954 &0.1086 && $k_0$ & 1.8 & 1.8318 & 0.1577&0.0876 \\
        & $k_1$ & 2.1 & 2.1100 & 0.1516& 0.0722 && $k_1$ & 2.1 & 2.1258 & 0.1835 &0.0874\\
        & $\lambda_0$ & 1.2 & 1.2051 & 0.1769 &0.1474&& $\lambda_0$ & 1.2 & 1.1928 & 0.1469 &0.1224\\
        & $\lambda_1$ & 1.5 & 1.4981 & 0.1278& 0.0852 && $\lambda_1$ & 1.5 & 1.5133 & 0.1690 &0.1127\\
        & $\alpha_0$ & 2 & 2.0473 & 0.6185 &0.3093&& $\alpha_0$ & 2 & 2.0079 & 0.4824 &0.2412\\
        & $\alpha_1$ & 2 & 2.0054 & 0.3528&0.1764 && $\alpha_1$ & 2 & 2.0316 & 0.4646 &0.2323\\
        \bottomrule
\end{tabular}%
}
\label{tab:copulaestimation3}
\end{table}

\begin{table}[htbp!]
\centering
\caption{The simulation results of the Clayton copula at $\alpha=8$  for different settings of $\tau$ over various time periods.}
\resizebox{1.01\textwidth}{0.185\textheight}{%
\begin{tabular}{*{12}{c}}
   \toprule
        \textbf{($\boldsymbol{\alpha_0}, \boldsymbol{\alpha_{01}}, \boldsymbol{\alpha_1}$)} & \textbf{Parameter} & \textbf{True value} & \textbf{$\boldsymbol{\hat{\mathrm{E}}(\cdot)}$} & \textbf{RMSE}$\boldsymbol{(\cdot)}$ &
        \textbf{RE}$\boldsymbol{(\cdot)}$ &
        \textbf{($\boldsymbol{\alpha_0}, \boldsymbol{\alpha_{01}}, \boldsymbol{\alpha_1}$)} & \textbf{Parameter} & \textbf{True value} & \textbf{$\boldsymbol{\hat{\mathrm{E}}(\cdot)}$} & \textbf{RMSE}$\boldsymbol{(\cdot)}$ &
        \textbf{RE}$\boldsymbol{(\cdot)}$ 
        \\
        \hline
        (8,8,8) & $\tau$ & 25 & 24.53 & 1.8735&0.0749 &(8,8,8) & $\tau$ & 50 & 49.67 & 1.4036 &0.0281\\
        & $k_0$ & 1.8 & 2.5857 & 2.6980&1.4989 && $k_0$ & 1.8 & 2.0309 & 1.3880 &0.7711\\
        & $k_1$ & 2.1 & 2.1081 & 0.4474&0.2130 && $k_1$ & 2.1 & 2.2328 & 0.4843 &0.2306\\
        & $\lambda_0$ & 1.2 & 1.1225 & 0.4603&0.3836 && $\lambda_0$ & 1.2 & 1.1451 & 0.4312 &0.3593\\
        & $\lambda_1$ & 1.5 & 1.5237 & 0.2851&0.1901 && $\lambda_1$ & 1.5 & 1.5317 & 0.2801 &0.1867\\
        & $\alpha_0$ & 8 & 7.0462 & 5.3786&0.6723 && $\alpha_0$ & 8 & 7.4623 & 4.3433 &0.5429\\
        & $\alpha_1$ & 8 & 9.1058 & 5.9327&0.7416 && $\alpha_1$ & 8 & 8.0903 & 3.3648 &0.4206\\
        \hline
        (8,8,8) & $\tau$ & 83 & 82.88 & 1.4491&0.0175 &(8,8,8) & $\tau$ & 125 & 124.66 & 1.6125 &0.0129\\
        & $k_0$ & 1.8 & 1.9063 & 0.8175& 0.4542&& $k_0$ & 1.8 & 1.9041 & 0.5741&0.3189 \\
        & $k_1$ & 2.1 & 2.288 & 0.6307&0.3003 && $k_1$ & 2.1 & 2.4262 & 0.7456& 0.3550 \\
        & $\lambda_0$ & 1.2 & 1.3287 & 0.6535 &0.5446&& $\lambda_0$ & 1.2 & 1.2047 & 0.3937 &0.3281\\
        & $\lambda_1$ & 1.5 & 1.5747 & 0.6389&0.4259 && $\lambda_1$ & 1.5 & 1.5481 & 0.2956 &0.1971\\
        & $\alpha_0$ & 8 & 8.0425 & 4.3691&0.5461 && $\alpha_0$ & 8 & 8.8051 & 4.4219 &0.5527\\
        & $\alpha_1$ & 8 & 8.1918 & 4.3615 &0.5452&& $\alpha_1$ & 8 & 7.6146 & 5.3206 &0.6651\\
        \bottomrule
\end{tabular}%
}
\label{tab:copulaestimation4}
\end{table}

\begin{table}[htbp!]
\centering
\caption{The simulation results of the Joe copula at $\alpha=8$  for different settings of $\tau$ over various time periods.}
\resizebox{1.01\textwidth}{0.185\textheight}{%
\begin{tabular}{*{12}{c}}
   \toprule
        \textbf{($\boldsymbol{\alpha_0}, \boldsymbol{\alpha_{01}}, \boldsymbol{\alpha_1}$)} & \textbf{Parameter} & \textbf{True value} & \textbf{$\boldsymbol{\hat{\mathrm{E}}(\cdot)}$} & \textbf{RMSE}$\boldsymbol{(\cdot)}$ &
        \textbf{RE}$\boldsymbol{(\cdot)}$ &
        \textbf{($\boldsymbol{\alpha_0}, \boldsymbol{\alpha_{01}}, \boldsymbol{\alpha_1}$)} & \textbf{Parameter} & \textbf{True value} & \textbf{$\boldsymbol{\hat{\mathrm{E}}(\cdot)}$} & \textbf{RMSE}$\boldsymbol{(\cdot)}$ &
        \textbf{RE}$\boldsymbol{(\cdot)}$ 
        \\
        \hline
        (8,8,8) & $\tau$ & 25 & 25.01 & 1.473 &0.0589& (8,8,8) & $\tau$ & 50 & 50.46 & 1.5556& 0.0311\\
        & $k_0$ & 1.8 & 2.5502 & 3.1625&1.7569 && $k_0$ & 1.8 & 2.0821 & 0.8704&0.4836 \\
        & $k_1$ & 2.1 & 2.1272 & 0.2035&0.0969 && $k_1$ & 2.1 & 2.1239 & 0.2140&0.1019 \\
        & $\lambda_0$ & 1.2 & 1.0661 & 0.5196&0.4330 && $\lambda_0$ & 1.2 & 1.1293 & 0.5330 &0.4442\\
        & $\lambda_1$ & 1.5 & 1.4543 & 0.3207&0.2138 && $\lambda_1$ & 1.5 & 1.4201 & 0.3208 &0.2139\\
        & $\alpha_0$ & 8 & 6.6829 & 7.5038 &0.9380&& $\alpha_0$ & 8 & 8.4361 & 9.0974 &1.1372\\
        & $\alpha_1$ & 8 & 7.9564 & 3.9159 &0.4895&& $\alpha_1$ & 8 & 7.5707 & 3.8232 &0.4779\\
        \hline
        (8,8,8) & $\tau$ & 83 & 83.19 & 1.3892&0.0167 &(8,8,8) & $\tau$ & 125 & 125.31 & 1.4526 &0.0116\\
        & $k_0$ & 1.8 & 1.8958 & 0.4262&0.2368 && $k_0$ & 1.8 & 1.8361 & 0.2878&0.1599 \\
        & $k_1$ & 2.1 & 2.1394 & 0.2423&0.1154 && $k_1$ & 2.1 & 2.1421 & 0.2699 &0.1285\\
        & $\lambda_0$ & 1.2 & 1.1527 & 0.4564&0.3803 && $\lambda_0$ & 1.2 & 1.1656 & 0.4649 &0.3874\\
        & $\lambda_1$ & 1.5 & 1.4172 & 0.3249 &0.2166&& $\lambda_1$ & 1.5 & 1.4118 & 0.3601 &0.2401\\
        & $\alpha_0$ & 8 & 8.7261 & 7.4880 &0.9360 && $\alpha_0$ & 8 & 8.8307 & 8.2639 &1.0330\\
        & $\alpha_1$ & 8 & 7.5329 & 3.7293 &0.4662 && $\alpha_1$ & 8 & 7.6063 & 4.1442 &0.5180\\
        \bottomrule
\end{tabular}%
}
\label{tab:copulaestimation5}
\end{table}

\begin{table}[htbp!]
\centering
\caption{The simulation results of the Clayton copula at $\alpha=2$ under different states of $\alpha_{01}^{as}$ }
\resizebox{1.01\textwidth}{0.185\textheight}{%
\begin{tabular}{*{12}{c}}
   \toprule
        \textbf{($\boldsymbol{\alpha_0}, \boldsymbol{\alpha_{01}^{as}}, \boldsymbol{\alpha_1}$)} & \textbf{Parameter} & \textbf{True value} & \textbf{$\boldsymbol{\hat{\mathrm{E}}(\cdot)}$} & \textbf{RMSE}$\boldsymbol{(\cdot)}$ &
        \textbf{RE}$\boldsymbol{(\cdot)}$ &
        \textbf{($\boldsymbol{\alpha_0}, \boldsymbol{\alpha_{01}^{as}}, \boldsymbol{\alpha_1}$)} & \textbf{Parameter} & \textbf{True value} & \textbf{$\boldsymbol{\hat{\mathrm{E}}(\cdot)}$} & \textbf{RMSE}$\boldsymbol{(\cdot)}$&
        \textbf{RE}$\boldsymbol{(\cdot)}$ 
        \\
        \hline
        (2,1,2) & $\tau$ & 125 & 124.78 & 1.5673& 0.0125 &(2,2,2) & $\tau$ & 125 & 124.75 & 1.5588 &0.0125\\
        & $k_0$ & 1.8 & 1.7924 & 0.2460& 0.1367 && $k_0$ & 1.8 & 1.7843 & 0.2571&0.1428 \\
        & $k_1$ & 2.1 & 2.2034 & 0.3548 &0.1690 && $k_1$ & 2.1 & 2.2184 & 0.3616&0.1722 \\
        & $\lambda_0$ & 1.2 & 1.1553 & 0.1517 & 0.1264&& $\lambda_0$ & 1.2 & 1.1493 & 0.1595 &0.1329\\
        & $\lambda_1$ & 1.5 & 1.5199 & 0.1599&0.1066 && $\lambda_1$ & 1.5 & 1.5276 & 0.1635 & 0.1090\\
        & $\alpha_0$ & 2 & 2.1364 & 0.5966 &0.2983 && $\alpha_0$ & 2 & 2.1556 & 0.6248  &0.3124\\
        & $\alpha_1$ & 2 & 1.9580 & 0.5874 &0.2937 && $\alpha_1$ & 2 & 1.9370 & 0.6095 & 0.3048\\
        \hline
        (2,4,2) & $\tau$ & 125 & 124.83 & 1.4595& 0.0117 &(2,8,2) & $\tau$ & 125 & 124.79 & 1.7176  &0.0137\\
        & $k_0$ & 1.8 & 1.7757 & 0.2689&0.1494 && $k_0$ & 1.8 & 1.7807 & 0.2707 &0.1504 \\
        & $k_1$ & 2.1 & 2.2324 & 0.3634 & 0.1730&& $k_1$ & 2.1 & 2.2408 & 0.3811 &0.1815 \\
        & $\lambda_0$ & 1.2 & 1.1458 & 0.1673 &0.1394 && $\lambda_0$ & 1.2 & 1.1491 & 0.1654  &0.1378\\
        & $\lambda_1$ & 1.5 & 1.5343 & 0.1646 &0.1097 && $\lambda_1$ & 1.5 & 1.5369 & 0.1768  &0.1179\\
        & $\alpha_0$ & 2 & 2.1859 & 0.7037 &0.3519&& $\alpha_0$ & 2 & 2.1652 & 0.6487 & 0.3244\\
        & $\alpha_1$ & 2 & 1.9142 & 0.6134 &0.3067 && $\alpha_1$ & 2 & 1.9094 & 0.6549 & 0.3275\\
        \bottomrule
\end{tabular}%
}
\label{tab:copulaestimation6}
\end{table}

\begin{table}[htbp!]
\centering
\caption{The simulation results of the Joe copula at $\alpha=2$ under different states of $\alpha_{01}^{as}$ }
\resizebox{1.01\textwidth}{0.185\textheight}{%
\begin{tabular}{*{12}{c}}
   \toprule
        \textbf{($\boldsymbol{\alpha_0}, \boldsymbol{\alpha_{01}^{as}}, \boldsymbol{\alpha_1}$)} & \textbf{Parameter} & \textbf{True value} & \textbf{$\boldsymbol{\hat{\mathrm{E}}(\cdot)}$} & \textbf{RMSE}$\boldsymbol{(\cdot)}$ &
        \textbf{RE}$\boldsymbol{(\cdot)}$ &
        \textbf{($\boldsymbol{\alpha_0}, \boldsymbol{\alpha_{01}^{as}}, \boldsymbol{\alpha_1}$)} & \textbf{Parameter} & \textbf{True value} & \textbf{$\boldsymbol{\hat{\mathrm{E}}(\cdot)}$} & \textbf{RMSE}$\boldsymbol{(\cdot)}$&
        \textbf{RE}$\boldsymbol{(\cdot)}$ 
        \\
        \hline
        (2,1,2) & $\tau$ & 125 & 125.15 & 1.2767&0.0102 &(2,2,2) & $\tau$ & 125 & 125.11 & 1.1958  &0.0096\\
        & $k_0$ & 1.8 & 1.8276 & 0.1559 &0.0866 && $k_0$ & 1.8 & 1.8318 & 0.1577 &0.0876 \\
        & $k_1$ & 2.1 & 2.1278 & 0.1831 & 0.0872 && $k_1$ & 2.1 & 2.1258 & 0.1835 &0.0874 \\
        & $\lambda_0$ & 1.2 & 1.1978 & 0.1322 &0.1102 && $\lambda_0$ & 1.2 & 1.1928 & 0.1469  &0.1224\\
        & $\lambda_1$ & 1.5 & 1.4923 & 0.1417 &0.0945 && $\lambda_1$ & 1.5 & 1.5133 & 0.1690  &0.1127\\
        & $\alpha_0$ & 2 & 2.0191 & 0.4121 &0.2061 && $\alpha_0$ & 2 & 2.0079 & 0.4824 & 0.2412\\
        & $\alpha_1$ & 2 & 1.9797 & 0.3329 &0.1665 && $\alpha_1$ & 2 & 2.0316 & 0.4646 & 0.2323\\
        \hline
        (2,4,2) & $\tau$ & 125 & 125.19 & 1.3453 & 0.0108&(2,8,2) & $\tau$ & 125 & 125.43 & 1.7000 & 0.0136\\
        & $k_0$ & 1.8 & 1.8330 & 0.1584 &0.0880 && $k_0$ & 1.8 & 1.8368 & 0.1572 &0.0873 \\
        & $k_1$ & 2.1 & 2.1289 & 0.1845 & 0.0879 && $k_1$ & 2.1 & 2.1215 & 0.1775 &0.0845  \\
        & $\lambda_0$ & 1.2 & 1.2034 & 0.1745 &0.1454 && $\lambda_0$ & 1.2 & 1.1871 & 0.1357 & 0.1131\\
        & $\lambda_1$ & 1.5 & 1.5021 & 0.1976 & 0.1317&& $\lambda_1$ & 1.5 & 1.5142 & 0.2324 & 0.1549\\
        & $\alpha_0$ & 2 & 2.0293 & 0.5949 &0.2975 && $\alpha_0$ & 2 & 1.9711 & 0.4250 & 0.2125\\
        & $\alpha_1$ & 2 & 2.0076 & 0.5675 & 0.2838&& $\alpha_1$ & 2 & 2.0601 & 0.8009  & 0.4005\\
        \bottomrule
\end{tabular}%
}
\label{tab:copulaestimation7}
\end{table}

\begin{table}[htbp!]
\centering
\caption{The simulation results of the Clayton copula at $\alpha=8$ under different states of $\alpha_{01}^{as}$ }
\resizebox{1.01\textwidth}{0.185\textheight}{%
\begin{tabular}{*{12}{c}}
   \toprule
        \textbf{($\boldsymbol{\alpha_0}, \boldsymbol{\alpha_{01}^{as}}, \boldsymbol{\alpha_1}$)} & \textbf{Parameter} & \textbf{True value} & \textbf{$\boldsymbol{\hat{\mathrm{E}}(\cdot)}$} & \textbf{RMSE}$\boldsymbol{(\cdot)}$ &
        \textbf{RE}$\boldsymbol{(\cdot)}$&
        \textbf{($\boldsymbol{\alpha_0}, \boldsymbol{\alpha_{01}^{as}}, \boldsymbol{\alpha_1}$)} & \textbf{Parameter} & \textbf{True value} & \textbf{$\boldsymbol{\hat{\mathrm{E}}(\cdot)}$} & \textbf{RMSE}$\boldsymbol{(\cdot)}$&
        \textbf{RE}$\boldsymbol{(\cdot)}$
        \\
        \hline
        (8,1,8) & $\tau$ & 125 & 124.68 & 1.6032 & 0.0128&(8,2,8) & $\tau$ & 125 & 124.63 & 1.6101 & 0.0129\\
        & $k_0$ & 1.8 & 1.9085 & 0.5586& 0.3103 && $k_0$ & 1.8 & 1.9022 & 0.5985& 0.3325 \\
        & $k_1$ & 2.1 & 2.3786 & 0.7517& 0.358 && $k_1$ & 2.1 & 2.4011 & 0.7429& 0.3538 \\
        & $\lambda_0$ & 1.2 & 1.1884 & 0.3307& 0.2756 && $\lambda_0$ & 1.2 & 1.2264 & 0.3931 &0.3276 \\
        & $\lambda_1$ & 1.5 & 1.5048 & 0.282& 0.1880 && $\lambda_1$ & 1.5 & 1.5430 & 0.3459  & 0.2306\\
        & $\alpha_0$ & 8 & 8.7792 & 4.6308& 0.5789 && $\alpha_0$ & 8 & 8.7630 & 4.3912 & 0.5489\\
        & $\alpha_1$ & 8 & 7.8808 & 3.9276 & 0.4910&& $\alpha_1$ & 8 & 7.6166 & 3.5148 & 0.4394\\
        \hline
        (8,4,8) & $\tau$ & 125 & 124.56 & 1.6425& 0.0131 &(8,8,8) & $\tau$ & 125 & 124.66 & 1.6125 &0.0129\\
        & $k_0$ & 1.8 & 1.9133 & 0.6251&0.3473 && $k_0$ & 1.8 & 1.9041 & 0.5741& 0.3189 \\
        & $k_1$ & 2.1 & 2.3987 & 0.6781&0.3229 && $k_1$ & 2.1 & 2.4262 & 0.7456 & 0.3550\\
        & $\lambda_0$ & 1.2 & 1.2208 & 0.4050&0.3375 && $\lambda_0$ & 1.2 & 1.2047 & 0.3937 & 0.3281\\
        & $\lambda_1$ & 1.5 & 1.5264 & 0.2758&0.1839 && $\lambda_1$ & 1.5 & 1.5481 & 0.2956 & 0.1971\\
        & $\alpha_0$ & 8 & 8.8334 & 4.4737& 0.5592 && $\alpha_0$ & 8 & 8.8051 & 4.4219 & 0.5527\\
        & $\alpha_1$ & 8 & 7.6232 & 4.6115& 0.5764 && $\alpha_1$ & 8 & 7.6146 & 5.3206 &0.6651 \\
        \bottomrule
\end{tabular}%
}
\label{tab:copulaestimation8}
\end{table}

\begin{table}[htbp!]
\centering
\caption{The simulation results of the Joe copula at $\alpha=8$  under different states of $\alpha_{01}^{as}$ }
\resizebox{1.01\textwidth}{0.185\textheight}{%
\begin{tabular}{*{12}{c}}
   \toprule
        \textbf{($\boldsymbol{\alpha_0}, \boldsymbol{\alpha_{01}^{as}}, \boldsymbol{\alpha_1}$)} & \textbf{Parameter} & \textbf{True value} & \textbf{$\boldsymbol{\hat{\mathrm{E}}(\cdot)}$} & \textbf{RMSE}$\boldsymbol{(\cdot)}$ &
        \textbf{RE}$\boldsymbol{(\cdot)}$&
        \textbf{($\boldsymbol{\alpha_0}, \boldsymbol{\alpha_{01}^{as}}, \boldsymbol{\alpha_1}$)} & \textbf{Parameter} & \textbf{True value} & \textbf{$\boldsymbol{\hat{\mathrm{E}}(\cdot)}$} & \textbf{RMSE}$\boldsymbol{(\cdot)}$&
        \textbf{RE}$\boldsymbol{(\cdot)}$
        \\
        \hline
        (8,1,8) & $\tau$ & 125 & 125.14 & 1.3341&0.0107 &(8,2,8) & $\tau$ & 125 & 125.11 & 1.2922 &0.0103\\
        & $k_0$ & 1.8 & 1.8345 & 0.2872& 0.1596 && $k_0$ & 1.8 & 1.8366 & 0.2818 &0.1566 \\
        & $k_1$ & 2.1 & 2.1417 & 0.2625&0.1250 && $k_1$ & 2.1 & 2.1415 & 0.2639 &0.1257 \\
        & $\lambda_0$ & 1.2 & 1.2062 & 0.3846&0.3205 && $\lambda_0$ & 1.2 & 1.1576 & 0.3192 & 0.2660\\
        & $\lambda_1$ & 1.5 & 1.4994 & 0.3976&0.2651 && $\lambda_1$ & 1.5 & 1.4390 & 0.3438 & 0.2292\\
        & $\alpha_0$ & 8 & 9.2725 & 7.0972&0.8872 && $\alpha_0$ & 8 & 8.4427 & 5.1568 & 0.6446\\
        & $\alpha_1$ & 8 & 8.7075 & 5.1904&0.6488 && $\alpha_1$ & 8 & 7.8928 & 4.3075 &0.5384\\
        \hline
        (8,4,8) & $\tau$ & 125 & 125.21 & 1.4248&0.0114 &(8,8,8) & $\tau$ & 125 & 125.31 & 1.4526 & 0.0116\\
        & $k_0$ & 1.8 & 1.8328 & 0.2889& 0.1605 && $k_0$ & 1.8 & 1.8361 & 0.2878&0.1599 \\
        & $k_1$ & 2.1 & 2.1434 & 0.2661& 0.1267 && $k_1$ & 2.1 & 2.1421 & 0.2699 & 0.1285\\
        & $\lambda_0$ & 1.2 & 1.1825 & 0.3999&0.3333 && $\lambda_0$ & 1.2 & 1.1656 & 0.4649  & 0.3874\\
        & $\lambda_1$ & 1.5 & 1.4237 & 0.3226&0.2151 && $\lambda_1$ & 1.5 & 1.4118 & 0.3601 & 0.2401\\
        & $\alpha_0$ & 8 & 8.9355 & 6.3297&0.7912 && $\alpha_0$ & 8 & 8.8307 & 8.2639 &1.0330\\
        & $\alpha_1$ & 8 & 7.6546 & 3.7183&0.4648 && $\alpha_1$ & 8 & 7.6063 & 4.1442& 0.5180\\
        \bottomrule
\end{tabular}%
}
\label{tab:copulaestimation9}
\end{table}

\begin{table}[htbp!]
\centering
\caption{The simulation results of the Clayton copula under different sample sizes.}
\resizebox{1.01\textwidth}{0.185\textheight}{%
\begin{tabular}{*{14}{c}}
   \toprule
        \textbf{($\boldsymbol{\alpha_0}, \boldsymbol{\alpha_{01}}, \boldsymbol{\alpha_1}$)} & \textbf{Sample size} & \textbf{Parameter} & \textbf{True value} & \textbf{$\boldsymbol{\hat{\mathrm{E}}(\cdot)}$} & \textbf{RMSE}$\boldsymbol{(\cdot)}$ &\textbf{RE}$\boldsymbol{(\cdot)}$&\textbf{($\boldsymbol{\alpha_0}, \boldsymbol{\alpha_{01}}, \boldsymbol{\alpha_1}$)} & \textbf{Sample size} & \textbf{Parameter} & \textbf{True value} & \textbf{$\boldsymbol{\hat{\mathrm{E}}(\cdot)}$} & \textbf{RMSE}$\boldsymbol{(\cdot)}$ &\textbf{RE}$\boldsymbol{(\cdot)}$
        \\
        \hline
        (2,2,2) & 100 & $\tau$ & 50 & 49.68 & 1.6912&0.0338 &(2,2,2) & 250 & $\tau$ & 125 & 124.75 & 1.5588 &0.0125\\
        & & $k_0$ & 1.8 & 1.9881 & 0.5066&0.2814 && & $k_0$ & 1.8 & 1.7843 & 0.2571 & 0.1428\\
        & & $k_1$ & 2.1 & 2.3163 & 0.5434& 0.2588 && & $k_1$ & 2.1 & 2.2184 & 0.3616 & 0.1722\\
        & & $\lambda_0$ & 1.2 & 1.1741 & 0.2529&0.2108 && & $\lambda_0$ & 1.2 & 1.1493 & 0.1595 & 0.1329\\
        & & $\lambda_1$ & 1.5 & 1.5300 & 0.2341&0.1561 && & $\lambda_1$ & 1.5 & 1.5276 & 0.1635  & 0.1090\\
        & & $\alpha_0$ & 2 & 1.9745 & 0.9156& 0.4575 && & $\alpha_0$ & 2 & 2.1556 & 0.6248 & 0.3124\\
        & & $\alpha_1$ & 2 & 1.9308 & 0.8456& 0.4228 && & $\alpha_1$ & 2 & 1.9370 & 0.6095 & 0.3048\\
        \hline
        (8,8,8) & 100 & $\tau$ & 50 & 49.63 & 1.995 & 0.0399 &(8,8,8) & 250 & $\tau$ & 125 & 124.66 & 1.6125 & 0.0129\\
        & & $k_0$ & 1.8 & 2.7239 & 1.4117& 0.7843 && & $k_0$ & 1.8 & 1.9041 & 0.5741 & 0.3189\\
        & & $k_1$ & 2.1 & 2.8480 & 1.3531& 0.6443 && & $k_1$ & 2.1 & 2.4262 & 0.7456 & 0.3550\\
        & & $\lambda_0$ & 1.2 & 1.3053 & 0.5542& 0.4618 && & $\lambda_0$ & 1.2 & 1.2047 & 0.3937 & 0.3281\\
        & & $\lambda_1$ & 1.5 & 1.5202 & 0.4569& 0.3046 && & $\lambda_1$ & 1.5 & 1.5481 & 0.2956& 0.1971 \\
        & & $\alpha_0$ & 8 & 6.7081 & 4.9274& 0.6159 && & $\alpha_0$ & 8 & 8.8051 & 4.4219 & 0.5527\\
        & & $\alpha_1$ & 8 & 7.3796 & 7.1606& 0.8951 && & $\alpha_1$ & 8 & 7.6146 & 5.3206 & 0.6651\\
        \bottomrule
\end{tabular}%
}
\label{tab:copulaestimation10}
\end{table}

\begin{table}[htbp!]
\centering
\caption{The simulation results of the Joe copula under different sample sizes.}
\resizebox{1.01\textwidth}{0.185\textheight}{%
\begin{tabular}{*{14}{c}}
   \toprule
        \textbf{($\boldsymbol{\alpha_0}, \boldsymbol{\alpha_{01}}, \boldsymbol{\alpha_1}$)} & \textbf{Sample size} & \textbf{Parameter} & \textbf{True value} & \textbf{$\boldsymbol{\hat{\mathrm{E}}(\cdot)}$} & \textbf{RMSE}$\boldsymbol{(\cdot)}$ &\textbf{RE}$\boldsymbol{(\cdot)}$&\textbf{($\boldsymbol{\alpha_0}, \boldsymbol{\alpha_{01}}, \boldsymbol{\alpha_1}$)} & \textbf{Sample size} & \textbf{Parameter} & \textbf{True value} & \textbf{$\boldsymbol{\hat{\mathrm{E}}(\cdot)}$} & \textbf{RMSE}$\boldsymbol{(\cdot)}$ &\textbf{RE}$\boldsymbol{(\cdot)}$
        \\
        \hline
        (2,2,2) & 100 & $\tau$ & 50 & 50.16 & 1.4142&0.0283 &(2,2,2) & 250 & $\tau$ & 125 & 125.11 & 1.1958 & 0.0096\\
        & & $k_0$ & 1.8 & 1.8655 & 0.2684& 0.1491 && & $k_0$ & 1.8 & 1.8318 & 0.1577 &0.0876 \\
        & & $k_1$ & 2.1 & 2.2448 & 0.3876& 0.1846 && & $k_1$ & 2.1 & 2.1258 & 0.1835 & 0.0874\\
        & & $\lambda_0$ & 1.2 & 1.2356 & 0.2681& 0.2234 && & $\lambda_0$ & 1.2 & 1.1928 & 0.1469 & 0.1224\\
        & & $\lambda_1$ & 1.5 & 1.5231 & 0.2622& 0.1748 & & & $\lambda_1$ & 1.5 & 1.5133 & 0.1690 &0.1127\\
        & & $\alpha_0$ & 2 & 2.1153 & 0.8516& 0.4258 && & $\alpha_0$ & 2 & 2.0079 & 0.4824 & 0.2412\\
        & & $\alpha_1$ & 2 & 1.9765 & 0.7361& 0.3681 && & $\alpha_1$ & 2 & 2.0316 & 0.4646 & 0.2323\\
        \hline
        (8,8,8) & 100 & $\tau$ & 50 & 50.14 & 1.2961& 0.0259 &(8,8,8) & 250 & $\tau$ & 125 & 125.31 & 1.4526 & 0.0116\\
        & & $k_0$ & 1.8 & 2.0247 & 0.6827&0.3818 & && $k_0$ & 1.8 & 1.8361 & 0.2878& 0.1599\\
        & & $k_1$ & 2.1 & 2.5443 & 0.9949& 0.4738 & && $k_1$ & 2.1 & 2.1421 & 0.2699 & 0.1285\\
        & & $\lambda_0$ & 1.2 & 1.1402 & 0.4588& 0.3823 & && $\lambda_0$ & 1.2 & 1.1656 & 0.4649& 0.3874\\
        & & $\lambda_1$ & 1.5 & 1.4374 & 0.4776 & 0.3184&& & $\lambda_1$ & 1.5 & 1.4118 & 0.3601& 0.2401 \\
        & & $\alpha_0$ & 8 & 7.9298 & 6.5711& 0.8214 && & $\alpha_0$ & 8 & 8.8307 & 8.2639 & 1.0330\\
        & & $\alpha_1$ & 8 & 7.2307 & 5.2488& 0.6561 & && $\alpha_1$ & 8 & 7.6063 & 4.1442& 0.5180\\
        \bottomrule
\end{tabular}%
}
\label{tab:copulaestimation11}
\end{table}

\subsection{Model misspecification test}

In this section, we investigate the effects of copula misspecification through a cross-comparison study between the Clayton and Joe copulas. Specifically, we first generate data using either the Clayton copula or the Joe copula, and then fit both copula models to the generated data in order to evaluate the impact of model misspecification. Table~\ref{tab:copulaestimation12} summarizes the estimation results together with the corresponding RMSE and RE values.

Under the correctly specified models, the parameter estimates are generally close to their true values, with relatively small RMSE and RE values. For example, under the Clayton copula, the RMSE and RE of $\hat{\tau}$ are 1.5588 and 0.0125, respectively, while under the Joe copula they are 1.1958 and 0.0096. These results indicate accurate and stable estimation when the copula model is correctly specified.

Under model misspecification, the RMSE values for $\hat{\tau}$ increase, although the corresponding RE values remain relatively small. For instance, when data generated from the Joe copula are analyzed using the Clayton copula model, the RMSE and RE of $\hat{\tau}$ become 1.9287 and 0.0154, respectively. Similarly, when data generated from the Clayton copula are fitted using the Joe copula model, the RMSE and RE increase to 3.5846 and 0.0287, respectively. Although the RMSE in the latter case is noticeably larger than that obtained under the correctly specified models, the associated RE remains below 0.03. Moreover, the estimated values of $\hat{\tau}$ remain close to the true change point value of 125 in all cases. These findings suggest that the proposed procedure remains reasonably robust to copula misspecification in estimating the change point.

On the other hand, the dependence parameters $\alpha_0$ and $\alpha_1$, which characterize the tail dependence structure of the copula, are more sensitive to model misspecification. In the misspecified settings, the RE values for these parameters increase substantially. For example, when the Joe copula data are fitted using the Clayton copula model, the RE values for $\alpha_0^C$ and $\alpha_1^C$ are 0.7715 and 0.7382, respectively. Likewise, when the Clayton copula data are analyzed using the Joe copula model, the RE values for $\alpha_0^J$ and $\alpha_1^J$ increase further to 0.9770 and 1.2028. These results indicate that the tail dependence parameters are considerably more affected by copula misspecification than the change point estimator itself.

Overall, the results demonstrate that the proposed method provides relatively stable estimation for the change point $\tau$ even under copula misspecification, as evidenced by the consistently small RE values for $\hat{\tau}$. However, the estimation of the dependence parameters associated with the copula tail structure becomes less reliable when the copula family is misspecified. This highlights the importance of selecting an appropriate copula model when the primary interest lies in accurately capturing tail dependence behavior.

\begin{table}[htbp!]
\centering
\caption{The simulation Results for cross comparison Clayton and Joe copulas.}
\resizebox{0.9\textwidth}{!}{%
\begin{tabular}{*{8}{c}}
   \toprule
         \textbf{($\boldsymbol{\alpha_0}, \boldsymbol{\alpha_{01}}, \boldsymbol{\alpha_1}$)} & \textbf{True model} & \textbf{Proposed model}  & \textbf{Parameter} & \textbf{True value} & \textbf{$\boldsymbol{\hat{\mathrm{E}}(\cdot)}$} & \textbf{RMSE}$\boldsymbol{(\cdot)}$ &
         \textbf{RE}$\boldsymbol{(\cdot)}$\\
        \hline
         (2,2,2) & Clayton& Clayton& $\tau$ & 125 & 124.75 & 1.5588 & 0.0125 \\
        & & & $k_0$ & 1.8 & 1.7843 & 0.2571 & 0.1428\\
        & & & $k_1$ & 2.1 & 2.2184 & 0.3616 & 0.1722\\
        & & & $\lambda_0$ & 1.2 & 1.1493 & 0.1595 & 0.1329\\
        & & & $\lambda_1$ & 1.5 & 1.5276 & 0.1635 & 0.1090\\
        & & & $\alpha_0^C$ & 2 & 2.1556 & 0.6248 & 0.3124\\
        & & & $\alpha_1^C$ & 2 & 1.9370 & 0.6095 & 0.3048\\
        \hline
        (2,2,2) & Joe & Clayton&$\tau$ & 125 & 124.64 & 1.9287 & 0.0154 \\
        & & & $k_0$ & 1.8 & 1.9237 & 0.2597 & 0.1443\\
        & & & $k_1$ & 2.1 & 2.2577 & 0.3248 & 0.1547\\
        & & & $\lambda_0$ & 1.2 & 1.2438 & 0.1810 & 0.1508\\
        & & & $\lambda_1$ & 1.5 & 1.5956 & 0.2441 & 0.1627\\
        & & & $\alpha_0^C$ & 2 & 1.4784 & 1.5429 & 0.7715\\
        & & & $\alpha_1^C$ & 2 & 1.5503 & 1.4763 & 0.7382\\
        \hline
        (2,2,2) & Joe & Joe& $\tau$ & 125 & 125.11 & 1.1958 & 0.0096 \\
        & & & $k_0$ & 1.8 & 1.8318 & 0.1577 & 0.0876\\
        & & & $k_1$ & 2.1 & 2.1258 & 0.1835 & 0.0874\\
        & & & $\lambda_0$ & 1.2 & 1.1928 & 0.1469 & 0.1224\\
        & & & $\lambda_1$ & 1.5 & 1.5133 & 0.1690 & 0.1127\\
        & & & $\alpha_0^J$ & 2 & 2.0079 & 0.4824 & 0.2412\\
        & & & $\alpha_1^J$ & 2 & 2.0316 & 0.4646 & 0.2323\\
        \hline
        (2,2,2) & Clayton & Joe & $\tau$ & 125 & 124.61 & 3.5846 & 0.0287 \\
        & & & $k_0$ & 1.8 & 1.9551 & 0.2525 & 0.1403\\
        & & & $k_1$ & 2.1 & 2.3605 & 0.4039 & 0.1923\\
        & & & $\lambda_0$ & 1.2 & 1.3727 & 0.2075 & 0.1729\\
        & & & $\lambda_1$ & 1.5 & 1.7512 & 0.2609 & 0.1739\\
        & & & $\alpha_0^J$ & 2 & 2.6941 & 1.9539 & 0.9770\\
        & & & $\alpha_1^J$ & 2 & 3.1591 & 2.4055 & 1.2028\\
        \bottomrule
\end{tabular}%
}
\label{tab:copulaestimation12}
\end{table}

\section{Empirical Study}\label{Sec: Em}
In this section, we consider two datasets for detecting structural change and then identifying the change point. 

\subsection{VIX Analysis}\label{VIX}

In this section, we investigate the structural change in the financial market during the global COVID-19 outbreak in 2020. To measure market uncertainty, we consider the VIX index, which is a widely used indicator of the market's expectation of the 30-day volatility of the S\&P 500 index. The study period ranges from September 1, 2019, to August 31, 2020, covering the onset and early development of the COVID-19 pandemic. The data set was obtained from Yahoo Finance (\texttt{finance.yahoo.com}). We apply the proposed copula-based Markov models with both the Clayton and Joe copulas, and the corresponding maximum likelihood estimators are computed using the Newton--Raphson (NR) method. Model selection is performed using the AIC, where the preferred model is chosen by minimizing the AIC value.

\begin{figure}[htbp!]
    \centering
    \includegraphics[width=0.7\linewidth]{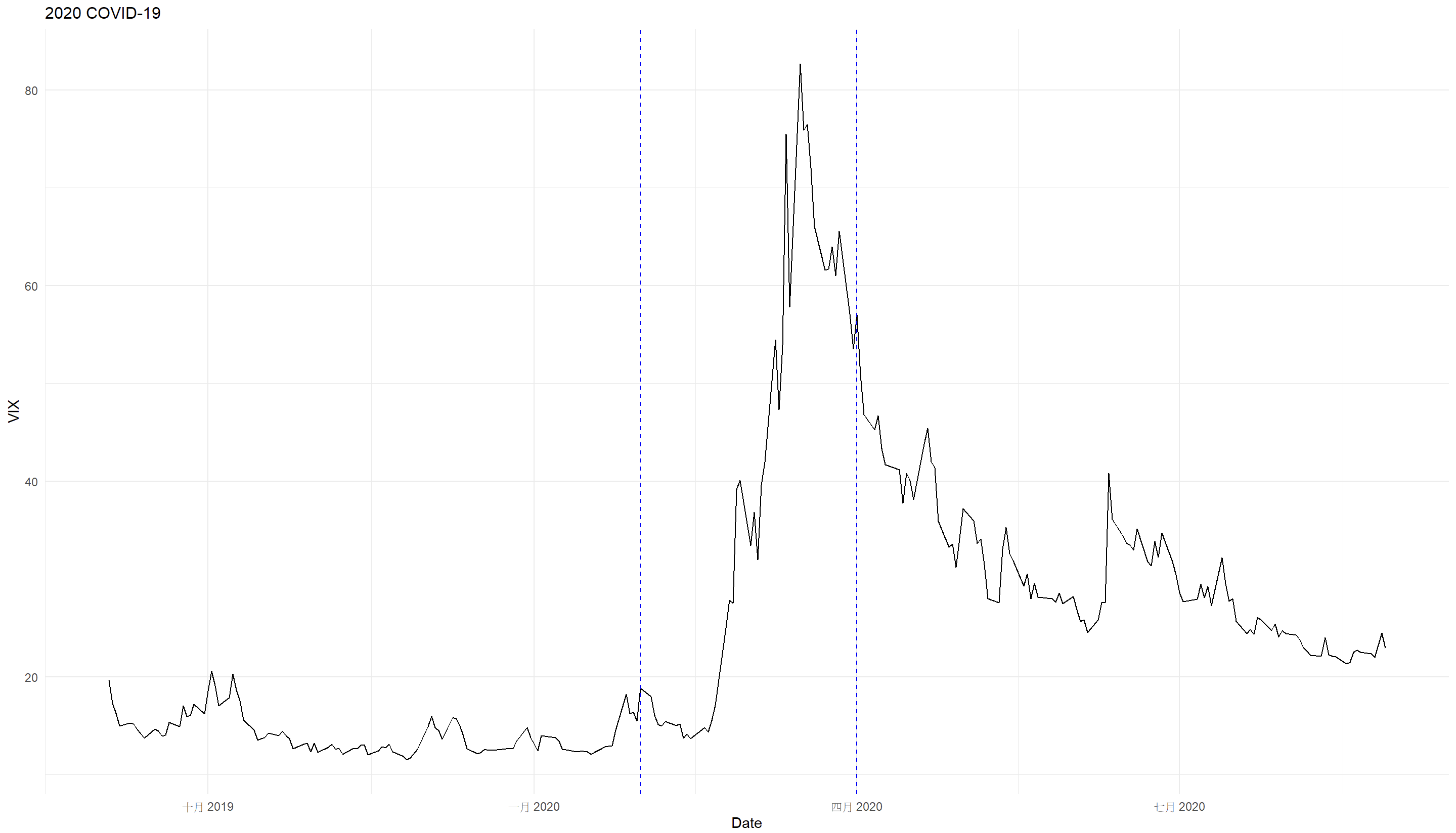}
    \caption{The VIX index from 2019/09/01 to 2020/08/31 during the 2020 COVID-19.}
    \label{fig:VIXindex}
\end{figure}

A structural change in the VIX series is visually evident around March 2020 due to the dramatic increase in market volatility; see Figure~\ref{fig:VIXindex}. The VIX data contain 251 observations during the study period. Figure~\ref{fig:VIXindex} suggests that the change point may occur between time points 105 and 147. During this period, the World Health Organization (WHO) declared the COVID-19 outbreak a ``Public Health Emergency of International Concern'' on January 30, 2020, followed by rapid global market reactions. Consequently, a substantial structural change in the volatility dynamics is expected.

Tables~\ref{tab:copulaestimation13} and \ref{tab:copulaestimation14} summarize the parameter estimates, estimated change points, and AIC values. The estimated change points remain relatively stable under different assumed values of $\alpha_{01}^{as}$, which is consistent with the findings from the simulation studies. Based on the minimum AIC criterion, the estimated change point is $\hat{\tau}=121$ under the Clayton copula model and $\hat{\tau}=120$ under the Joe copula model. Furthermore, the confidence interval for the change point under the Clayton copula is $(119,124)$, while the corresponding interval under the Joe copula is $(116,123)$. In addition, since the confidence intervals for the distribution parameters before and after the change point do not overlap, the results indicate that the structural change is jointly driven by shifts in these parameters.

Although the estimated change points are similar under both copula models, the Clayton copula-based Markov model yields a smaller AIC value, indicating a better fit for the VIX data. This suggests that the VIX series exhibits stronger lower-tail dependence rather than upper-tail dependence during the COVID-19 period. Therefore, we conclude that the structural change associated with the COVID-19 outbreak began around February 25, 2020, which is consistent with the timing of major pandemic-related announcements and the subsequent escalation in financial market uncertainty.

The detected change point reflects not only a shift in the volatility level, but also a change in the underlying dependence structure. In particular, the Clayton copula is capable of capturing stronger lower-tail dependence, which became more pronounced as the VIX surged in response to widespread market sell-offs and heightened downside risk during the pandemic.
  
\begin{figure}[htbp!]
    \centering
    \includegraphics[width=0.7\linewidth]{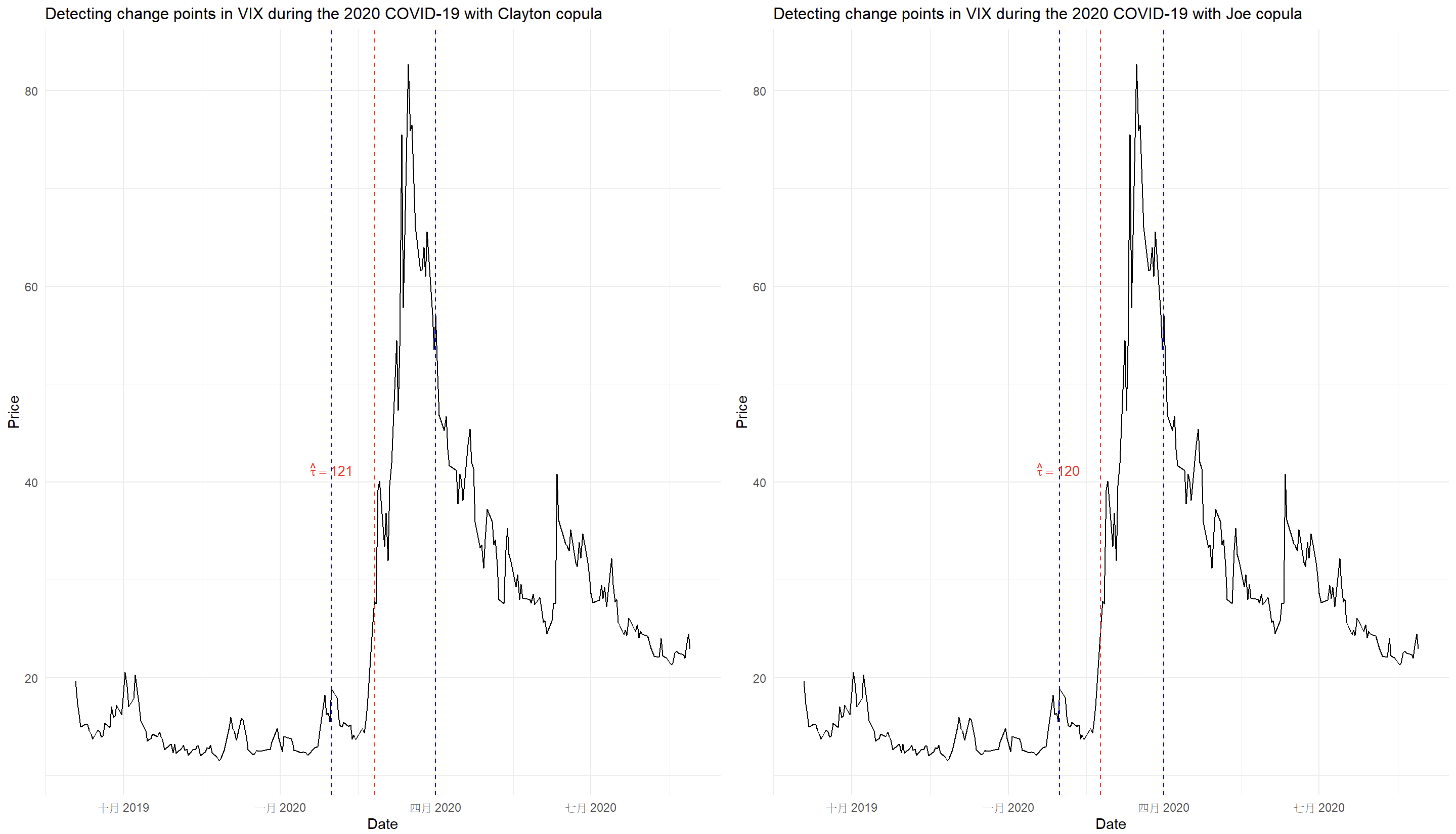}
    \caption{The VIX index during the 2020 COVID-19 under change point detection from 2019/09/01 to 2020/08/31.}
    \label{fig:enterlabel}
\end{figure}

\begin{table}[htbp!]
\centering
\caption{Parameter estimates of proposed model for  2020 COVID-19 under Clayton copula.}
\resizebox{1.01\textwidth}{0.185\textheight}{%
\begin{tabular}{*{10}{c}}
   \toprule
         $\boldsymbol{\alpha_{01}^{as}}$
       & \textbf{Parameter} & \textbf{Estiamtion} & \textbf{95\% CI} & \textbf{AIC} &  $\boldsymbol{\alpha_{01}^{as}}$
       & \textbf{Parameter} & \textbf{Estiamtion} & \textbf{95\% CI} & \textbf{AIC}
        \\
        \hline
        1 & $\tau$ & 121 & (119,124) & 1215.6 &2 & $\tau$ & 120 & (115,122) & 1223.7 \\
        & $k_0$ & 8.4292 & (6.2797,11.6014) &  && $k_0$ & 8.4915 & (6.1112,11.2432) &  \\
        & $k_1$ & 2.6127 & (2.3636,4.2084) &  && $k_1$ & 2.6706 & (2.4771,4.3814) &  \\
        & $\lambda_0$ & 14.7426 & (14.2776,15.6305) &  && $\lambda_0$ & 15.0603 & (14.0889,15.7472) &   \\
        & $\lambda_1$ & 36.5852 & (32.9747,43.6509) &  && $\lambda_1$ & 36.9278 & (34.1271,44.4948) &   \\
        & $\alpha_0$ & 2.2035 & (1.1428,3.4832) &  && $\alpha_0$ & 2.2633 & (1.482,3.5088) &  \\
        & $\alpha_1$ & 2.5334 & (1.2294,4.0044) &  && $\alpha_1$ & 2.4411 & (1.1486,3.9389) &  \\
        \hline
        4 & $\tau$ & 122 & (116,124) & 1229.25 &8 & $\tau$ & 121 & (116,126) & 1237.78 \\
        & $k_0$ & 8.5881 & (6.0412,12.6219)&  && $k_0$ & 8.6087 &(6.1695,12.7079) &  \\
        & $k_1$ & 2.5601 & (2.3514,4.3137) &  && $k_1$ & 2.5307 & (2.2438,4.3706) &  \\
        & $\lambda_0$ & 14.5974 & (14.1604,15.4452)&  && $\lambda_0$ & 14.7906 & (14.3267,15.6693)&  \\
        & $\lambda_1$ & 35.9475 &(32.7155,42.8871) &  && $\lambda_1$ & 35.6501 & (31.8384,42.8142) &  \\
        & $\alpha_0$ & 2.2686 & (1.1847,3.5609) &  && $\alpha_0$ & 2.3615 & (1.2348,3.7944) &  \\
        & $\alpha_1$ & 2.4314 & (1.17044,4.2901) &  && $\alpha_1$ & 2.5746 & (1.2164,4.2111) &  \\
        \bottomrule
\end{tabular}%
}
\label{tab:copulaestimation13}
\end{table}

\begin{table}[htbp!]
\centering
\caption{Parameter estimates of proposed model for  2020 COVID-19 under Joe copula.}
\resizebox{1.01\textwidth}{0.185\textheight}{%
\begin{tabular}{*{10}{c}}
   \toprule
       $\boldsymbol{\alpha_{01}^{as}}$ & \textbf{Parameter} & \textbf{Estiamtion} & \textbf{95\% CI} & \textbf{AIC} & $\boldsymbol{\alpha_{01}^{as}}$ & \textbf{Parameter} & \textbf{Estiamtion} & \textbf{95\% CI} & \textbf{AIC}
        \\
        \hline
        1 & $\tau$ & 122 & (119,126) & 1423.97 &2 & $\tau$ & 120 & (116,123) & 1413.54 \\
        & $k_0$ & 8.8519 & (8.2919,11.3165) &  && $k_0$ & 8.7972 & (8.1996,11.2017) &  \\
        & $k_1$ & 2.4926 & (1.858,2.869) &  && $k_1$ & 2.4346 & (1.8025,2.8249) &  \\
        & $\lambda_0$ & 14.9374 & (14.3125,15.5461) &  && $\lambda_0$ & 15.1171 & (14.3881,15.5124) &   \\
        & $\lambda_1$ & 35.5120 & (25.002,64.8539) &  && $\lambda_1$ & 35.0213 & (25.9564,64.0358) &   \\
        & $\alpha_0$ & 2.6730 & (1.4724,3.0615) &  && $\alpha_0$ & 2.7115 & (1.4892,3.0856) &  \\
        & $\alpha_1$ & 5.2230 & (3.0353,23.402) &  && $\alpha_1$ & 5.1877 & (3.2883,23.0474) &  \\
        \hline
        4 & $\tau$ & 120 & (116,124) & 1424.94 &8 & $\tau$ & 122 & (117,129) & 1436.85 \\
        & $k_0$ & 8.7618 & (8.3019,11.9403)&  && $k_0$ & 8.7498 & (7.9316,11.1136) &  \\
        & $k_1$ & 2.4536 & (1.8848,2.8185) &  && $k_1$ & 2.4558 & (1.8563,2.8997) &  \\
        & $\lambda_0$ & 14.9829 & (14.241,15.6379) &  && $\lambda_0$ & 15.0321 & (14.5758,16.3488) &  \\
        & $\lambda_1$ & 35.4578 & (25.7716,65.9046) &  && $\lambda_1$ & 34.0728 & (24.9675,66.4435) &  \\
        & $\alpha_0$ & 2.7269 & (1.4524,3.1531) &  && $\alpha_0$ & 2.7378 & (1.3698,3.1832) &  \\
        & $\alpha_1$ &5.1679 & (2.9879,23.467) &  && $\alpha_1$ & 4.9487 & (3.0011,25.0849) &  \\
        \bottomrule
\end{tabular}%
}
\label{tab:copulaestimation14}
\end{table}

\subsection{Interarrival time}
\begin{figure}[htbp!]
    \centering
    \includegraphics[width=0.9\linewidth]{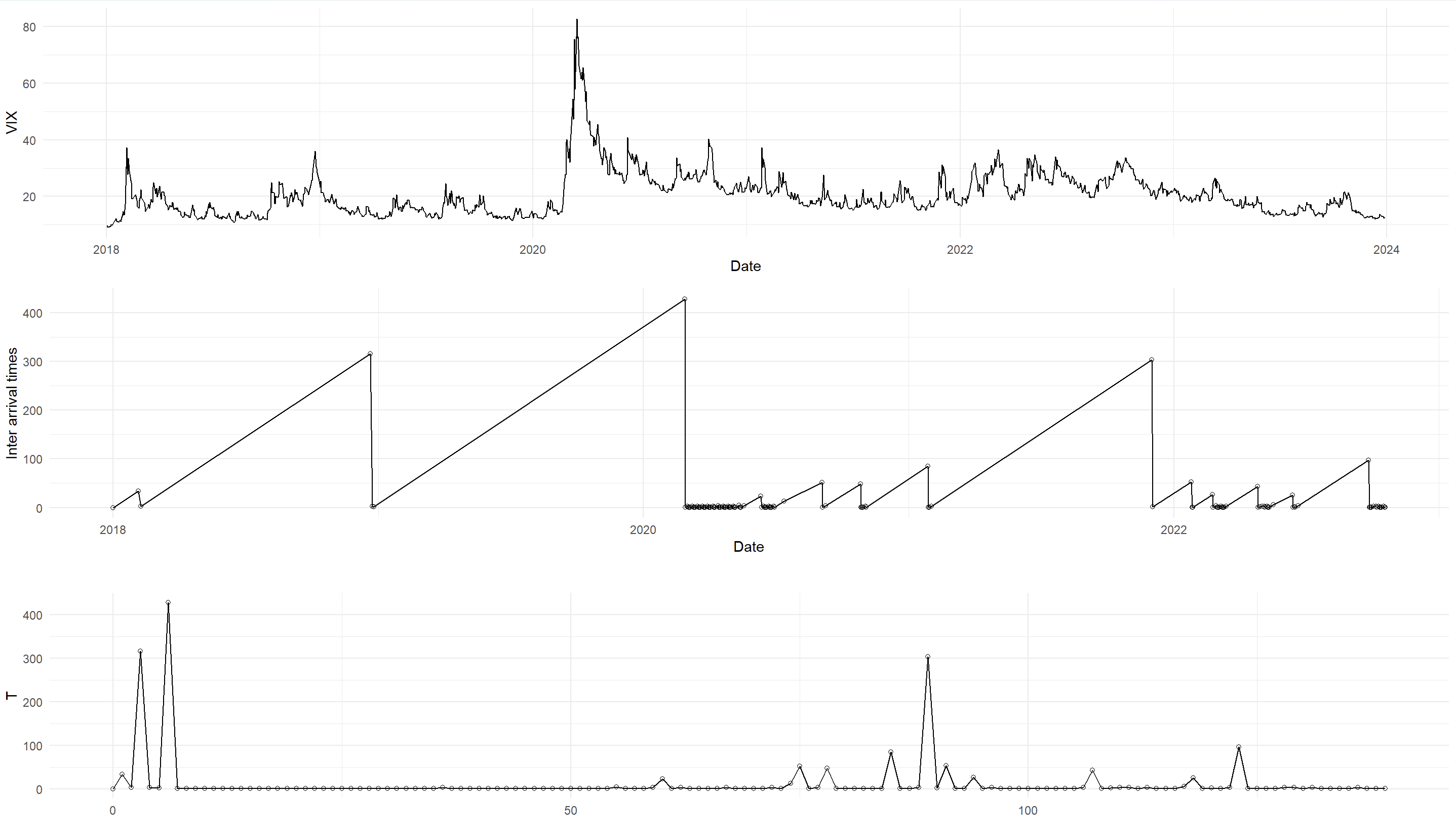}
    \caption{Time series trend of the VIX index.(top) Time series of interarrival times.(middle) Sequential $T$ (Bottom).}
    \label{fig:interarrivaltimes}
\end{figure}

In practice, a VIX value exceeding 30 is commonly interpreted as an indication of heightened market fear and increasing financial risk \citep{JustinKuepper}. It is often regarded as a warning signal that the market has entered a panic regime characterized by elevated uncertainty and substantial volatility. In such situations, investors may anticipate continued increases in market risk, which can lead to large-scale sell-offs and potential market instability. Figure~\ref{fig:interarrivaltimes} illustrates the VIX series from January 1, 2018, to December 31, 2024. The top panel displays the original VIX series, the middle panel shows the temporal trend of the interarrival times $T$, and the bottom panel presents the observed interarrival data points.

Motivated by this interpretation, we define a VIX value exceeding 30 as an indication that the market expects substantially greater future volatility. Based on the VIX data from January 1, 2018, to December 31, 2024, we record the occurrence times at which the VIX exceeds the threshold value of 30. If the VIX remains below 30, the elapsed time continues to accumulate until the threshold is exceeded, at which point the waiting time is recorded as an interarrival interval $T$, defined by
\begin{eqnarray}
	T=\inf\{t:\mathrm{VIX}>30\}. \nonumber
\end{eqnarray}
A total of 139 interarrival times are obtained from the data.

\begin{figure}[htbp!]
    \centering
    \includegraphics[width=0.9\linewidth]{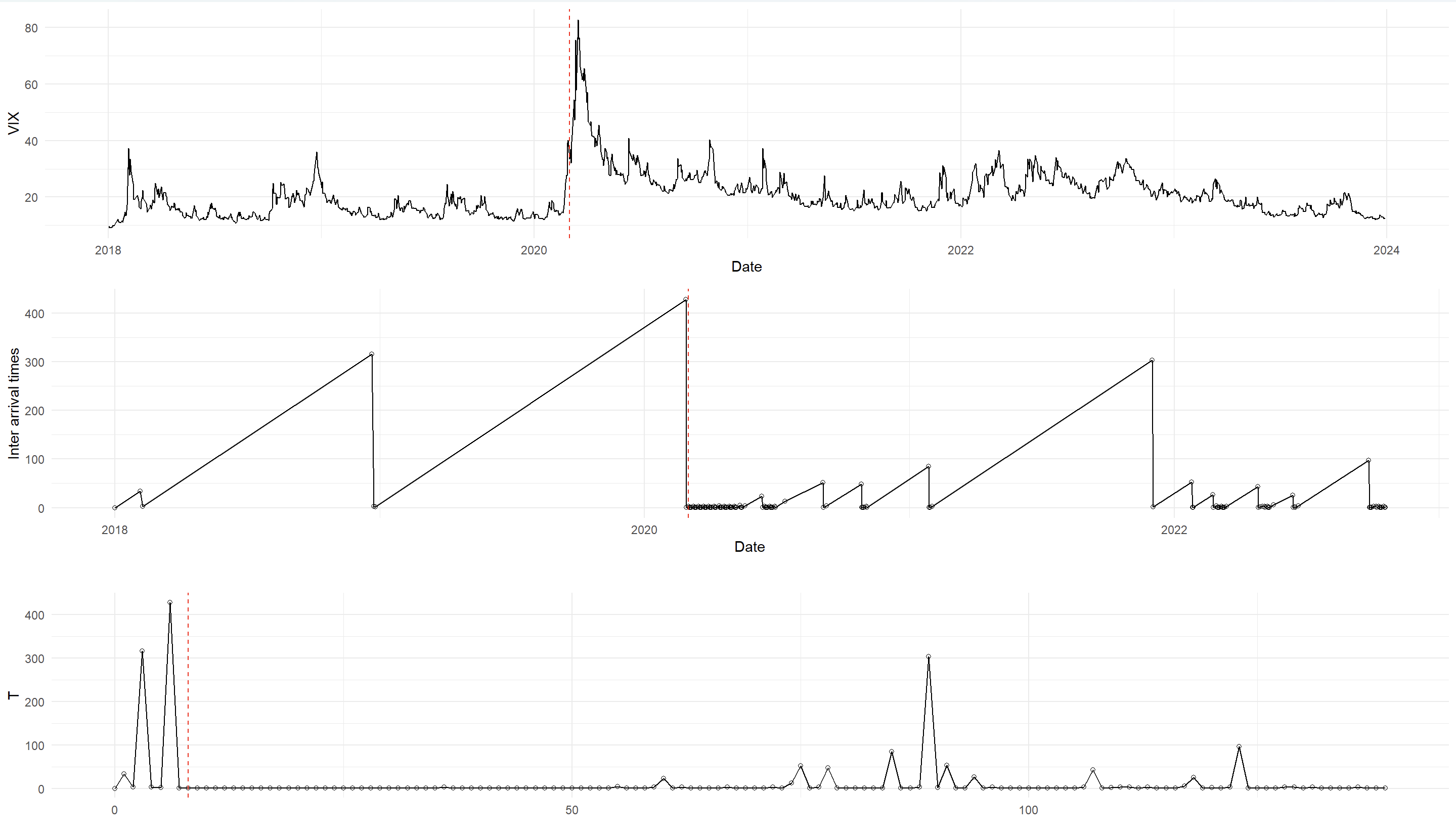}
    \caption{Detection of structural change in the VIX (Top) in inter arrival times (Middle), and sequential $T$ (Bottom).}
    \label{fig:interarrivaltimes2}
\end{figure}

Using the AIC for model selection, Table~16 shows that the estimated structural change point occurs at $\hat{\tau}=8$ under the Clayton copula model and at $\hat{\tau}=7$ under the Joe copula model. The corresponding confidence interval for the change point under the Clayton copula model is $(3,31)$, whereas the interval under the Joe copula model is $(4,39)$. Consistent with the findings in Section~\ref{VIX}, the Clayton copula-based Markov model produces a smaller AIC value, indicating a better fit for the interarrival time data. In Figure~\ref{fig:interarrivaltimes2}, the red vertical line represents the estimated change point $\hat{\tau}$.

Based on these results, we conclude that the structural change associated with the COVID-19 outbreak began around March 2, 2020, which is broadly consistent with the findings in Section~\ref{VIX}. The detected structural change is likely related to the World Health Organization (WHO) announcements concerning the global spread of COVID-19 and the resulting escalation in financial market uncertainty.

\begin{table}[htbp!]
\centering
\caption{Parameter estimates of proposed model for 2020 COVID-19.}
\resizebox{1.01\textwidth}{0.3\textheight}{%
\begin{tabular}{*{5}{c}|*{5}{c}}
   \toprule
    &  & \textbf{Clayton copula} &  &  & &  & \textbf{Joe copula} &  &  \\
   \hline
        $\boldsymbol{\alpha_{01}^{as}}$ & \textbf{Parameter} & \textbf{Estiamtion} & \textbf{95\% CI} & \textbf{AIC} & $\boldsymbol{\alpha_{01}^{as}}$& \textbf{Parameter} & \textbf{Estiamtion} & \textbf{95\% CI} & \textbf{AIC}
        \\
        \hline
        0.5 & $\tau$ & 7 & (3,33) & 434.43 &1 & $\tau$ & 6 & (3,43) & 753.45 \\
        & $k_0$ & 0.3319 & (0.3261,0.3378) &  && $k_0$ & 0.4233 & (0.3299,0.5432) &  \\
        & $k_1$ & 0.6571 & (0.6553,0.657) &  && $k_1$ & 0.5956 & (0.553,0.6656) &  \\
        & $\lambda_0$ & 8.8597 & (7.3637,10.6596) &  && $\lambda_0$ & 7.8106 & (4.5699,13.3495) &   \\
        & $\lambda_1$ & 19.9971 & (19.9575,20.0367) &  && $\lambda_1$ & 3.8316 & (2.8776,5.1018) &   \\
        & $\alpha_0$ & 0.0189 & (0.001348,0.2652) &  && $\alpha_0$ & 1.0021 & (1.0014,1.097) &  \\
        & $\alpha_1$ & 0.047 & (0.04605,0.0479) &  && $\alpha_1$ & 1.00069 & (1.0004,1.0013) &  \\
        \hline
        1 & $\tau$ & 7 & (3,33) & 435.45 &1.5 & $\tau$ & 6 & (3,45) & 762.17 \\
        & $k_0$ & 0.2687 & (0.2667,0.2706)&  && $k_0$ & 0.4231 &(0.3299,0.5427) &  \\
        & $k_1$ & 0.6567 & (0.6549,0.6586) &  && $k_1$ & 0.595 & (0.5325,0.6649) &  \\
        & $\lambda_0$ & 7.2436 & (7.1647,7.3233)&  && $\lambda_0$ & 7.8621 & (5.0611,13.4346)&  \\
        & $\lambda_1$ & 19.9989 &(19.9594,20.0386) &  && $\lambda_1$ & 3.7967 & (2.8514,5.0553) &  \\
        & $\alpha_0$ & 0.0144 & (0.01407,0.01476) &  && $\alpha_0$ & 1.0049 & (1.0027,1.0916) &  \\
        & $\alpha_1$ & 0.0469 & (0.046004,0.0478) &  && $\alpha_1$ & 1.00021 & (1.000114,1.00392) &  \\
        \hline
        2 & $\tau$ & 8 & (3,31) & 412.66 & 2 & $\tau$ & 7 & (4,39) & 746.53 \\
        & $k_0$ & 0.3567 & (0.3126,0.4068)&  && $k_0$ & 0.4213 &(0.3127,0.4568) &  \\
        & $k_1$ & 0.6343 & (0.6322,0.6364) &  && $k_1$ & 0.5968 & (0.5009,0.6746) &  \\
        & $\lambda_0$ & 8.7576 & (7.9661,9.0097)&  && $\lambda_0$ & 9.1286 & (4.8303,13.7978)&  \\
        & $\lambda_1$ & 13.0062 &(12.8379,13.1766) &  && $\lambda_1$ & 3.8402 & (2.7407,5.6415) &  \\
        & $\alpha_0$ & 0.0206 & (0.00314,0.13571) &  && $\alpha_0$ & 1.0283 & (1.003139,1.03571) &  \\
        & $\alpha_1$ & 0.0508 & (0.04971,0.05191) &  && $\alpha_1$ & 1.0042 & (1.002568,1.05191) &  \\
        \bottomrule
\end{tabular}%
}
\label{tab:copulaestimation16}
\end{table}

\section{Conclusion and Future Research}\label{Sec: Con}
In this paper, we investigated offline change-point estimation for time series data exhibiting nonlinear serial dependence. To address this problem, we proposed a copula-based Markov chain model with Weibull marginal distributions, incorporating the Clayton and Joe copulas to capture asymmetric lower- and upper-tail dependence structures. The corresponding likelihood function was derived, and the maximum likelihood estimators for the change point and model parameters were obtained using the Newton--Raphson algorithm.

Extensive simulation studies were conducted to evaluate the finite-sample performance and robustness of the proposed method under a variety of parameter settings, including different sample sizes, change-point locations, dependence strengths, and copula misspecification scenarios. The simulation results demonstrated that the proposed estimators perform well in terms of RMSE and RE, particularly for the estimation of the change point, where the relative errors remained small even under moderate model misspecification. Sensitivity analysis with respect to the dependence parameter $\alpha_{01}$ further showed that the proposed procedure is reasonably robust to misspecification of the serial dependence parameter. In addition, the cross-comparison study between the Clayton and Joe copulas revealed that, although misspecification affects the estimation of the tail dependence parameters, the estimation of the change point itself remains relatively stable.

An empirical application to the VIX index during the COVID-19 pandemic further illustrated the effectiveness of the proposed framework in detecting meaningful structural changes in financial market volatility. The estimated change points were consistent with major pandemic-related events and announcements, and the Clayton copula-based Markov model provided the best fit according to the AIC. The analysis also demonstrated the importance of accounting for asymmetric tail dependence when modeling financial time series during periods of market stress.

Overall, the results indicate that copula-based Markov chain models provide a flexible and reliable framework for change-point estimation in Weibull time series with nonlinear dependence structures. Several directions for future research are worth pursuing. First, the proposed framework may be extended to accommodate multiple change points by modifying the model structure accordingly. Second, alternative copula families, such as Gaussian, Gumbel, or survival-Gumbel copulas, may be considered to capture different forms of dependence and tail behavior. Third, more flexible marginal distributions may also be incorporated to improve model adaptability in practical applications. However, likelihood-based estimation under more complicated copula structures may become computationally demanding. In such settings, Bayesian approaches or simulation-based methods may provide useful alternatives for parameter estimation and statistical inference \cite{jammalamadaka2018multivariate,jammalamadaka2019predicting,qiu2020multivariate}. These extensions represent promising directions for future investigation.

\appendix

\section{Derivatives of the Transformed Log-Likelihood Function}

The first- and second-order derivatives of the transformed log-likelihood function are obtained by applying the chain rule to the re-parameterized variables introduced in Section~\ref{Sec: PE}. The corresponding expressions are given as follows:
\begin{equation*}
	\begin{aligned}
		\frac{\partial\tilde{\ell}}{\partial K_0} &=\frac{\partial\ell}{\partial k_0}\times\frac{\partial}{\partial K_0}(e^{K_0})=e^{K_0}\frac{\partial\ell}{\partial k_0},\\
		\frac{\partial\tilde{\ell}}{\partial K_1} &=\frac{\partial\ell}{\partial k_1}\times\frac{\partial}{\partial K_1}(e^{K_1})=e^{K_1}\frac{\partial\ell}{\partial k_1},\\
		\frac{\partial\tilde{\ell}}{\partial\Lambda_0} &=\frac{\partial\ell}{\partial\lambda_0}\times\frac\partial{\partial\Lambda_0}\left(e^{\Lambda_0}\right)=e^{\Lambda_0}\frac{\partial\ell}{\partial\lambda_0},\\
		\frac{\partial\tilde{\ell}}{\partial\Lambda_1} &=\frac{\partial\ell}{\partial\lambda_1}\times\frac\partial{\partial\Lambda_1}\left(e^{\Lambda_1}\right)=e^{\Lambda_1}\frac{\partial\ell}{\partial\lambda_1},\\
		\frac{\partial\tilde{\ell}}{\partial A_0} &=\frac{\partial\ell}{\partial\alpha_0}\times\frac\partial{\partial A_0}(e^{A_0}-1)=e^{A_0}\frac{\partial\ell}{\partial\alpha_0},\\
		\frac{\partial\tilde{\ell}}{\partial A_1} &=\frac{\partial\ell}{\partial\alpha_1}\times\frac\partial{\partial A_1}(e^{A_1}-1)=e^{A_1}\frac{\partial\ell}{\partial\alpha_1},
	\end{aligned}
\end{equation*}
,\\[-1ex] %
\begin{equation*}
	\begin{aligned}
		\frac{\partial^2\tilde{\ell}}{\partial K_0\partial K_0} &=\frac{\partial^2\ell}{\partial k_0^2}\times(\frac\partial{\partial K_0}(e^{K_0}))^2+\frac\partial{\partial k_0}\times\frac{\partial^2}{\partial K_0^2}(e^{K_0})=e^{2K_0}\frac{\partial^2\ell}{\partial k_0^2}+e^{K_0}\frac{\partial\ell}{\partial k_0},\\
		\frac{\partial^2\tilde{\ell}}{\partial K_0\partial K_1} &=\frac{\partial^2\ell}{\partial k_0\partial k_1}\times\frac\partial{\partial K_0}(e^{K_0})\times\frac\partial{\partial K_1}(e^{K_1})=0,\\
		\frac{\partial^2\tilde{\ell}}{\partial K_0\partial\Lambda_0} &=\frac{\partial^2\ell}{\partial k_0\partial\lambda_0}\times\frac{\partial}{\partial K_0}\big(e^{K_0}\big)\times\frac{\partial}{\partial\Lambda_0}\big(e^{\Lambda_0}\big)=e^{(K_0+\Lambda_0)}\frac{\partial^2\ell}{\partial k_0\partial\lambda_0},\\
		\frac{\partial^2\tilde{\ell}}{\partial K_0\partial\Lambda_1} &=\frac{\partial^2\ell}{\partial k_0\partial\lambda_1}\times\frac{\partial}{\partial K_0}\big(e^{K_0}\big)\times\frac{\partial}{\partial\Lambda_1}\big(e^{\Lambda_1}\big)=0,\\
		\frac{\partial^2\tilde{\ell}}{\partial K_0\partial\mathcal{A}_0} &=\frac{\partial^2\ell}{\partial k_0\partial\alpha_0}\times\frac\partial{\partial K_0}\left(e^{K_0}\right)\times\frac\partial{\partial\mathcal{A}_0}\left(e^{\mathcal{A}_0}-1\right)=e^{(K_0+\mathcal{A}_0)}\frac{\partial^2\ell}{\partial k_0\partial\alpha_0},\\
		\frac{\partial^2\tilde{\ell}}{\partial K_0\partial\mathcal{A}_1} &=\frac{\partial^2\ell}{\partial k_0\partial\alpha_1}\times\frac\partial{\partial K_0}\left(e^{K_0}\right)\times\frac\partial{\partial\mathcal{A}_1}\left(e^{\mathcal{A}_1}-1\right)=0,\\
	\end{aligned}
\end{equation*}
,\\[-1ex] %

\begin{equation*}
	\begin{aligned}
		\frac{\partial^2\tilde{\ell}}{\partial K_1\partial K_0} &=\frac{\partial^2\ell}{\partial k_1\partial k_0}\times\frac\partial{\partial K_1}(e^{K_1})\times\frac\partial{\partial K_0}(e^{K_0})=0,\\
		\frac{\partial^2\tilde{\ell}}{\partial K_1\partial K_1} &=\frac{\partial^2\ell}{\partial k_1^2}\times(\frac\partial{\partial K_1}(e^{K_1}))^2+\frac\partial{\partial k_1}\times\frac{\partial^2}{\partial K_1^2}(e^{K_1})=e^{2K_1}\frac{\partial^2\ell}{\partial k_1^2}+e^{K_1}\frac{\partial\ell}{\partial k_1},\\
		\frac{\partial^2\tilde{\ell}}{\partial K_1\partial\Lambda_0} &=\frac{\partial^2\ell}{\partial k_1\partial\lambda_0}\times\frac{\partial}{\partial K_1}\big(e^{K_1}\big)\times\frac{\partial}{\partial\Lambda_0}\big(e^{\Lambda_0}\big)=0,\\
		\frac{\partial^2\tilde{\ell}}{\partial K_1\partial\Lambda_1} &=\frac{\partial^2\ell}{\partial k_1\partial\lambda_1}\times\frac{\partial}{\partial K_1}\big(e^{K_1}\big)\times\frac{\partial}{\partial\Lambda_1}\big(e^{\Lambda_1}\big)=e^{(K_1+\Lambda_1)}\frac{\partial^2\ell}{\partial k_1\partial\lambda_1},\\
		\frac{\partial^2\tilde{\ell}}{\partial K_1\partial\mathcal{A}_0} &=\frac{\partial^2\ell}{\partial k_1\partial\alpha_0}\times\frac\partial{\partial K_1}\left(e^{K_1}\right)\times\frac\partial{\partial\mathcal{A}_0}\left(e^{\mathcal{A}_0}-1\right)=0,\\
		\frac{\partial^2\tilde{\ell}}{\partial K_1\partial\mathcal{A}_1} &=\frac{\partial^2\ell}{\partial k_1\partial\alpha_1}\times\frac\partial{\partial K_1}\left(e^{K_1}\right)\times\frac\partial{\partial\mathcal{A}_1}\left(e^{\mathcal{A}_1}-1\right)=e^{(K_1+\mathcal{A}_1)}\frac{\partial^2\ell}{\partial k_1\partial\alpha_1},\\
	\end{aligned}
\end{equation*}
,\\[-1ex] %

\begin{equation*}
	\begin{aligned}
		\frac{\partial^2\tilde{\ell}}{\partial\Lambda_0\partial\Lambda_0} &=\frac{\partial^2\ell}{\partial\lambda_0^2}\times(\frac\partial{\partial\Lambda_0}(e^{\Lambda_0}))^2+\frac\partial{\partial\lambda_0}\times\frac{\partial^2}{\partial\Lambda_0^2}(e^{\Lambda_0})=e^{2\Lambda_0}\frac{\partial^2\ell}{\partial\lambda_0^2}+e^{\Lambda_0}\frac{\partial\ell}{\partial\lambda_0},\\
		\frac{\partial^2\tilde{\ell}}{\partial\Lambda_0\partial\Lambda_1} &=\frac{\partial^2\ell}{\partial\lambda_0\partial\lambda_1}\times\frac\partial{\partial\Lambda_0}(e^{\Lambda_0})\times\frac\partial{\partial\Lambda_1}(e^{\Lambda_1})=0,\\
		\frac{\partial^2\tilde{\ell}}{\partial\Lambda_0\partial K_0} &=\frac{\partial^2\ell}{\partial\lambda_0\partial k_0}\times\frac{\partial}{\partial\Lambda_0}\big(e^{\Lambda_0}\big)\times\frac{\partial}{\partial K_0}\big(e^{K_0}\big)=e^{(\Lambda_0+K_0)}\frac{\partial^2\ell}{\partial\lambda_0\partial k_0},\\
		\frac{\partial^2\tilde{\ell}}{\partial\Lambda_0\partial K_1} &=\frac{\partial^2\ell}{\partial\lambda_0\partial k_1}\times\frac{\partial}{\partial\Lambda_0}\big(e^{\Lambda_0}\big)\times\frac{\partial}{\partial K_1}\big(e^{K_1}\big)=0,\\
		\frac{\partial^2\tilde{\ell}}{\partial \Lambda_0\partial\mathcal{A}_0} &=\frac{\partial^2\ell}{\partial \lambda_0\partial\alpha_0}\times\frac\partial{\partial \Lambda_0}\left(e^{\Lambda_0}\right)\times\frac\partial{\partial\mathcal{A}_0}\left(e^{\mathcal{A}_0}-1\right)=e^{(\Lambda_0+\mathcal{A}_0)}\frac{\partial^2\ell}{\partial \lambda_0\partial\alpha_0},\\
		\frac{\partial^2\tilde{\ell}}{\partial \Lambda_0\partial\mathcal{A}_1} &=\frac{\partial^2\ell}{\partial \lambda_0\partial\alpha_1}\times\frac\partial{\partial \Lambda_0}\left(e^{\Lambda_0}\right)\times\frac\partial{\partial\mathcal{A}_1}\left(e^{\mathcal{A}_1}-1\right)=0,\\
	\end{aligned}
\end{equation*}

\begin{equation*}
	\begin{aligned}
		\frac{\partial^2\tilde{\ell}}{\partial\Lambda_1\partial\Lambda_0} &=\frac{\partial^2\ell}{\partial\lambda_1\partial\lambda_0}\times\frac\partial{\partial\Lambda_1}(e^{\Lambda_1})\times\frac\partial{\partial\Lambda_0}(e^{\Lambda_0})=0,\\
		\frac{\partial^2\tilde{\ell}}{\partial\Lambda_1\partial\Lambda_1} &=\frac{\partial^2\ell}{\partial\lambda_1^2}\times(\frac\partial{\partial\Lambda_1}(e^{\Lambda_1}))^2+\frac\partial{\partial\lambda_1}\times\frac{\partial^2}{\partial\Lambda_1^2}(e^{\Lambda_1})=e^{2\Lambda_1}\frac{\partial^2\ell}{\partial\lambda_1^2}+e^{\Lambda_1}\frac{\partial\ell}{\partial\lambda_1},\\
		\frac{\partial^2\tilde{\ell}}{\partial\Lambda_1\partial K_0} &=\frac{\partial^2\ell}{\partial\lambda_1\partial k_0}\times\frac{\partial}{\partial\Lambda_1}\big(e^{\Lambda_1}\big)\times\frac{\partial}{\partial K_0}\big(e^{K_0}\big)=0,\\
		\frac{\partial^2\tilde{\ell}}{\partial\Lambda_1\partial K_1} &=\frac{\partial^2\ell}{\partial\lambda_1\partial k_1}\times\frac{\partial}{\partial\Lambda_1}\big(e^{\Lambda_1}\big)\times\frac{\partial}{\partial K_1}\big(e^{K_1}\big)=e^{(\Lambda_1+K_1)}\frac{\partial^2\ell}{\partial\lambda_1\partial k_1},\\
		\frac{\partial^2\tilde{\ell}}{\partial \Lambda_1\partial\mathcal{A}_0} &=\frac{\partial^2\ell}{\partial \lambda_1\partial\alpha_0}\times\frac\partial{\partial \Lambda_1}\left(e^{\Lambda_1}\right)\times\frac\partial{\partial\mathcal{A}_0}\left(e^{\mathcal{A}_0}-1\right)=0,\\
		\frac{\partial^2\tilde{\ell}}{\partial \Lambda_1\partial\mathcal{A}_1} &=\frac{\partial^2\ell}{\partial \lambda_1\partial\alpha_1}\times\frac\partial{\partial \Lambda_1}\left(e^{\Lambda_1}\right)\times\frac\partial{\partial\mathcal{A}_1}\left(e^{\mathcal{A}_1}-1\right)=e^{(\Lambda_1+\mathcal{A}_1)}\frac{\partial^2\ell}{\partial \lambda_1\partial\alpha_1},\\
	\end{aligned}
\end{equation*}
,\\[-1ex] %

\begin{equation*}
	\begin{aligned}
		\frac{\partial^2\tilde{\ell}}{\partial\mathcal{A}_0\partial K_0} &=\frac{\partial^2\ell}{\partial\alpha_0\partial k_0}\times\frac\partial{\partial\mathcal{A}_0}\left(e^{\mathcal{A}_0}-1\right)\times\frac\partial{\partial K_0}\left(e^{K_0}\right)=e^{(\mathcal{A}_0+K_0)}\frac{\partial^2\ell}{\partial\alpha_0\partial k_0},\\
		\frac{\partial^2\tilde{\ell}}{\partial\mathcal{A}_0\partial K_1} &=\frac{\partial^2\ell}{\partial\alpha_0\partial k_1}\times\frac\partial{\partial\mathcal{A}_0}\left(e^{\mathcal{A}_0}-1\right)\times\frac\partial{\partial K_1}\left(e^{K_1}\right)=0,\\
		\frac{\partial^2\tilde{\ell}}{\partial\mathcal{A}_0\partial \Lambda_0} &=\frac{\partial^2\ell}{\partial\alpha_0\partial \lambda_0}\times\frac\partial{\partial\mathcal{A}_0}\left(e^{\mathcal{A}_0}-1\right)\times\frac\partial{\partial \Lambda_0}\left(e^{\Lambda_0}\right)=e^{(\mathcal{A}_0+\Lambda_0)}\frac{\partial^2\ell}{\partial\alpha_0\partial \lambda_0},\\
		\frac{\partial^2\tilde{\ell}}{\partial\mathcal{A}_0\partial \Lambda_1} &=\frac{\partial^2\ell}{\partial\alpha_0\partial \lambda_1}\times\frac\partial{\partial\mathcal{A}_0}\left(e^{\mathcal{A}_0}-1\right)\times\frac\partial{\partial \Lambda_1}\left(e^{\Lambda_1}\right)=0,\\
		\frac{\partial^2\tilde{\ell}}{\partial\mathcal{A}_0\partial\mathcal{A}_0} &=\frac{\partial^2\ell}{\partial \alpha_0^2}\times(\frac\partial{\partial \mathcal{A}_0}\left(e^{\mathcal{A}_0}-1\right))^2+\frac\partial{\partial \alpha_0}\times\frac{\partial^2}{\partial \mathcal{A}_0^2}(e^{\mathcal{A}_0}-1)=e^{2\mathcal{A}_0}\frac{\partial^2\ell}{\partial \alpha_0^2}+e^{\mathcal{A}_0}\frac{\partial\ell}{\alpha_0},\\
		\frac{\partial^2\tilde{\ell}}{\partial \mathcal{A}_0\partial\mathcal{A}_1} &=\frac{\partial^2\ell}{\partial \alpha_0\partial \alpha_1}\times\frac\partial{\partial \mathcal{A}_0}(e^{\mathcal{A}_0}-1)\times\frac\partial{\partial \mathcal{A}_1}(e^{\mathcal{A}_1}-1)=0,\\
	\end{aligned}
\end{equation*}
,\\[-1ex] %

\begin{equation*}
	\begin{aligned}
		\frac{\partial^2\tilde{\ell}}{\partial\mathcal{A}_1\partial K_0} &=\frac{\partial^2\ell}{\partial\alpha_1\partial k_1}\times\frac\partial{\partial\mathcal{A}_1}\left(e^{\mathcal{A}_1}-1\right)\times\frac\partial{\partial K_0}\left(e^{K_0}\right)=0,\\
		\frac{\partial^2\tilde{\ell}}{\partial\mathcal{A}_1\partial K_1} &=\frac{\partial^2\ell}{\partial\alpha_1\partial k_0}\times\frac\partial{\partial\mathcal{A}_1}\left(e^{\mathcal{A}_1}-1\right)\times\frac\partial{\partial K_1}\left(e^{K_1}\right)=e^{(\mathcal{A}_1+K_1)}\frac{\partial^2\ell}{\partial\alpha_1\partial k_1},\\
		\frac{\partial^2\tilde{\ell}}{\partial\mathcal{A}_1\partial \Lambda_0} &=\frac{\partial^2\ell}{\partial\alpha_1\partial \lambda_0}\times\frac\partial{\partial\mathcal{A}_1}\left(e^{\mathcal{A}_1}-1\right)\times\frac\partial{\partial \Lambda_0}\left(e^{\Lambda_0}\right)=0,\\
		\frac{\partial^2\tilde{\ell}}{\partial\mathcal{A}_1\partial \Lambda_1} &=\frac{\partial^2\ell}{\partial\alpha_1\partial \lambda_1}\times\frac\partial{\partial\mathcal{A}_1}\left(e^{\mathcal{A}_1}-1\right)\times\frac\partial{\partial \Lambda_1}\left(e^{\Lambda_1}\right)=e^{(\mathcal{A}_1+\Lambda_1)}\frac{\partial^2\ell}{\partial\alpha_1\partial \lambda_1},\\
		\frac{\partial^2\tilde{\ell}}{\partial \mathcal{A}_1\partial\mathcal{A}_0} &=\frac{\partial^2\ell}{\partial \alpha_1\partial \alpha_0}\times\frac\partial{\partial \mathcal{A}_1}(e^{\mathcal{A}_1}-1)\times\frac\partial{\partial \mathcal{A}_0}(e^{\mathcal{A}_0}-1)=0,\\
		\frac{\partial^2\tilde{\ell}}{\partial\mathcal{A}_1\partial\mathcal{A}_1} &=\frac{\partial^2\ell}{\partial \alpha_1^2}\times(\frac\partial{\partial \mathcal{A}_1}\left(e^{\mathcal{A}_1}-1\right))^2+\frac\partial{\partial \alpha_1}\times\frac{\partial^2}{\partial \mathcal{A}_1^2}(e^{\mathcal{A}_1}-1)=e^{2\mathcal{A}_1}\frac{\partial^2\ell}{\partial \alpha_1^2}+e^{\mathcal{A}_1}\frac{\partial\ell}{\alpha_1}.
	\end{aligned}
\end{equation*}

\bigskip

\section{Derivatives of the Copula Log-Likelihood Function}

\subsection{First- and Second-Order Derivatives for the Clayton Copula}

In this appendix, we present the first- and second-order derivatives of the copula log-likelihood function associated with the Clayton copula. These derivatives are required for implementing the Newton--Raphson algorithm described in Section~\ref{Sec: PE}.
The first-order derivatives of the log-likelihood function are given by
\begin{equation*}
	\begin{aligned}
		\partial_{\alpha_0}\ell=&\sum_{t=1}^{\tau-1}\left\{\frac{1}{1+\alpha_0}-\log\bigl(H_0(x_t)\bigr)-\log\bigl(H_0(x_{t+1})\bigr)+\frac{\log\bigl(H_0(x_t)^{-\alpha_0}+H_0(x_{t+1})^{-\alpha_0}-1\bigr)}{\alpha_0^2}\right.\\&+\left(\frac{1}{\alpha_0}+2\right)\frac{H_0(x_t)^{-\alpha_0}\log\bigl(H_0(x_t)\bigr)+H_0(x_{t+1})^{-\alpha_0}\log\bigl(H_0(x_{t+1})\bigr)}{H_0(x_t)^{-\alpha_0}+H_0(x_{t+1})^{-\alpha_0}-1}\biggr\}
	,\end{aligned}
\end{equation*}

\begin{equation*}
	\begin{aligned}
		\partial_{\alpha_1}\ell=&\sum_{t=\tau+1}^{T-1}\left\{\frac{1}{1+\alpha_1}-\log\bigl(H_1(x_t)\bigr)-\log\bigl(H_1(x_{t+1})\bigr)+\frac{\log\bigl(H_1(x_t)^{-\alpha_1}+H_1(x_{t+1})^{-\alpha_1}-1\bigr)}{\alpha_1^2}\right.\\&+\left(\frac{1}{\alpha_1}+2\right)\frac{H_1(x_t)^{-\alpha_1}\log\bigl(H_1(x_t)\bigr)+H_1(x_{t+1})^{-\alpha_1}\log\bigl(H_1(x_{t+1})\bigr)}{H_1(x_t)^{-\alpha_1}+H_1(x_{t+1})^{-\alpha_1}-1}\biggr\}\
,	\end{aligned}
\end{equation*}

\begin{equation*}
	\begin{aligned}
		\partial_{k_0}\ell & \begin{aligned}~=\sum_{t=1}^{\tau}\frac{\partial_{k_0}h_0(x_t)}{h_0(x_t)}+\sum_{t=1}^{\tau-1}\left\{-(1+\alpha_0)\frac{\partial_{k_0}H_0(x_t)}{H_0(x_t)}-(1+\alpha_0)\frac{\partial_{k_0}H_0(x_{t+1})}{H_0(x_{t+1})}\right.\end{aligned} \\
		&\begin{aligned}+ (1+2\alpha_0)\frac{H_0(x_t)^{-\alpha_0-1}\partial_{k_0}H_0(x_t)+H_0(x_{t+1})^{-\alpha_0-1}\partial_{k_0}H_0(x_{t+1})}{H_0(x_t)^{-\alpha_0}+H_0(x_{t+1})^{-\alpha_0}-1}\Bigg\}-(1+\alpha_{01})\frac{\partial_{k_0}H_0(x_{\tau})}{H_0(x_{\tau})}\end{aligned} \\
		&\begin{aligned}&+(1+2\alpha_{01})\frac{H_{0}(x_{\tau})^{-\alpha_{01}-1}\partial_{k_0}H_{0}(x_{\tau})}{H_{0}(x_{\tau})^{-\alpha_{01}}+H_{1}(x_{\tau+1})^{-\alpha_{01}}-1}
,		\end{aligned}
	\end{aligned}
\end{equation*}

\begin{equation*}
	\begin{aligned}
		\partial_{k_1}\ell &
		\begin{aligned}~=\sum_{t=\tau+1}^T\frac{\partial_{k_1}h_1(x_t)}{h_1(x_t)}+\sum_{t=\tau+1}^{T-1}\Bigg\{-(1+\alpha_1)\frac{\partial_{k_1}H_1(x_t)}{H_1(x_t)}-(1+\alpha_1)\frac{\partial_{k_1}H_1(x_{t+1})}
			{H_1(x_{t+1})}\end{aligned} \\
		&\begin{aligned}&+(1+2\alpha_1)\frac{H_1(x_t)^{-\alpha_1-1}\partial_{k_1}H_1(x_t)+H_1(x_{t+1})^{-\alpha_1-1}\partial_{k_1}H_1(x_{t+1})}{H_1(x_t)^{-\alpha_1}+H_1(x_{t+1})^{-\alpha_1}-1}\Bigg\}-(1+\alpha_{01})\frac{\partial_{k_1}H_1(x_{\tau+1})}{H_1(x_{\tau+1})}\end{aligned} \\
		&\begin{aligned}&+(1+2\alpha_{01})\frac{H_{1}(x_{\tau+1})^{-\alpha_{01}-1}\partial_{k_{1}}H_{1}(x_{\tau+1})}{H_{0}(x_{\tau})^{-\alpha_{01}}+H_{1}(x_{\tau+1})^{-\alpha_{01}}-1}
,		\end{aligned}
	\end{aligned}
\end{equation*}

\begin{equation*}
	\begin{aligned}
		\partial_{\lambda_0}\ell & \begin{aligned}=&\sum_{t=1}^{\tau}\frac{\partial_{\lambda_0}h_0(x_t)}{h_0(x_t)}+\sum_{t=1}^{\tau-1}\Bigg\{-(1+\alpha_0)\frac{\partial_{\lambda_0}H_0(x_t)}{H_0(x_t)}-(1+\alpha_0)\frac{\partial_{\lambda_0}H_0(x_{t+1})}{H_0(x_{t+1})}\end{aligned} \\
		&\begin{aligned}+ (1+2\alpha_0)\frac{H_0(x_t)^{-\alpha_0-1}\partial_{\lambda_0}H_0(x_t)+H_0(x_{t+1})^{-\alpha_0-1}\partial_{\lambda_0}H_0(x_{t+1})}{H_0(x_t)^{-\alpha_0}+H_0(x_{t+1})^{-\alpha_0}-1}\Bigg\}-(1+\alpha_{01})\frac{\partial_{\lambda_0}H_0(x_{\tau})}{H_0(x_{\tau})}\end{aligned} \\
		&\begin{aligned}&+(1+2\alpha_{01})\frac{H_{0}(x_{\tau})^{-\alpha_{01}-1}\partial_{\lambda_{0}}H_{0}(x_{\tau})}{H_{0}(x_{\tau})^{-\alpha_{01}}+H_{1}(x_{\tau+1})^{-\alpha_{01}}-1}
,		\end{aligned}
	\end{aligned}
\end{equation*}

\begin{equation*}
	\begin{aligned}
		\partial_{\lambda_1}\ell & \begin{aligned}=&\sum_{t=\tau+1}^T\frac{\partial_{\lambda_1}h_1(x_t)}{h_1(x_t)}+\sum_{t=\tau+1}^{T-1}\left\{-(1+\alpha_1)\frac{\partial_{\lambda_1}H_1(x_t)}{H_1(x_t)}-(1+\alpha_1)\frac{\partial_{\lambda_1}H_1(x_{t+1})}{H_1(x_{t+1})}\right.\end{aligned} \\
		&\begin{aligned}+(1+2\alpha_1)\frac{H_1(x_t)^{-\alpha_1-1}\partial_{\lambda_1}H_1(x_t)+H_1(x_{t+1})^{-\alpha_1-1}\partial_{\lambda_1}H_1(x_{t+1})}{H_1(x_t)^{-\alpha_1}+H_1(x_{t+1})^{-\alpha_1}-1}\Bigg\}-(1+\alpha_{01})\frac{\partial_{\lambda_1}H_1(x_{\tau+1})}{H_1(x_{\tau+1})}\end{aligned} \\
		&\begin{aligned}&+(1+2\alpha_{01})\frac{H_{1}(x_{\tau+1})^{-\alpha_{01}-1}\partial_{\lambda_{1}}H_{1}(x_{\tau+1})}{H_{1}(x_{\tau})^{-\alpha_{01}}+H_{1}(x_{\tau+1})^{-\alpha_{01}}-1}
,		\end{aligned}
	\end{aligned}
\end{equation*}
where
\begin{equation*}
	\partial_{k_0}h_0(x)=\frac1{\lambda_0}\left(\frac x{\lambda_0}\right)^{k_0-1}e^{-\left(\frac x{\lambda_0}\right)^{k_0}}\left[1+k_0\ln\left(\frac x{\lambda_0}\right)-k_0\left(\frac x{\lambda_0}\right)^{k_0}\ln\left(\frac x{\lambda_0}\right)\right],
\end{equation*}
\begin{equation*}
	\partial_{k_1}h_1(x)=\frac1{\lambda_1}\left(\frac x{\lambda_1}\right)^{k_1-1}e^{-\left(\frac x{\lambda_1}\right)^{k_1}}\left[1+k_1\ln\left(\frac x{\lambda_1}\right)-k_1\left(\frac x{\lambda_1}\right)^{k_1}\ln\left(\frac x{\lambda_1}\right)\right],
\end{equation*}
\begin{equation*}
	\partial_{\lambda_0}h_0(x)=\frac{k_0^2}{\lambda_0^2}\left(\frac{x}{\lambda_0}\right)^{k_0-1}e^{-\left(\frac{x}{\lambda_0}\right)^{k_0}}\left[\left(\frac{x}{\lambda_0}\right)^{k_0}-1\right],
\end{equation*}
\begin{equation*}
	\quad\partial_{\lambda_1}h_1(x)=\frac{k_1^2}{\lambda_1^2}\left(\frac{x}{\lambda_1}\right)^{k_1-1}e^{-\left(\frac{x}{\lambda_1}\right)^{k_1}}\left[\left(\frac{x}{\lambda_1}\right)^{k_1}-1\right],
\end{equation*}
\begin{equation*}
	\partial_{k_0}H_0(x)=e^{-\left(\frac x{\lambda_0}\right)^{k_0}}\left(\frac x{\lambda_0}\right)^{k_0}\ln\left(\frac x{\lambda_0}\right),\quad\partial_{k_1}H_1(y)=e^{-\left(\frac x{\lambda_1}\right)^{k_1}}\left(\frac x{\lambda_1}\right)^{k_1}\ln\left(\frac x{\lambda_1}\right),
\end{equation*}
\begin{equation*}
	\partial_{\lambda_0}H_0(x)=e^{-\left(\frac x{\lambda_0}\right)^{k_0}}\left(\frac x{\lambda_0}\right)^{k_0}\left(\frac{k_0}{\lambda_0}\right),\quad\partial_{\lambda_1}H_1(x)=e^{-\left(\frac x{\lambda_1}\right)^{k_1}}\left(\frac x{\lambda_1}\right)^{k_1}\left(\frac{k_1}{\lambda_1}\right).
\end{equation*}

The second-order derivatives of log-likelihood function  are given by
\begin{equation*}
	\begin{aligned}
		\partial_{\alpha_{0},\alpha_{0}}\ell =&\sum_{t=1}^{\tau-1}\Big\{ -\frac{1}{(1+\alpha_0)^2}-\frac{2}{\alpha_0^3}\mathrm{log}(H_0(x_t)^{-\alpha_0}+H_0(x_{t+1})^{-\alpha_0}-1) \\
		&-\frac2{\alpha_0^2}\frac{H_0(x_t)^{-\alpha_0}\mathrm{log}(H_0(x_t))+H_0(x_{t+1})^{-\alpha_0}\mathrm{log}(H_0(x_{t+1}))}{H_0(x_t)^{-\alpha_0}+H_0(x_{t+1})^{-\alpha_0}-1} \\
		&-(\frac1{\alpha_0}+2)\frac{(H_0(x_t)^{-\alpha_0}(\log(H_0(x_t)))^2}{(H_0(x_t)^{-\alpha_0}+H_0(x_{t+1})^{-\alpha_0}-1)^2} \\
		&-(\frac1{\alpha_0}+2)\frac{H_0(x_{t+1})^{-\alpha_0}(log(H_0(x_{t+1})))^2)(H_0(x_t)^{-\alpha_0}+H_0(x_{t+1})^{-\alpha_0}-1)}{(H_0(x_t)^{-\alpha_0}+H_0(x_{t+1})^{-\alpha_0}-1)^2} \\
		&\begin{aligned}+(\frac{1}{\alpha_{0}}+2)\frac{(H_{0}(x_{t})^{-\alpha_{0}}\mathrm{log}(H_{0}(x_{t}))+H_{0}(x_{t+1})^{-\alpha_{0}}\mathrm{log}(H_{0}(x_{t+1})))^{2}}{(H_{0}(x_{t})^{-\alpha_{0}}+H_{0}(x_{t+1})^{-\alpha_{0}}-1)^{2}}\end{aligned}\biggr\} \\
		\partial_{\alpha_{0},\alpha_{1}}\ell =&0 \\
		\partial_{\alpha_{0},{k_0}}\ell =&\sum_{t=1}^{\tau-1}\left\{-\frac{\partial_{k_0}H_0(x_t)}{H_0(x_t)}-\frac{\partial_{k_0}H_0(x_{t+1})}{H_0(x_{t+1})}\right. \\
		&+2\frac{H_{0}(y_{t})^{-\alpha_{0}-1}\partial_{k_0}H_{0}(x_{t})+H_{0}(x_{t+1})^{-\alpha_{0}-1}\partial_{k_0}H_{0}(x_{t+1})}{H_{0}(x_{t})^{-\alpha_{0}}+H_{0}(x_{t+1})^{-\alpha_{0}}-1} \\
		&+(1+2\alpha_{0})[-\frac{H_{0}(x_{t})^{-\alpha_{0}-1}\partial_{k_0}H_{0}(x_{t})\operatorname{log}H_{0}(x_{t})+H_{0}(x_{t+1})^{-\alpha_{0}-1}\partial_{k_0}H_{0}(x_{t+1})\operatorname{log}H_{0}(x_{t+1})}{H_{0}(x_{t})^{-\alpha_{0}}+H_{0}(x_{t+1})^{-\alpha_{0}}-1}] \\
		&\begin{aligned}+(1+2\alpha_{0})[\frac{(H_{0}(x_{t})^{-\alpha_{0}-1}\partial_{k_0}H_{0}(x_{t})+H_{0}(x_{t+1})^{-\alpha_{0}-1}\partial_{k_0}H_{0}(x_{t+1}))}{(H_{0}(x_{t})^{-\alpha_{0}}+H_{0}(x_{t+1})^{-\alpha_{0}}-1)^{2}}\end{aligned} \\
		&\times(H_{0}(x_{t})^{-\alpha_{0}}\log H_{0}(x_{t})+H_{0}(x_{t+1})^{-\alpha_{0}}\log H_{0}(x_{t+1}))]\biggr\} \\
		\partial_{\alpha_{0},k_1}\ell =&0 \\
		\partial_{\alpha_{0},\lambda_{0}}\ell =&\sum_{t=1}^{\tau-1}\left\{-\frac{\partial_{\lambda_{0}}H_0(x_t)}{H_0(x_t)}-\frac{\partial_{\lambda_{0}}H_0(x_{t+1})}{H_0(x_{t+1})}\right. \\
		&+2\frac{H_{0}(x_{t})^{-\alpha_{0}-1}\partial_{\lambda_{0}}H_{0}(x_{t})+H_{0}(x_{t+1})^{-\alpha_{0}-1}\partial_{\lambda_{0}}H_{0}(x_{t+1})}{H_{0}(x_{t})^{-\alpha_{0}}+H_{0}(x_{t+1})^{-\alpha_{0}}-1} \\
		&+(1+2\alpha_{0})[-\frac{H_{0}(x_{t})^{-\alpha_{0}-1}\partial_{\lambda_{0}}H_{0}(x_{t})\operatorname{log}H_{0}(x_{t})+H_{0}(x_{t+1})^{-\alpha_{0}-1}\partial_{\lambda_{0}}H_{0}(x_{t+1})\operatorname{log}H_{0}(x_{t+1})}{H_{0}(x_{t})^{-\alpha_{0}}+H_{0}(x_{t+1})^{-\alpha_{0}}-1}] \\
		&\begin{aligned}+(1+2\alpha_{0})[\frac{(H_{0}(x_{t})^{-\alpha_{0}-1}\partial_{\lambda_{0}}H_{0}(x_{t})+H_{0}(x_{t+1})^{-\alpha_{0}-1}\partial_{\lambda_{0}}H_{0}(x_{t+1}))}{(H_{0}(x_{t})^{-\alpha_{0}}+H_{0}(x_{t+1})^{-\alpha_{0}}-1)^{2}}\end{aligned} \\
		&\times(H_{0}(x_{t})^{-\alpha_{0}}\log H_{0}(x_{t})+H_{0}(x_{t+1})^{-\alpha_{0}}\log H_{0}(x_{t+1}))]\biggr\}, \\
		\partial_{\alpha_{0},\lambda_{1}}\ell =&0, \\
	\end{aligned}
\end{equation*}

\begin{equation*}
	\begin{aligned}
		\partial_{\alpha_{1},\alpha_{1}}\ell =&\sum_{t=\tau+1}^T\left\{-\frac{1}{(1+\alpha_1)^2}-\frac{2}{\alpha_1^3}\mathrm{log}(H_1(x_t)^{-\alpha_1}+H_1(x_{t+1})^{-\alpha_1}-1)\right. \\
		&-\frac2{\alpha_1^2}\frac{H_1(x_t)^{-\alpha_1}\mathrm{log}(H_1(x_t))+H_1(x_{t+1})^{-\alpha_1}\mathrm{log}(H_1(x_{t+1}))}{H_1(x_t)^{-\alpha_1}+H_1(x_{t+1})^{-\alpha_1}-1} \\
		&-(\frac1{\alpha_1}+2)\frac{(H_1(x_t)^{-\alpha_1}(\log(H_1(x_t)))^2}{(H_1(x_t)^{-\alpha_1}+H_1(x_{t+1})^{-\alpha_1}-1)^2} \\
		&-(\frac{1}{\alpha_{1}}+2)\frac{H_{1}(x_{t+1})^{-\alpha_{1}}(\log(H_{1}(x_{t+1})))^{2})(H_{1}(x_{t})^{-\alpha_{1}}+H_{1}(x_{t+1})^{-\alpha_{1}}-1)}{(H_{1}(x_{t})^{-\alpha_{1}}+H_{1}(x_{t+1})^{-\alpha_{1}}-1)^{2}} \\
		&\begin{aligned}+(\frac{1}{\alpha_1}+2)\frac{(H_1(x_t)^{-\alpha_1}\mathrm{log}(H_1(x_t))+H_1(x_{t+1})^{-\alpha_1}\mathrm{log}(H_1(x_{t+1})))^2}{(H_1(x_t)^{-\alpha_1}+H_1(x_{t+1})^{-\alpha_1}-1)^2}\end{aligned}\biggr\} \\
		\partial_{\alpha_{1},k_0}\ell =&0, \\
		\partial_{\alpha_{1},k_1}\ell =&\sum_{t=\tau+1}^T\left\{-\frac{\partial_{k_1}H_1(x_t)}{H_1(x_t)}-\frac{\partial_{k_1}H_1(x_{t+1})}{H_1(x_{t+1})}\right. \\
		&\begin{aligned}+2\frac{H_1(x_t)^{-\alpha_1-1}\partial_{k_1}H_1(x_t)+H_1(x_{t+1})^{-\alpha_1-1}\partial_{k_1}H_1(x_{t+1})}{H_1(x_t)^{-\alpha_1}+H_1(x_{t+1})^{-\alpha_1}-1}\end{aligned} \\
		&\begin{aligned}+(1+2\alpha_1)[-\frac{H_1(x_t)^{-\alpha_1-1}\partial_{k_1}H_1(x_t)\log H_1(x_t)+H_1(x_{t+1})^{-\alpha_1-1}\partial_{k_1}H_1(x_{t+1})\log H_1(x_{t+1})}{H_1(x_t)^{-\alpha_1}+H_1(x_{t+1})^{-\alpha_1}-1}]\end{aligned} \\
		&\begin{aligned}+(1+2\alpha_{1})[\frac{(H_{1}(x_{t})^{-\alpha_{1}-1}\partial_{k_1}H_{1}(x_{t})+H_{1}(x_{t+1})^{-\alpha_{1}-1}\partial_{k_1}H_{1}(x_{t+1}))}{(H_{1}(x_{t})^{-\alpha_{1}}+H_{1}(x_{t+1})^{-\alpha_{1}}-1)^{2}}\end{aligned} \\
		&\times(H_{1}(x_{t})^{-\alpha_{1}}\operatorname{log}H_{1}(x_{t})+H_{1}(x_{t+1})^{-\alpha_{1}}\operatorname{log}H_{1}(x_{t+1}))]\Biggr\} \\
		\partial_{\alpha_1,\lambda_0}\ell =&0 \\
		\partial_{\alpha_1,\lambda_1}\ell =&\sum_{t=\tau+1}^T\left\{-\frac{\partial_{\lambda_1}H_1(x_t)}{H_1(x_t)}-\frac{\partial_{\lambda_1}H_1(x_{t+1})}{H_1(x_{t+1})}\right. \\
		&\begin{aligned}+2\frac{H_{1}(x_{t})^{-\alpha_{1}-1}\partial_{\lambda_{1}}H_{1}(x_{t})+H_{1}(x_{t+1})^{-\alpha_{1}-1}\partial_{\lambda_{1}}H_{1}(x_{t+1})}{H_{1}(x_{t})^{-\alpha_{1}}+H_{1}(x_{t+1})^{-\alpha_{1}}-1}\end{aligned} \\
		&\begin{aligned}+(1+2\alpha_1)[-\frac{H_1(x_t)^{-\alpha_1-1}\partial_{\lambda_1}H_1(x_t)\log H_1(x_t)+H_1(x_{t+1})^{-\alpha_1-1}\partial_{\lambda_1}H_1(x_{t+1})\log H_1(x_{t+1})}{H_1(x_t)^{-\alpha_1}+H_1(x_{t+1})^{-\alpha_1}-1}]\end{aligned} \\
		&\begin{aligned}+(1+2\alpha_1)[\frac{(H_1(x_t)^{-\alpha_1-1}\partial_{\lambda_1}H_1(x_t)+H_1(x_{t+1})^{-\alpha_1-1}\partial_{\lambda_1}H_1(x_{t+1}))}{(H_1(x_t)^{-\alpha_1}+H_1(x_{t+1})^{-\alpha_1}-1)^2}\end{aligned} \\
		&\times(H_{1}(x_{t})^{-\alpha_{1}}\operatorname{log}H_{1}(x_{t})+H_{1}(x_{t+1})^{-\alpha_{1}}\operatorname{log}H_{1}(x_{t+1}))]\Biggr\}
	\end{aligned}
\end{equation*}
\begin{equation*}
	\begin{aligned}
		\partial_{k_0,k_0}\ell =&\sum_{t=1}^{\tau}\left\{\frac{\partial_{k_0,k_0}h_0(x_t)}{h_0(x_t)}-\frac{\partial_{k_0}h_0(x_t)^2}{h_0(x_t)^2}\right\}+\sum_{t=1}^{\tau-1}\left\{-(1+\alpha_0)(\frac{\partial_{k_0,k_0}H_0(x_t)}{H_0(x_t)}-\frac{\partial_{k_0}H_0(x_t)^2}{H_0(x_t)^2})\right. \\
		&-(1+\alpha_{0})(\frac{\partial_{{k_0,k_0}}H_{0}(x_{t+1})}{H_{0}(x_{t+1})}-\frac{\partial_{{k_0}}H_{0}(x_{t+1})^{2}}{H_{0}(x_{t+1})^{2}}) \\
		&\begin{aligned}&-(1+2\alpha_0)(1+\alpha_0)\frac{(H_0(x_t)^{-\alpha_0-2}\partial_{k_0}H_0(x_t)^2+H_0(x_{t+1})^{-\alpha_0-2}\partial_{k_0}H_0(x_{t+1})^2)}{H_0(x_t)^{\alpha_0}+H_0(x_{t+1})^{\alpha_0}-1}\end{aligned} \\
		&+(1+2\alpha_0)\frac{(H_0(x_t)^{-\alpha_0-1}\partial_{k_0,k_0}H_0(x_t)+H_0(x_{t+1})^{-\alpha_0-1}\partial_{k_0,k_0}H_0(x_{t+1}))}{H_0(x_t)^{\alpha_0}+H_0(x_{t+1})^{\alpha_0}-1} \\
		&+\alpha_0(1+2\alpha_0)\frac{(H_0(x_t)^{-\alpha_0-1}\partial_{k_0}H_0(x_t)+H_0(x_{t+1})^{-\alpha_0-1}\partial_{k_0}H_0(x_{t+1}))^2}{(H_0(x_t)^{\alpha_0}+H_0(x_{t+1})^{\alpha_0}-1)^2} \\
		&-(1+\alpha_{01})(\frac{\partial_{{k_0,k_0}}H_{0}(x_{\tau})}{H_{0}(x_{\tau})}-\frac{\partial_{{k_0}}H_{0}(x_{\tau})^{2}}{H_{0}(x_{\tau})^{2}}) \\
		&-(1+2\alpha_{01})(1+\alpha_{01})\frac{H_{0}(x_{\tau})^{-\alpha_{01}-2}\partial_{k_0}H_{0}(x_{\tau})^{2}}{H_{0}(x_{\tau})^{-\alpha_{01}}+H_{1}(x_{\tau+1})^{-\alpha_{01}}-} \\
		&+(1+2\alpha_{01})\frac{H_0(x_\tau)^{-\alpha_{01}-1}\partial_{k_0,k_0}H_0(x_\tau)}{H_0(x_\tau)^{-\alpha_{01}}+H_1(x_{\tau+1})^{-\alpha_{01}}-1} \\
		&+\alpha_{01}(1+2\alpha_{01})\frac{(H_{0}(x_{\tau})^{-\alpha_{01}-1}\partial_{k_0}H_{0}(x_{\tau}))^{2}}{(H_{0}(x_{\tau})^{-\alpha_{01}}+H_{1}(x_{\tau+1})^{-\alpha_{01}}-1)^{2}} \\
		\partial_{k_0,k_1}\ell =&\alpha_{01}(1+2\alpha_{01})\frac{H_{0}(x_{\tau})^{-\alpha_{01}-1}\partial_{k_0}H_{0}(x_{\tau})H_{1}(x_{\tau+1})^{-\alpha_{01}-1}\partial_{k_1}H_{1}(x_{\tau+1})}{(H_{0}(x_{\tau})^{-\alpha_{01}}+H_{1}(x_{\tau+1})^{-\alpha_{01}}-1)^{2}} \\
		\partial_{k_0,\lambda_{0}}\ell & =\sum_{t=1}^\tau\left\{\frac{\partial_{k_0,\lambda_{0}}h_0(x_t)}{h_0(x_t)}-\frac{\partial_{k_0}h_0(x_t)\partial_{\lambda_0}h_0(x_t)}{h_0(x_t)^2}\right\} \\
		&+\sum_{t=1}^{\tau-1}\bigg\{-(1+\alpha_0)(\frac{\partial_{k_0,\lambda_{0}}H_0(x_t)}{H_0(x_t)}-\frac{\partial_{k_0}H_0(x_t)\partial_{\lambda_0}H_0(x_t)}{H_0(x_t)^2}) \\
		&-(1+\alpha_{0})(\frac{\partial_{{k_0,\lambda_{0}}}H_{0}(x_{t+1})}{H_{0}(x_{t+1})}-\frac{\partial_{{k_0}}H_{0}(x_{t+1})\partial_{{\lambda_{0}}}H_{0}(x_{t+1})}{H_{0}(x_{t+1})^{2}}) \\
		&\begin{aligned}&- (1+2\alpha_0)(1+\alpha_0)\frac{(H_0(x_t)^{-\alpha_0-2}\partial_{k_0}H_0(x_t)\partial_{\lambda_0}H_0(x_t)+H_0(x_{t+1})^{-\alpha_0-2}\partial_{k_0}H_0(x_{t+1})\partial_{\lambda_0}H_0(x_t))}{H_0(x_t)^{\alpha_0}+H_0(x_{t+1})^{\alpha_0}-1}\end{aligned} \\
		&+(1+2\alpha_{0})\frac{(H_{0}(x_{t})^{{-\alpha_{0}-1}}\partial_{{k_0,\lambda_{0}}}H_{0}(x_{t})+H_{0}(x_{t+1})^{{-\alpha_{0}-1}}\partial_{{k_0,\lambda_{0}}}H_{0}(x_{t+1}))}{H_{0}(x_{t})^{{\alpha_{0}}}+H_{0}(x_{t+1})^{{\alpha_{0}}}-1} \\
		&+\alpha_0(1+2\alpha_0)\frac{(H_0(x_t)^{-\alpha_0-1}\partial_{k_0}H_0(x_t)+H_0(x_{t+1})^{-\alpha_0-1}\partial_{k_0}H_0(x_{t+1}))}{(H_0(x_t)^{\alpha_0}+H_0(x_{t+1})^{\alpha_0}-1)^2} \\
		&\times\left(H_{0}(x_{t})^{-\alpha_{0}-1}\partial_{\lambda_{0}}H_{0}(x_{t})+H_{0}(x_{t+1})^{-\alpha_{0}-1}\partial_{\lambda_{0}}H_{0}(x_{t+1}))\right\} \\
		&- (1+\alpha_{01})(\frac{\partial_{{k_0,\lambda_{0}}}H_{0}(x_{\tau})}{H_{0}(y_{\tau})}-\frac{\partial_{{k_0}}H_{0}(x_{\tau})\partial_{{\lambda_{0}}}H_{0}(x_{\tau})}{H_{0}(x_{\tau})^{2}}) \\
		&-(1+2\alpha_{01})(1+\alpha_{01})\frac{H_{0}(x_{\tau})^{{-\alpha_{01}-2}}\partial_{k_0}H_{0}(x_{\tau})\partial_{\lambda_{0}}H_{0}(x_{\tau})}{H_{0}(x_{\tau})^{{-\alpha_{01}}}+H_{1}(x_{\tau+1})^{{-\alpha_{01}}}-1} \\
		&\begin{aligned}&+(1+2\alpha_{01})\frac{H_{0}(x_{\tau})^{-\alpha_{01}-1}\partial_{k_0,\lambda_{0}}H_{0}(x_{\tau})}{H_{0}(x_{\tau})^{-\alpha_{01}}+H_{1}(x_{\tau+1})^{-\alpha_{01}}-1}\end{aligned} + \alpha_{01}(1+2\alpha_{01})\frac{H_{0}(x_{\tau})^{-2\alpha_{01}-2}\partial_{k_0}H_{0}(x_{\tau})\partial_{\lambda_{0}}H_{0}(x_{\tau})}{(H_{0}(x_{\tau})^{-\alpha_{01}}+H_{1}(x_{\tau+1})^{-\alpha_{01}}-1)^{2}}\\
	\end{aligned}
\end{equation*}

\begin{equation*}
	\begin{aligned}
		\partial_{k_0,\lambda_{1}}\ell =&\alpha_{01}(1+2\alpha_{01})\frac{H_{0}(x_{\tau})^{-\alpha_{01}-1}\partial_{k_0}H_{0}(x_{\tau})H_{1}(x_{\tau+1})^{-\alpha_{01}-1}\partial_{\lambda_{1}}H_{1}(x_{\tau+1})}{(H_{0}(x_{\tau})^{-\alpha_{01}}+H_{1}(x_{\tau+1})^{-\alpha_{01}}-1)^{2}} \\
		\partial_{k_1,k_1}\ell =&\sum_{t=\tau+1}^{T}\left\{\frac{\partial_{k_1,k_1}h_1(x_t)}{h_1(x_t)}-\frac{\partial_{k_1}h_1(x_t)^2}{h_1(x_t)^2}\right\}+\sum_{t=\tau+1}^{T-1}\left\{-(1+\alpha_1)(\frac{\partial_{k_1,k_1}H_1(x_t)}{H_1(x_t)}-\frac{\partial_{k_1}H_1(x_t)^2}{H_1(x_t)^2})\right. \\
		&-(1+\alpha_{1})(\frac{\partial_{{k_1,k_1}}H_{1}(x_{t+1})}{H_{1}(x_{t+1})}-\frac{\partial_{{k_1}}H_{1}(x_{t+1})^{2}}{H_{1}(x_{t+1})^{2}}) \\
		&\begin{aligned}&-(1+2\alpha_1)(1+\alpha_1)\frac{(H_1(x_t)^{-\alpha_1-2}\partial_{k_1}H_1(x_t)^2+H_1(x_{t+1})^{-\alpha_1-2}\partial_{k_1}H_1(x_{t+1})^2)}{H_1(x_t)^{\alpha_1}+H_1(x_{t+1})^{\alpha_1}-1}\end{aligned} \\
		&+(1+2\alpha_1)\frac{(H_1(x_t)^{-\alpha_1-1}\partial_{k_1,k_1}H_1(x_t)+H_1(x_{t+1})^{-\alpha_1-1}\partial_{k_1,k_1}H_1(x_{t+1}))}{H_1(x_t)^{\alpha_1}+H_1(x_{t+1})^{\alpha_1}-1} \\
		&+\alpha_1(1+2\alpha_1)\frac{(H_1(x_t)^{-\alpha_1-1}\partial_{k_1}H_1(x_t)+H_1(x_{t+1})^{-\alpha_1-1}\partial_{k_1}H_1(x_{t+1}))^2}{(H_1(x_t)^{\alpha_1}+H_1(x_{t+1})^{\alpha_1}-1)^2} \\
		&-(1+\alpha_{01})(\frac{\partial_{{k_1,k_1}}H_{1}(x_{\tau+1})}{H_{1}(x_{\tau+1})}-\frac{\partial_{{k_1}}H_{1}(x_{\tau+1})^{2}}{H_{1}(x_{\tau+1})^{2}}) \\
		&-(1+2\alpha_{01})(1+\alpha_{01})\frac{H_{1}(x_{\tau+1})^{-\alpha_{01}-2}\partial_{k_1}H_{1}(x_{\tau+1})^{2}}{H_{0}(x_{\tau})^{-\alpha_{01}}+H_{1}(x_{\tau+1})^{-\alpha_{01}}-1} \\
		&+(1+2\alpha_{01})\frac{H_1(x_{\tau+1})^{-\alpha_{01}-1}\partial_{k_1,k_1}H_1(x_{\tau+1})}{H_0(x_\tau)^{-\alpha_{01}}+H_1(x_{\tau+1})^{-\alpha_{01}}-1} \\
		&+\alpha_{01}(1+2\alpha_{01})\frac{(H_{1}(x_{\tau+1})^{-\alpha_{01}-1}\partial_{k_1}H_{1}(x_{\tau+1}))^{2}}{(H_{0}(x_{\tau})^{-\alpha_{01}}+H_{1}(x_{\tau+1})^{-\alpha_{01}}-1)^{2}} \\
		\partial_{k_1,\lambda_{0}}\ell =&\alpha_{01}(1+2\alpha_{01})\frac{H_{1}(x_{\tau+1})^{-\alpha_{01}-1}\partial_{k_1}H_{1}(x_{\tau+1})H_{0}(x_{\tau})^{-\alpha_{01}-1}\partial_{\lambda_{0}}H_{0}(x_{\tau})}{(H_{0}(x_{\tau})^{-\alpha_{01}}+H_{1}(x_{\tau+1})^{-\alpha_{01}}-1)^{2}} 
	\end{aligned}
\end{equation*}

\begin{equation*}
	\begin{aligned}
		\partial_{k_1,\lambda_{1}}\ell & \begin{aligned}=\sum_{t=\tau+1}^T\left\{\frac{\partial_{k_1,\lambda_{1}}h_1(x_t)}{h_1(x_t)}-\frac{\partial_{k_1}h_1(x_t)\partial_{\lambda_1}h_1(x_t)}{h_1(x_t)^2}\right\}\end{aligned} \\
		&+\sum_{t=\tau+1}^{T-1}\bigg\{-(1+\alpha_1)(\frac{\partial_{k_1,\lambda_{1}}H_1(x_t)}{H_1(x_t)}-\frac{\partial_{k_1}H_1(x_t)\partial_{\lambda_1}H_1(x_t)}{H_1(x_t)^2}) \\
		&-(1+\alpha_{1})(\frac{\partial_{k_1,\lambda_{1}}H_{1}(x_{t+1})}{H_{1}(x_{t+1})}-\frac{\partial_{k_{1}}H_{1}(x_{t+1})\partial_{\lambda_{1}}H_{1}(x_{t+1})}{H_{1}(x_{t+1})^{2}}) \\
		&-(1+2\alpha_{1})(1+\alpha_{1})\frac{(H_{1}(x_{t})^{-\alpha_{1}-2}\partial_{k_1}H_{1}(x_{t})\partial_{\lambda_{1}}H_{1}(x_{t})+H_{1}(x_{t+1})^{-\alpha_{1}-2}\partial_{k_1}H_{1}(x_{t+1})\partial_{\lambda_{1}}H_{1}(x_{t}))}{H_{1}(x_{t})^{\alpha_{1}}+H_{1}(x_{t+1})^{\alpha_{1}}-1} \\
		&+(1+2\alpha_1)\frac{(H_1(x_t)^{-\alpha_1-1}\partial_{k_1,\lambda_{1}}H_1(x_t)+H_1(x_{t+1})^{-\alpha_1-1}\partial_{k_1,\lambda_{1}}H_1(x_{t+1}))}{H_1(x_t)^{\alpha_1}+H_1(x_{t+1})^{\alpha_1}-1} \\
		&+\alpha_{1}(1+2\alpha_{1})\frac{(H_{1}(x_{t})^{-\alpha_{1}-1}\partial_{k_1}H_{1}(x_{t})+H_{1}(x_{t+1})^{-\alpha_{1}-1}\partial_{k_1}H_{1}(x_{t+1}))}{(H_{1}(x_{t})^{\alpha_{1}}+H_{1}(x_{t+1})^{\alpha_{1}}-1)^{2}} \\
		&\times\left(H_1(x_t)^{-\alpha_1-1}\partial_{\lambda_1}H_1(x_t)+H_1(x_{t+1})^{-\alpha_1-1}\partial_{\lambda_1}H_1(x_{t+1}))\right\} \\
		&-(1+\alpha_{01})(\frac{\partial_{k_1,\lambda_{1}}H_1(x_{\tau+1})}{H_1(x_{\tau+1})}-\frac{\partial_{k_1}H_1(x_{\tau+1})\partial_{\lambda_1}H_1(x_{\tau+1})}{H_1(x_{\tau+1})^2}) \\
		&-(1+2\alpha_{01})(1+\alpha_{01})\frac{H_{1}(x_{\tau+1})^{-\alpha_{01}-2}\partial_{k_1}H_{1}(x_{\tau+1})\partial_{\lambda_{1}}H_{1}(x_{\tau+1})}{H_{0}(x_{\tau})^{-\alpha_{01}}+H_{1}(x_{\tau+1})^{-\alpha_{01}}-1} \\
		&+(1+2\alpha_{01})\frac{H_{1}(x_{\tau+1})^{-\alpha_{01}-1}\partial_{{k_1,\lambda_{1}}}H_{1}(x_{\tau+1})}{H_{0}(x_{\tau})^{-\alpha_{01}}+H_{1}(x_{\tau+1})^{-\alpha_{01}}-1} \\
		&+ \alpha_{01}(1+2\alpha_{01})\frac{H_{1}(x_{\tau+1})^{-2\alpha_{01}-2}\partial_{k_1}H_{1}(x_{\tau+1})\partial_{\lambda_{1}}H_{1}(x_{\tau+1})}{(H_{0}(x_{\tau})^{-\alpha_{01}}+H_{1}(x_{\tau+1})^{-\alpha_{01}}-1)^{2}}\\
		\partial_{\lambda_{0},\lambda_{0}}\ell =&\sum_{t=1}^{\tau}\left\{\frac{\partial_{\lambda_{0},\lambda_{0}}h_0(x_t)}{h_0(x_t)}-\frac{\partial_{\lambda_{0}}h_0(x_t)^2}{h_0(x_t)^2}\right\}+\sum_{t=1}^{\tau-1}\left\{-(1+\alpha_0)(\frac{\partial_{\lambda_{0},\lambda_{0}}H_0(x_t)}{H_0(x_t)}-\frac{\partial_{\lambda_{0}}H_0(x_t)^2}{H_0(x_t)^2})\right. \\
		&-(1+\alpha_{0})(\frac{\partial_{{\lambda_{0},\lambda_{0}}}H_{0}(x_{t+1})}{H_{0}(x_{t+1})}-\frac{\partial_{{\lambda_{0}}}H_{0}(x_{t+1})^{2}}{H_{0}(x_{t+1})^{2}}) \\
		&\begin{aligned}&-(1+2\alpha_0)(1+\alpha_0)\frac{(H_0(x_t)^{-\alpha_0-2}\partial_{\lambda_{0}}H_0(x_t)^2+H_0(x_{t+1})^{-\alpha_0-2}\partial_{\lambda_{0}}H_0(x_{t+1})^2)}{H_0(x_t)^{\alpha_0}+H_0(x_{t+1})^{\alpha_0}-1}\end{aligned} \\
		&+(1+2\alpha_0)\frac{(H_0(x_t)^{-\alpha_0-1}\partial_{\lambda_{0},\lambda_{0}}H_0(x_t)+H_0(x_{t+1})^{-\alpha_0-1}\partial_{\lambda_{0},\lambda_{0}}H_0(x_{t+1}))}{H_0(x_t)^{\alpha_0}+H_0(x_{t+1})^{\alpha_0}-1} \\
		&+\alpha_0(1+2\alpha_0)\frac{(H_0(x_t)^{-\alpha_0-1}\partial_{\lambda_{0}}H_0(x_t)+H_0(x_{t+1})^{-\alpha_0-1}\partial_{\lambda_{0}}H_0(x_{t+1}))^2}{(H_0(x_t)^{\alpha_0}+H_0(x_{t+1})^{\alpha_0}-1)^2} \\
		&-(1+\alpha_{01})(\frac{\partial_{{\lambda_{0},\lambda_{0}}}H_{0}(x_{\tau})}{H_{0}(x_{\tau})}-\frac{\partial_{{\lambda_{0}}}H_{0}(x_{\tau})^{2}}{H_{0}(x_{\tau})^{2}}) \\
		&-(1+2\alpha_{01})(1+\alpha_{01})\frac{H_{0}(x_{\tau})^{-\alpha_{01}-2}\partial_{\lambda_{0}}H_{0}(x_{\tau})^{2}}{H_{0}(x_{\tau})^{-\alpha_{01}}+H_{1}(x_{\tau+1})^{-\alpha_{01}}-} \\
		&+(1+2\alpha_{01})\frac{H_0(x_\tau)^{-\alpha_{01}-1}\partial_{\lambda_{0},\lambda_{0}}H_0(x_\tau)}{H_0(x_\tau)^{-\alpha_{01}}+H_1(x_{\tau+1})^{-\alpha_{01}}-1} \\
		&+\alpha_{01}(1+2\alpha_{01})\frac{(H_{0}(x_{\tau})^{-\alpha_{01}-1}\partial_{\lambda_{0}}H_{0}(x_{\tau}))^{2}}{(H_{0}(x_{\tau})^{-\alpha_{01}}+H_{1}(x_{\tau+1})^{-\alpha_{01}}-1)^{2}} 
	\end{aligned}
\end{equation*}

\begin{equation*}
	\begin{aligned}
	\partial_{\lambda_{0},\lambda_{1}}\ell =&\alpha_{01}(1+2\alpha_{01})\frac{G_{0}(x_{\tau})^{-\alpha_{01}-1}\partial_{\lambda_{0}}H_{0}(x_{\tau})H_{1}(x_{\tau+1})^{-\alpha_{01}-1}\partial_{\lambda_{1}}H_{1}(x_{\tau+1})}{(H_{0}(x_{\tau})^{-\alpha_{01}}+H_{1}(x_{\tau+1})^{-\alpha_{01}}-1)^{2}} \\
		\partial_{\lambda_{1},\lambda_{1}}\ell =&\sum_{t=\tau+1}^{T}\left\{\frac{\partial_{\lambda_{1},\lambda_{1}}h_1(x_t)}{h_1(x_t)}-\frac{\partial_{\lambda_{1}}h_1(x_t)^2}{h_1(x_t)^2}\right\}+\sum_{t=\tau+1}^{T-1}\left\{-(1+\alpha_1)(\frac{\partial_{\lambda_{1},\lambda_{1}}H_1(x_t)}{H_1(x_t)}-\frac{\partial_{\lambda_{1}}H_1(x_t)^2}{H_1(x_t)^2})\right. \\
		&-(1+\alpha_{1})(\frac{\partial_{{\lambda_{1},\lambda_{1}}}H_{1}(x_{t+1})}{H_{1}(x_{t+1})}-\frac{\partial_{{\lambda_{1}}}H_{1}(x_{t+1})^{2}}{H_{1}(x_{t+1})^{2}}) \\
		&\begin{aligned}&-(1+2\alpha_1)(1+\alpha_1)\frac{(H_1(x_t)^{-\alpha_1-2}\partial_{\lambda_{1}}H_1(x_t)^2+H_1(x_{t+1})^{-\alpha_1-2}\partial_{\lambda_{1}}H_1(x_{t+1})^2)}{H_1(x_t)^{\alpha_1}+H_1(x_{t+1})^{\alpha_1}-1}\end{aligned} \\
		&+(1+2\alpha_1)\frac{(H_1(x_t)^{-\alpha_1-1}\partial_{\lambda_{1},\lambda_{1}}H_1(x_t)+H_1(x_{t+1})^{-\alpha_1-1}\partial_{\lambda_{1},\lambda_{1}}H_1(x_{t+1}))}{H_1(x_t)^{\alpha_1}+H_1(x_{t+1})^{\alpha_1}-1} \\
		&+\alpha_1(1+2\alpha_1)\frac{(H_1(x_t)^{-\alpha_1-1}\partial_{\lambda_{1}}H_1(x_t)+H_1(x_{t+1})^{-\alpha_1-1}\partial_{\lambda_{1}}H_1(x_{t+1}))^2}{(H_1(x_t)^{\alpha_1}+H_1(x_{t+1})^{\alpha_1}-1)^2} \\
		&-(1+\alpha_{01})(\frac{\partial_{{\lambda_{1},\lambda_{1}}}H_{1}(x_{\tau+1})}{H_{1}(x_{\tau+1})}-\frac{\partial_{{\lambda_{1}}}H_{1}(x_{\tau+1})^{2}}{H_{1}(x_{\tau+1})^{2}}) \\
		&-(1+2\alpha_{01})(1+\alpha_{01})\frac{H_{1}(x_{\tau+1})^{-\alpha_{01}-2}\partial_{\lambda_{1}}H_{1}(x_{\tau+1})^{2}}{H_{0}(x_{\tau})^{-\alpha_{01}}+H_{1}(x_{\tau+1})^{-\alpha_{01}}-1} \\
		&+(1+2\alpha_{01})\frac{H_1(x_{\tau+1})^{-\alpha_{01}-1}\partial_{\lambda_{1},\lambda_{1}}H_1(x_{\tau+1})}{H_0(x_\tau)^{-\alpha_{01}}+H_1(x_{\tau+1})^{-\alpha_{01}}-1} \\
		&+\alpha_{01}(1+2\alpha_{01})\frac{(H_{1}(x_{\tau+1})^{-\alpha_{01}-1}\partial_{\lambda_{1}}H_{1}(x_{\tau+1}))^{2}}{(H_{0}(x_{\tau})^{-\alpha_{01}}+H_{1}(x_{\tau+1})^{-\alpha_{01}}-1)^{2}} \\
	\end{aligned}
\end{equation*}

where
\begin{equation*}
	\begin{aligned}
		\partial_{k_0,k_0}h_0(x)=&\frac1{\lambda_0}\left(\frac x{\lambda_0}\right)^{k_0-1}e^{-\left(\frac x{\lambda_0}\right)^{k_0}}\ln\left(\frac x{\lambda_0}\right)\left\{2+k_0\ln\left(\frac x{\lambda_0}\right)\left[1-\left(\frac x{\lambda_0}\right)^{2k_0}-\left(\frac x{\lambda_0}\right)^{k_0}\right]\right\}\\
		\partial_{k_1,k_1}h_1(x)=&\frac1{\lambda_1}\left(\frac x{\lambda_1}\right)^{k_1-1}e^{-\left(\frac x{\lambda_1}\right)^{k_1}}\ln\left(\frac x{\lambda_1}\right)\left\{2+k_1\ln\left(\frac x{\lambda_1}\right)\left[1-\left(\frac x{\lambda_1}\right)^{2k_1}-\left(\frac x{\lambda_1}\right)^{k_1}\right]\right\}\\
		\partial_{k_0,\lambda_0}h_0(x)=&\left(\frac{x}{\lambda_0}\right)^{k_0-1}e^{-\left(\frac{x}{\lambda_0}\right)^{k_0}}\left\{1+k_0\ln\left(\frac{x}{\lambda_0}\right)+\frac{1-2k_0}{\lambda_0^2}\right.+k_0\left(\frac{x}{\lambda_0}\right)^{k_0}\left[-\ln\left(\frac{x}{\lambda_0}\right)+\frac{k_0}{\lambda_0^2}\right]\\
		&\quad+\frac{k_0}{\lambda_0^2}\ln\left(\frac{x}{\lambda_0}\right)\left[\left(\frac{x}{\lambda_0}\right)^{k_0}(1+k_0)+(1-k_0)\right]-\frac{k_0^2}{\lambda_0^2}\biggl(\frac{x}{\lambda_0}\biggr)^{2k_0}\ln\biggl(\frac{x}{\lambda_0}\biggr)\biggr\}\\
		\partial_{k_1,\lambda_1}h_1(x)& =\left(\frac{x}{\lambda_1}\right)^{k_1-1}e^{-\left(\frac{x}{\lambda_1}\right)^{k_1}}\left\{1+k_1\ln\left(\frac{x}{\lambda_1}\right)+\frac{1-2k_1}{\lambda_1^2}\right.+k_1\left(\frac{x}{\lambda_1}\right)^{k_1}\left[-\ln\left(\frac{x}{\lambda_1}\right)+\frac{k_1}{\lambda_1^2}\right]\\
		&\quad+\frac{k_1}{\lambda_1^2}\ln\left(\frac{x}{\lambda_1}\right)\left[\left(\frac{x}{\lambda_1}\right)^{k_1}(1+k_1)+(1-k_1)\right]-\frac{k_1^2}{\lambda_1^2}\biggl(\frac{x}{\lambda_1}\biggr)^{2k_1}\ln\biggl(\frac{x}{\lambda_1}\biggr)\biggr\}\\
		\partial_{\lambda_0,\lambda_0}h_0(x)=&\frac{k_0^2}{\lambda_0^3}e^{-\left(\frac{x}{\lambda_0}\right)^{k_0}}\left\{-(k_0+1)\left[\left(\frac{x}{\lambda_0}\right)^{k_0}-1\right]+k_0\left(\frac{x}{\lambda_0}\right)^{k_0-1}\left[\left(\frac{x}{\lambda_0}\right)^{2k_0-1}-\left(\frac{x}{\lambda_0}\right)^{k_0}-1\right]\right\}\\
		\partial_{\lambda_1,\lambda_1}h_1(x)=&\frac{k_1^2}{\lambda_1^3}e^{-\left(\frac{x}{\lambda_1}\right)^{k_1}}\left\{-(k_1+1)\left[\left(\frac{x}{\lambda_1}\right)^{k_1}-1\right]+k_1\left(\frac{x}{\lambda_1}\right)^{k_1-1}\left[\left(\frac{x}{\lambda_1}\right)^{2k_1-1}-\left(\frac{x}{\lambda_1}\right)^{k_1}-1\right]\right\}.
	\end{aligned}
\end{equation*}

\subsection{First- and Second-Order Derivatives for the Joe Copula}

The Joe copula is defined by
\[
C_\alpha(u,v)=1-J(u,v,\alpha)^{1/\alpha},
\]
where
\[
J(u,v,\alpha)=(1-u)^{\alpha}+(1-v)^{\alpha}-(1-u)^{\alpha}(1-v)^{\alpha}.
\]
The corresponding copula density function is
\[
c_\alpha(u,v)=J(u,v,\alpha)^{\frac{1}{\alpha}-2}(1-u)^{\alpha-1}(1-v)^{\alpha-1}
\left[(\alpha-1)+J(u,v,\alpha)\right].
\]
Hence, the log-density function can be written as
\[
\log c_\alpha(u,v)
=
\left(\frac{1}{\alpha}-2\right)\log J(u,v,\alpha)
+
(\alpha-1)\log(1-u)
+
(\alpha-1)\log(1-v)
+
\log\left[(\alpha-1)+J(u,v,\alpha)\right].
\]
For notational simplicity, let
\[
H_k^{t}=H_k(x_t), \qquad H_k^{t+1}=H_k(x_{t+1}), \qquad k=0,1.
\]
The corresponding first- and second-order derivatives are given as follows:
\begin{equation*}
	\begin{aligned}
		& \partial_\alpha\log c_\alpha(H_0^t,H_0^{t+1})=-\frac{1}{\alpha^2}\log\mathcal{J}+\left(\frac{1}{\alpha}-2\right)\frac{\partial_\alpha J}{J}+\log(1-H_0^t)+\log(1-H_0^{t+1})+
		\frac{1+\partial_\alpha J}{\alpha-1+J}
	\end{aligned}
\end{equation*}

\begin{equation*}
	\begin{aligned}
		\partial_k\log c_\alpha(H_0^t,H_0^{t+1})=&\left(\frac{1}{\alpha}-2\right)\frac{\partial_kJ(\partial_kH_0^t+\partial_kH_0^{t+1})}{J}\\
		&+(\alpha-1)\left(\frac{\partial_kH_0^t}{1-H_0^t}+\frac{\partial_kH_0^{t+1}}{1-H_0^{t+1}}\right)+
		\frac{\partial_kJ(\partial_kH_0^t+\partial_kH_0^{t+1})}{\alpha-1+J}
	\end{aligned}
\end{equation*}

\begin{equation*}
	\begin{aligned}
		\partial_\lambda\log c_\alpha(H_0^t,H_0^{t+1})=&\left(\frac{1}{\alpha}-2\right)\frac{\partial_\lambda J(\partial_\lambda H_0^t+\partial_\lambda H_0^{t+1})}{J}\\
		&+(\alpha-1)\left(\frac{\partial_\lambda H_0^t}{1-H_0^t}+\frac{\partial_\lambda H_0^{t+1}}{1-H_0^{t+1}}\right)+ 
		\frac{\partial_\lambda J(\partial_\lambda H_0^t+\partial_\lambda H_0^{t+1})}{\alpha-1+J}
	\end{aligned}
\end{equation*}

\begin{equation*}
	\begin{aligned}
		\partial_{\alpha\alpha}\log c_\alpha(H_0^t,H_0^{t+1})=&\frac{2}{\alpha^3}\log J-\frac{1}{\alpha^2}\frac{\partial_\alpha J}{J}+\left(\frac{1}{\alpha}-2\right)\frac{\partial_{\alpha\alpha}JJ-\partial_\alpha J^2}{J^2}\\
		&+
		\frac{\partial_{\alpha\alpha}J(\alpha-1+J)-\partial_{\alpha}J(1+\partial_{\alpha}J)}{(\alpha-1+J)^{2}}
	\end{aligned}
\end{equation*}

\begin{equation*}
	\begin{aligned}
		\partial_{\alpha k}\log c_{\alpha}(H_{0}^{t},H_{0}^{t+1})=&-\frac{1}{\alpha^{2}}\frac{\partial_{k}J(\partial_{k}H_{0}^{t}+\partial_{k}H_{0}^{t+1})}{J}-\frac{\partial_kH_0^t}{1-H_0^t}-\frac{\partial_kH_0^{t+1}}{1-H_0^{t+1}}\\
		&+ \frac{\partial_{\alpha k}J\left(\partial_{\alpha k}H_0^t+\partial_{\alpha k}H_0^{t+1}\right)J-\partial_\alpha J\partial_kJ\left(\partial_kH_0^t+\partial_kH_0^{t+1}\right)}{J^2} \\
		&+ 
		\frac{\partial_{\alpha k}J\left(\partial_{\alpha k}H_0^t+\partial_{\alpha k}H_0^{t+1}\right)(\alpha-1+J)-(1+\partial_\alpha J)\partial_kJ\left(\partial_kH_0^t+\partial_kH_0^{t+1}\right)}{(\alpha-1+J)^2}
	\end{aligned}
\end{equation*}

\begin{equation*}
	\begin{aligned}
		\partial_{\alpha\lambda}\log c_{\alpha}(H_{0}^{t},H_{0}^{t+1})=&-\frac{1}{\alpha^{2}}\frac{\partial_{\lambda}J(\partial_{\lambda}H_{0}^{l}+\partial_{\lambda}H_{0}^{l+1})}{J}
		-\frac{\partial_{\lambda}H_{0}^{t}}{1-H_{0}^{t}}-\frac{\partial_{\lambda}H_{0}^{t+1}}{1-H_{0}^{t+1}}+ \\&+\frac{\partial_{\alpha\lambda}J(\partial_{\alpha\lambda}H_{0}^{t}+\partial_{\alpha\lambda}H_{0}^{t+1})J-\partial_{\alpha}J\partial_{\lambda}J(\partial_{\lambda}H_{0}^{t}+\partial_{\lambda}H_{0}^{t+1})}{J^{2}}\\
		&\frac{\partial_{\alpha\lambda}J\left(\partial_{\alpha\lambda}H_{0}^{t}+\partial_{\alpha\lambda}H_{0}^{t+1}\right)(\alpha-1+J)-(1+\partial_{\alpha}J)\partial_{\lambda}J\left(\partial_{\lambda}H_{0}^{t}+\partial_{\lambda}H_{0}^{t+1}\right)}{(\alpha-1+J)^{2}}
	\end{aligned}
\end{equation*}

\begin{equation*}
	\begin{aligned}
		&\partial_{k\lambda}\operatorname{log}c_{\alpha}(H_{0}^{t},H_{0}^{t+1}) =\left\{\frac{\partial_{k\lambda}J\left(\partial_{\lambda}G_{0}^{t}+\partial_{\lambda}H_{0}^{t+1}\right)\left(\partial_{k}H_{0}^{t}+\partial_{k}H_{0}^{t+1}\right)+\partial_{k}J\left(\partial_{k\lambda}H_{0}^{t}+\partial_{k\lambda}H_{0}^{t+1}\right)J}{J^{2}}\right. \\
		&-\left.\frac{\partial_{k}J\partial_{\lambda}J\left(\partial_{\lambda}H_{0}^{t}+\partial_{\lambda}H_{0}^{t+1}\right)\left(\partial_{k}H_{0}^{t}+\partial_{k}H_{0}^{t+1}\right)}{J^{2}}\right\}    \left(\frac{1}{\alpha}-2\right)\\
		& +(\alpha-1)\left(\frac{\partial_{k\lambda}H_{0}^{t}\left(1-H_{0}^{t}\right)+\partial_{k}H_{0}^{t}\partial_{\lambda}H_{0}^{t}}{\left(1-H_{0}^{t}\right)^{2}}+\right.
		\left.\frac{\partial_{k\lambda}H_0^{t+1}\left(1-H_0^{t+1}\right)+\partial_kH_0^{t+1}\partial_\lambda H_0^{t+1}}{\left(1-H_0^{t+1}\right)^2}\right) \\
		& + \frac{\{\partial_{k\lambda}J\left(\partial_{\lambda}H_{0}^{t}+\partial_{\lambda}H_{0}^{t+1}\right)\left(\partial_{k}H_{0}^{t}+\partial_{k}H_{0}^{t+1}\right)+\partial_{k}J\left(\partial_{k\lambda}H_{0}^{t}+\partial_{k\lambda}H_{0}^{t+1}\right)\}(\alpha-1+J)}{(\alpha-1+J)^2} \\
		& -\frac{\partial_{k}J\partial_{\lambda}J\left(\partial_{\lambda}H_{0}^{t}+\partial_{\lambda}H_{0}^{t+1}\right)\left(\partial_{k}H_{0}^{t}+\partial_{k}H_{0}^{t+1}\right)}{(\alpha-1+J)^{2}}
	\end{aligned}
\end{equation*}
where
\begin{equation*}
	\partial_{\alpha}J(H_{0}^{t},H_{0}^{t+1},\alpha)=\alpha\{(1-H_{0}^{t})^{\alpha-1}+(1-H_{0}^{t+1})^{\alpha-1}-[(1-H_{0}^{t})(1-H_{0}^{t+1})]^{\alpha-1}\}
\end{equation*}
\begin{equation*}
	\begin{aligned}
		\partial_k\mathcal{J}(H_0^t,H_0^{t+1},\alpha) =&-\alpha(1-H_0^t)^{\alpha-1}\partial_kH_0^t-\alpha(1-H_0^{t+1})^{\alpha-1}\partial_kH_0^{t+1}\\
		& +\alpha(1-H_0^t)^{\alpha-1}\partial_kH_0^t(1-H_0^{t+1})^\alpha+\alpha(1-H_0^t)^\alpha\partial_kH_0^{t+1}(1-H_0^{t+1})^{\alpha-1}
	\end{aligned}
\end{equation*}
\begin{equation*}
	\begin{aligned}
		\partial_\lambda\mathcal{J}(H_0^t,H_0^{t+1},\alpha) =&-\alpha(1-H_0^t)^{\alpha-1}\partial_\lambda H_0^t-\alpha(1-H_0^{t+1})^{\alpha-1}\partial_\lambda H_0^{t+1}\\
		& +\alpha(1-H_0^t)^{\alpha-1}\partial_\lambda H_0^t(1-H_0^{t+1})^\alpha+\alpha(1-H_0^t)^\alpha\partial_\lambda H_0^{t+1}(1-H_0^{t+1})^{\alpha-1}
	\end{aligned}
\end{equation*}
\begin{equation*}
	\partial_{\alpha\alpha}\mathcal{J}(H_{0}^{t},H_{0}^{t+1},\alpha)=\alpha(\alpha-1)\{(1-H_{0}^{t})^{\alpha-2}+(1-H_{0}^{t+1})^{\alpha-2}-[(1-H_{0}^{t})(1-H_{0}^{t+1})]^{\alpha-2}\}
\end{equation*}
\begin{equation*}
	\begin{aligned}
		\partial_{\alpha k}\mathcal{J}(H_0^t,H_0^{t+1},\alpha) =&-\alpha(1-H_0^t)^{\alpha-1}\partial_kH_0^t-\alpha(1-H_0^{t+1})^{\alpha-1}\partial_kH_0^{t+1}\\
		& +\alpha(1-H_0^t)^{\alpha-1}\partial_kH_0^t(1-H_0^{t+1})^\alpha+\alpha(1-H_0^t)^\alpha\partial_kH_0^{t+1}(1-H_0^{t+1})^{\alpha-1}
	\end{aligned}
\end{equation*}
\begin{equation*}
	\begin{aligned}
		\partial_{k\lambda}\mathcal{J}(H_0^t,H_0^{t+1},\alpha) & =\alpha(\alpha-1)(1-H_0^t)^{\alpha-2}\partial_kH_0^t\partial_\lambda H_0^t-\alpha(1-H_0^t)^{\alpha-1}\partial_{k\lambda}H_0^t \\
		& +\alpha(\alpha-1)(1-H_0^{t+1})^{\alpha-2}\partial_kH_0^{t+1}\partial_\lambda H_0^{t+1}-\alpha(1-H_0^{t+1})^{\alpha-1}\partial_{k\lambda}H_0^{t+1} \\
		& -\alpha(\alpha-1)(1-H_0^t)^{\alpha-2}\partial_kH_0^t\partial_\lambda H_0^t(1-H_0^{t+1})^\alpha \\
		& +\alpha(1-H_0^t)^{\alpha-1}\partial_{k\lambda}H_0^t(1-H_0^{t+1})^\alpha \\
		& -\alpha^2(1-H_0^t)^{\alpha-1}(1-H_0^{t+1})^{\alpha-1}(\partial_kH_0^t\partial_\lambda H_0^{t+1}+\partial_\lambda H_0^t\partial_kH_0^{t+1}) \\
		& -\alpha(\alpha-1)(1-H_0^t)^\alpha(1-H_0^{t+1})^{\alpha-2}\partial_kH_0^{t+1}\partial_\lambda H_0^{t+1} \\
		& +\alpha(1-H_0^t)^{\alpha-1}(1-H_0^{t+1})^{\alpha-1}\partial_{k\lambda}H_0^{t+1}
	\end{aligned}
\end{equation*}

According to the previous results, we can obtain the first and the second derivatives for $\ell(\theta)=\log\mathcal{L}(\theta)$ given by
\begin{equation*}
	\partial_{\alpha_0}\ell(\theta)=\sum_{t=1}^{\tau-1}\frac{\partial_{\alpha_0}c_{\alpha_0}\left(H_0(x_t),H_0(x_{t+1})\right)}{c_{\alpha_0}\left(H_0(x_t),H_0(x_{t+1})\right)}
\end{equation*}
\begin{equation*}
	\partial_{\alpha_1}\ell(\theta)=\sum_{t=\tau+1}^{T-1}\frac{\partial_{\alpha_1}c_{\alpha_1}\left(H_1(x_t),H_1(x_{t+1})\right)}{c_{\alpha_1}\left(H_1(x_t),H_1(x_{t+1})\right)}
\end{equation*}
\begin{equation*}
	\begin{aligned}
		\partial_{k_0}\ell(\theta) & =\sum_{t=1}^\tau\frac{\partial_{k_0}h_0(x_t)}{h_0(x_t)}+\sum_{t=1}^{\tau-1}\frac{\partial_{k_0}c_{\alpha_0}\left(H_0(x_t),H_0(x_{t+1})\right)}{c_{\alpha_0}\left(H_0(x_t),H_0(x_{t+1})\right)}  +\frac{\partial_{k_0}c_{\alpha_{01}}\left(H_0(x_\tau),H_1(x_{\tau+1})\right)}{c_{\alpha_{01}}\left(H_0(x_\tau),H_1(x_{\tau+1})\right)}
	\end{aligned}
\end{equation*}
\begin{equation*}
	\begin{aligned}
		\partial_{k_1}\ell(\theta) & =\sum_{t=\tau+1}^T\frac{\partial_{k_1}h_1(x_t)}{h_1(x_t)}+\sum_{t=\tau+1}^{T-1}\frac{\partial_{k_1}c_{\alpha_1}\left(H_1(x_t),H_1(x_{t+1})\right)}{c_{\alpha_1}\left(H_1(x_t),H_1(x_{t+1})\right)} 
		+\frac{\partial_{k_1}c_{\alpha_{01}}\left(H_0(x_\tau),H_1(x_{\tau+1})\right)}{c_{\alpha_{01}}\left(H_0(x_\tau),H_1(x_{\tau+1})\right)}
	\end{aligned}
\end{equation*}
\begin{equation*}
	\begin{aligned}
		\partial_{\lambda_0}\ell(\theta) & =\sum_{t=1}^\tau\frac{\partial_{\lambda_0}h_0(x_t)}{h_0(x_t)}+\sum_{t=1}^{\tau-1}\frac{\partial_{\lambda_0}c_{\alpha_0}\left(H_0(x_t),H_0(x_{t+1})\right)}{c_{\alpha_0}\left(H_0(x_t),H_0(x_{t+1})\right)} 
		+\frac{\partial_{\lambda_0}c_{\alpha_{01}}\left(H_0(x_\tau),H_1(x_{\tau+1})\right)}{c_{\alpha_{01}}\left(H_0(x_\tau),H_1(x_{\tau+1})\right)}
	\end{aligned}
\end{equation*}
\begin{equation*}
	\begin{aligned}
		\partial_{\lambda_1}\ell(\theta) & =\sum_{t=\tau+1}^T\frac{\partial_{\lambda_1}h_1(x_t)}{h_1(x_t)}+\sum_{t=\tau+1}^{T-1}\frac{\partial_{\lambda_1}c_{\alpha_1}\left(H_1(x_t),H_1(x_{t+1})\right)}{c_{\alpha_1}\left(H_1(x_t),H_1(x_{t+1})\right)} 
		+\frac{\partial_{\lambda_1}c_{\alpha_{01}}\left(H_0(x_\tau),H_1(x_{\tau+1})\right)}{c_{\alpha_{01}}\left(H_0(x_\tau),H_1(x_{\tau+1})\right)}
	\end{aligned}
\end{equation*}
\begin{equation*}
	\begin{aligned}
		& \partial_{\alpha_0\alpha_0}\ell(\theta) 
		=\sum_{t=1}^{\tau-1}\frac{\partial_{\alpha_0\alpha_0}c_{\alpha_0}(H_0(x_t),H_0(x_{t+1}))c_{\alpha_0}(H_0(x_t),H_0(x_{t+1}))-\left[\partial_{\alpha_0}c_{\alpha_0}(H_0(x_t),H_0(x_{t+1}))\right]^2}{\left[c_{\alpha_0}(H_0(x_t),H_0(x_{t+1}))\right]^2}
	\end{aligned}
\end{equation*}
\begin{equation*}
	\partial_{\alpha_1\alpha_1}\ell(\theta)=\sum_{t=\tau+1}^T\frac{\partial_{\alpha_1\alpha_1}c_{\alpha_1}(H_1(x_t),H_1(x_{t+1}))c_{\alpha_1}(H_1(x_t),H_1(x_{t+1}))-\left[\partial_{\alpha_1}c_{\alpha_1}(H_1(x_t),H_1(x_{t+1}))\right]^2}{\left[c_{\alpha_1}(H_1(x_t),H_1(x_{t+1}))\right]^2}
\end{equation*}
\begin{equation*}
	\partial_{\alpha_0\alpha_1}\ell(\theta) =0
\end{equation*}
\begin{equation*}
	\begin{aligned}
		\partial_{\alpha_0k_0}\ell(\theta) & =\sum_{t=1}^{\tau-1}\frac{\partial_{\alpha_0k_0}c_{\alpha_0}\left(H_0(x_t),H_0(x_{t+1})\right)c_{\alpha_0}\left(H_0(x_t),H_0(x_{t+1})\right)}{\left[c_{\alpha_0}\left(H_0(x_t),H_0(x_{t+1})\right)\right]^2} \\
		& -\sum_{t=1}^{\tau-1}\frac{\partial_{\alpha_0}c_{\alpha_0}\left(H_0(x_t),H_0(x_{t+1})\right)\partial_{k_0}c_{\alpha_0}\left(H_0(x_t),H_0(x_{t+1})\right)}{\left[c_{\alpha_0}\left(H_0(x_t),H_0(x_{t+1})\right)\right]^2}
	\end{aligned}
\end{equation*}
\begin{equation*}
	\begin{aligned}
		\partial_{\alpha_1k_1}\ell(\theta) & =\sum_{t=\tau+1}^{T}\frac{\partial_{\alpha_1k_1}c_{\alpha_1}\left(H_1(x_t),H_1(x_{t+1})\right)c_{\alpha_1}\left(H_1(x_t),H_1(x_{t+1})\right)}{\left[c_{\alpha_1}\left(H_1(x_t),H_1(x_{t+1})\right)\right]^2} \\
		& -\sum_{t=\tau+1}^{T}\frac{\partial_{\alpha_1}c_{\alpha_1}\left(H_1(x_t),H_1(x_{t+1})\right)\partial_{k_1}c_{\alpha_1}\left(H_1(x_t),H_1(x_{t+1})\right)}{\left[c_{\alpha_1}\left(H_1(x_t),H_1(x_{t+1})\right)\right]^2}
	\end{aligned}
\end{equation*}
\begin{equation*}
	\partial_{\alpha_0k_1}\ell(\theta) =0
\end{equation*}
\begin{equation*}
	\partial_{\alpha_1k_0}\ell(\theta) =0
\end{equation*}
\begin{equation*}
	\begin{aligned}
		\partial_{\alpha_0\lambda_0}\ell(\theta) & =\sum_{t=1}^{\tau-1}\frac{\partial_{\alpha_0\lambda_0}c_{\alpha_0}\left(H_0(x_t),H_0(x_{t+1})\right)c_{\alpha_0}\left(H_0(x_t),H_0(x_{t+1})\right)}{\left[c_{\alpha_0}\left(H_0(x_t),H_0(x_{t+1})\right)\right]^2} \\
		& -\sum_{t=1}^{\tau-1}\frac{\partial_{\alpha_0}c_{\alpha_0}\left(H_0(x_t),H_0(x_{t+1})\right)\partial_{\lambda_0}c_{\alpha_0}\left(H_0(x_t),H_0(x_{t+1})\right)}{\left[c_{\alpha_0}\left(H_0(x_t),H_0(x_{t+1})\right)\right]^2}
	\end{aligned}
\end{equation*}
\begin{equation*}
	\begin{aligned}
		\partial_{\alpha_1\lambda_1}\ell(\theta) & =\sum_{t=\tau+1}^{T}\frac{\partial_{\alpha_1\lambda_1}c_{\alpha_1}\left(H_1(x_t),H_1(x_{t+1})\right)c_{\alpha_1}\left(H_1(x_t),H_1(x_{t+1})\right)}{\left[c_{\alpha_1}\left(H_1(x_t),H_1(x_{t+1})\right)\right]^2} \\
		& -\sum_{t=\tau+1}^{T}\frac{\partial_{\alpha_1}c_{\alpha_1}\left(H_1(x_t),H_1(x_{t+1})\right)\partial_{\lambda_1}c_{\alpha_1}\left(H_1(x_t),H_1(x_{t+1})\right)}{\left[c_{\alpha_1}\left(H_1(x_t),H_1(x_{t+1})\right)\right]^2}
	\end{aligned}
\end{equation*}
\begin{equation*}
	\partial_{\alpha_0\lambda_1}\ell(\theta) =0
\end{equation*}
\begin{equation*}
	\partial_{\alpha_1\lambda_0}\ell(\theta) =0
\end{equation*}
\begin{equation*}
	\begin{aligned}
		\partial_{k_0k_0}\ell(\theta) 
		& =\sum_{t=1}^\tau\frac{\partial_{k_0k_0}h_0(x_t)h_0(x_t)-\left[\partial_{k_0}h_0(x_t)\right]^2}{[h_0(x_t)]^2} \\
		& +\sum_{t=1}^{\tau-1}\frac{\partial_{k_0k_0}c_{\alpha_0}\big(H_0(x_t),H_0(x_{t+1})\big)c_{\alpha_0}\big(H_0(x_t),H_0(x_{t+1})\big)-\big[\partial_{k_0}c_{\alpha_0}\big(H_0(x_t),H_0(x_{t+1})\big)\big]^2}{\left[c_{\alpha_0}\big(H_0(x_t),H_0(x_{t+1})\big)\big]^2\right]} \\
		& +\frac{\partial_{k_0k_0}c_{\alpha_{01}}\big(H_0(x_\tau),H_1(x_{\tau+1})\big)c_{\alpha_{01}}\big(H_0(x_\tau),H_1(x_{\tau+1})\big)-\big[\partial_{k_0}c_{\alpha_{01}}\big(H_0(x_\tau),H_1(x_{\tau+1})\big)\big]^2}{\big[c_{\alpha_{01}}\big(H_0(x_\tau),H_1(x_{\tau+1})\big)\big]^2}
	\end{aligned}
\end{equation*}
\begin{equation*}
	\begin{aligned}
		\partial_{k_0k_1}\ell(\theta) & =\frac{\partial_{k_0k_1}c_{\alpha_{01}}\left(H_0(x_\tau),H_1(x_{\tau+1})\right)c_{\alpha_{01}}\left(H_0(x_\tau),H_1(x_{\tau+1})\right)}{\left[c_{\alpha_{01}}\left(H_0(x_\tau),H_1(x_{\tau+1})\right)\right]^2} \\
		& -\frac{\partial_{k_0}c_{\alpha_{01}}\left(H_0(x_\tau),H_1(x_{\tau+1})\right)\partial_{k_1}c_{\alpha_{01}}\left(H_0(x_\tau),H_1(x_{\tau+1})\right)}{\left[c_{\alpha_{01}}\left(H_0(x_\tau),H_1(x_{\tau+1})\right)\right]^2}
	\end{aligned}
\end{equation*}
\begin{equation*}
	\begin{aligned}
		\partial_{k_1k_1}\ell(\theta) 
		& =\sum_{t=\tau+1}^T\frac{\partial_{k_1k_1}h_1(x_t)h_1(x_t)-\left[\partial_{k_1}h_1(x_t)\right]^2}{[h_1(x_t)]^2} \\
		& 
		\begin{aligned}
			& +\sum_{t=\tau+1}^{T-1}\frac{\partial_{k_1k_1}c_{\alpha_1}\big(H_1(x_t),H_1(x_{t+1})\big)c_{\alpha_1}\big(H_1(x_t),H_1(x_{t+1})\big)-\big[\partial_{k_1}c_{\alpha_1}\big(H_1(x_t),H_1(x_{t+1})\big)\big]^2}{\big[c_{\alpha_1}\big(H_1(x_t),H_1(x_{t+1})\big)\big]^2}
		\end{aligned} \\
		& +\frac{\partial_{k_1k_1}c_{\alpha_{01}}\big(H_0(x_\tau),H_1(x_{\tau+1})\big)c_{\alpha_{01}}\big(H_0(x_\tau),H_1(x_{\tau+1})\big)-\big[\partial_{k_1}c_{\alpha_{01}}\big(H_0(x_\tau),H_1(x_{\tau+1})\big)\big]^2}{\left[c_{\alpha_{01}}\big(H_0(x_\tau),H_1(x_{\tau+1})\big)\big]^2\right]^2}
	\end{aligned}
\end{equation*}
\begin{equation*}
	\begin{aligned}
		\partial_{\lambda_0\lambda_0}\ell(\theta) 
		& =\sum_{t=1}^\tau\frac{\partial_{\lambda_0\lambda_0}h_0(x_t)h_0(x_t)-\left[\partial_{\lambda_0}h_0(x_t)\right]^2}{[h_0(x_t)]^2} \\
		& +\sum_{t=1}^{\tau-1}\frac{\partial_{\lambda_0\lambda_0}c_{\alpha_0}\left(H_0(x_t),H_0(x_{t+1})\right)c_{\alpha_0}\left(H_0(x_t),H_0(x_{t+1})\right)-\left[\partial_{\lambda_0}c_{\alpha_0}\left(H_0(x_t),H_0(x_{t+1})\right)\right]^2}{\left[c_{\alpha_0}\left(H_0(x_t),H_0(x_{t+1})\right)\right]^2} \\
		& +\frac{\partial_{\lambda_0\lambda_0}c_{\alpha_{01}}\left(H_0(x_\tau),H_1(x_{\tau+1})\right)c_{\alpha_{01}}\left(H_0(x_\tau),H_1(x_{\tau+1})\right)-\left[\partial_{\lambda_0}c_{\alpha_{01}}\left(H_0(x_\tau),H_1(x_{\tau+1})\right)\right]^2}{\left[c_{\alpha_{01}}\left(H_0(x_\tau),H_1(x_{\tau+1})\right)\right]^2}
	\end{aligned}
\end{equation*}
\begin{equation*}
	\begin{aligned}
		\partial_{\lambda_0\lambda_1}\ell(\theta) & =\frac{\partial_{\lambda_0\lambda_1}c_{\alpha_{01}}\left(H_0(x_\tau),H_1(x_{\tau+1})\right)c_{\alpha_{01}}\left(H_0(x_\tau),H_1(x_{\tau+1})\right)}{\left[c_{\alpha_{01}}\left(H_0(x_\tau),H_1(x_{\tau+1})\right)\right]^2} \\
		& -\frac{\partial_{\lambda_0}c_{\alpha_{01}}\left(H_0(x_\tau),H_1(x_{\tau+1})\right)\partial_{\lambda_1}c_{\alpha_{01}}\left(H_0(x_\tau),H_1(x_{\tau+1})\right)}{\left[c_{\alpha_01}\left(H_0(x_\tau),H_1(x_{\tau+1})\right)\right]^2}
	\end{aligned}
\end{equation*}
\begin{equation*}
	\begin{aligned}
		\partial_{\lambda_1\lambda_1}\ell(\theta) 
		& =\sum_{t=\tau}^T\frac{\partial_{\lambda_1\lambda_1}h_1(x_t)h_1(x_t)-\left[\partial_{\lambda_1}h_1(x_t)\right]^2}{[h_1(x_t)]^2} \\
		& +\sum_{t=\tau+1}^{T-1}\frac{\partial_{\lambda_1\lambda_1}c_{\alpha_1}\left(H_1(x_t),H_1(x_{t+1})\right)c_{\alpha_1}\left(H_1(x_t),H_1(x_{t+1})\right)-\left[\partial_{\lambda_1}c_{\alpha_1}\left(H_1(x_t),H_1(x_{t+1})\right)\right]^2}{\left[c_{\alpha_1}\left(H_1(x_t),H_1(x_{t+1})\right)\right]^2} \\
		& +\frac{\partial_{\lambda_1\lambda_1}c_{\alpha_{01}}(H_0(x_\tau),H_1(x_{\tau+1}))c_{\alpha_{01}}(H_0(x_\tau),H_1(x_{\tau+1}))-\left[\partial_{\lambda_1}c_{\alpha_{01}}(H_0(x_\tau),H_1(x_{\tau+1}))\right]^2}{\left[c_{\alpha_{01}}(H_0(x_\tau),H_1(x_{\tau+1}))\right]^2}
	\end{aligned}
\end{equation*}

\begin{equation*}
	\begin{aligned}
		\partial_{k_0\lambda_0}\ell(\theta) & =\sum_{t=1}^{\tau}\frac{\partial_{k_0\lambda_0}h_0(x_t)h_0(x_t)-\partial_{k_0}h_0(x_t)\partial_{\lambda_0}h_0(x_t)}{[h_0(x_t)]^2} 
		+\\&\sum_{t=1}^{\tau-1}\frac{\partial_{k_0\lambda_0}c_{\alpha_0}(H_0(x_t),H_0(x_{t+1}))c_{\alpha_0}(H_0(x_t),H_0(x_{t+1}))}{\left[c_{\alpha_0}(H_0(x_t),H_0(x_{t+1}))\right]^2} \\
		& -\sum_{t=1}^{\tau-1}\frac{\partial_{k_0}c_{\alpha_0}\left(H_0(x_t),H_0(x_{t+1})\right)\partial_{\lambda_0}c_{\alpha_0}\left(H_0(x_t),H_0(x_{t+1})\right)}{\left[c_{\alpha_0}\left(H_0(x_t),H_0(x_{t+1})\right)\right]^2} \\
		& +\frac{\partial_{k_0\lambda_0}c_{\alpha_{01}}\left(H_0(x_\tau),H_1(x_{\tau+1})\right)c_{\alpha_{01}}\left(H_0(x_\tau),H_1(x_{\tau+1})\right)}{\left[c_{\alpha_{01}}\left(H_0(x_\tau),H_1(x_{\tau+1})\right)\right]^2} \\
		& -\frac{\partial_{k_0}c_{\alpha_{01}}\left(H_0(x_\tau),H_1(x_{\tau+1})\right)\partial_{\lambda_0}c_{\alpha_{01}}\left(H_0(x_\tau),H_1(x_{\tau+1})\right)}{\left[c_{\alpha_{01}}\left(H_0(x_\tau),H_1(x_{\tau+1})\right)\right]^2}
	\end{aligned}
\end{equation*}
\begin{equation*}
	\begin{aligned}
		\partial_{k_0\lambda_1}\ell(\theta) & =\frac{\partial_{k_0\lambda_1}c_{\alpha_{01}}\left(H_0(x_\tau),H_1(x_{\tau+1})\right)c_{\alpha_{01}}\left(H_0(x_\tau),H_1(x_{\tau+1})\right)}{\left[c_{\alpha_{01}}\left(H_0(x_\tau),H_1(x_{\tau+1})\right)\right]^2} \\
		& -\frac{\partial_{k_0}c_{\alpha_{01}}\left(H_0(x_\tau),H_1(x_{\tau+1})\right)\partial_{\lambda_1}c_{\alpha_{01}}\left(H_0(x_\tau),H_1(x_{\tau+1})\right)}{\left[c_{\alpha_01}\left(H_0(x_\tau),H_1(x_{\tau+1})\right)\right]^2}
	\end{aligned}
\end{equation*}
\begin{equation*}
	\begin{aligned}
		\partial_{k_1\lambda_0}\ell(\theta) & =\frac{\partial_{k_1\lambda_0}c_{\alpha_{01}}\left(H_0(x_\tau),H_1(x_{\tau+1})\right)c_{\alpha_{01}}\left(H_0(x_\tau),H_1(x_{\tau+1})\right)}{\left[c_{\alpha_{01}}\left(H_0(x_\tau),H_1(x_{\tau+1})\right)\right]^2} \\
		& -\frac{\partial_{k_1}c_{\alpha_{01}}\left(H_0(x_\tau),H_1(x_{\tau+1})\right)\partial_{\lambda_0}c_{\alpha_{01}}\left(H_0(x_\tau),H_1(x_{\tau+1})\right)}{\left[c_{\alpha_{01}}\left(H_0(x_\tau),H_1(x_{\tau+1})\right)\right]^2}
	\end{aligned}
\end{equation*}
\begin{equation*}
	\begin{aligned}
		\partial_{k_1\lambda_1}\ell(\theta) & =\sum_{t=\tau+1}^T\frac{\partial_{k_1\lambda_1}h_1(x_t)h_1(x_t)-\partial_{k_1}h_1(x_t)\partial_{\lambda_1}h_1(x_t)}{[h_1(x_t)]^2} 
		\\&+\sum_{t=\tau+1}^{T-1}\frac{\partial_{k_1\lambda_1}c_{\alpha_1}\left(H_1(x_t),H_1(x_{t+1})\right)c_{\alpha_1}\left(H_1(x_t),H_1(x_{t+1})\right)}{\left[c_{\alpha_1}\left(H_1(x_t),H_1(x_{t+1})\right)\right]^2} \\
		& -\sum_{t=\tau+1}^{T-1}\frac{\partial_{k_1}c_{\alpha_1}\left(H_1(x_t),H_1(x_{t+1})\right)\partial_{\lambda_1}c_{\alpha_1}\left(H_1(x_t),H_1(x_{t+1})\right)}{\left[c_{\alpha_1}\left(H_1(x_t),H_1(x_{t+1})\right)\right]^2} \\    & +\frac{\partial_{k_1\lambda_1}c_{\alpha_{01}}\left(H_0(x_\tau),H_1(x_{\tau+1})\right)c_{\alpha_{01}}\left(H_0(x_\tau),H_1(x_{\tau+1})\right)}{\left[c_{\alpha_{01}}\left(H_0(x_\tau),H_1(x_{\tau+1})\right)\right]^2} \\
		& -\frac{\partial_{k_1}c_{\alpha_{01}}\left(H_0(x_\tau),H_1(x_{\tau+1})\right)\partial_{\lambda_1}c_{\alpha_{01}}\left(H_0(x_\tau),H_1(x_{\tau+1})\right)}{\left[c_{\alpha_{01}}\left(H_0(x_\tau),H_1(x_{\tau+1})\right)\right]^2}
	\end{aligned}
\end{equation*}

\section{VIX Interarrival Times Above Threshold 30}

This section presents the interarrival times associated with VIX values exceeding the threshold level of 30 during the period from 2018 to 2024, which is commonly interpreted as a signal of elevated market fear and financial uncertainty.

\begin{table}[htbp!]
  \centering
  \scriptsize 
  \begin{tabular}{@{} ll ll ll ll @{}}
    \toprule
    Breakthrough date & T value  & Breakthrough date & T value  & Breakthrough date & T value  & Breakthrough date & T value  \\
    \midrule
    2018-02-05 &  34 & 2020-04-09 &   1 & 2020-06-25 &   1 & 2022-03-11 &   1 \\
    2018-02-08 &   3 & 2020-04-13 &   4 & 2020-06-26 &   1 & 2022-03-14 &   3 \\
    2018-12-21 & 316 & 2020-04-14 &   1 & 2020-06-29 &   3 & 2022-04-26 &  43 \\
    2018-12-24 &   3 & 2020-04-15 &   1 & 2020-06-30 &   1 & 2022-04-27 &   1 \\
    2018-12-26 &   2 & 2020-04-16 &   1 & 2020-07-13 &  13 & 2022-04-29 &   2 \\
    2020-02-27 & 428 & 2020-04-17 &   1 & 2020-09-03 &  52 & 2022-05-02 &   3 \\
    2020-02-28 &   1 & 2020-04-20 &   1 & 2020-09-04 &   1 & 2022-05-05 &   3 \\
    2020-03-02 &   1 & 2020-04-21 &   1 & 2020-09-08 &   4 & 2022-05-06 &   1 \\
    2020-03-03 &   1 & 2020-04-22 &   1 & 2020-10-26 &  48 & 2022-05-09 &   3 \\
    2020-03-04 &   1 & 2020-04-23 &   1 & 2020-10-27 &   1 & 2022-05-10 &   1 \\
    2020-03-05 &   1 & 2020-04-24 &   1 & 2020-10-28 &   1 & 2022-05-11 &   1 \\
    2020-03-06 &   1 & 2020-04-27 &   1 & 2020-10-29 &   1 & 2022-05-12 &   1 \\
    2020-03-09 &   1 & 2020-04-28 &   1 & 2020-10-30 &   1 & 2022-05-18 &   6 \\
    2020-03-10 &   1 & 2020-04-29 &   1 & 2020-11-02 &   1 & 2022-06-13 &  26 \\
    2020-03-11 &   1 & 2020-04-30 &   1 & 2020-11-03 &   1 & 2022-06-14 &   1 \\
    2020-03-12 &   1 & 2020-05-01 &   1 & 2021-01-27 &  85 & 2022-06-16 &   2 \\
    2020-03-13 &   1 & 2020-05-04 &   1 & 2021-01-28 &   1 & 2022-06-17 &   1 \\
    2020-03-16 &   1 & 2020-05-05 &   1 & 2021-01-29 &   1 & 2022-06-21 &   4 \\
    2020-03-17 &   1 & 2020-05-06 &   1 & 2021-02-01 &   3 & 2022-09-26 &  97 \\
    2020-03-18 &   1 & 2020-05-07 &   1 & 2021-12-01 & 303 & 2022-09-27 &   1 \\
    2020-03-19 &   1 & 2020-05-12 &   5 & 2021-12-03 &   2 & 2022-09-28 &   1 \\
    2020-03-20 &   1 & 2020-05-13 &   1 & 2022-01-25 &  53 & 2022-09-29 &   1 \\
    2020-03-23 &   1 & 2020-05-14 &   1 & 2022-01-26 &   1 & 2022-09-30 &   1 \\
    2020-03-24 &   1 & 2020-05-15 &   1 & 2022-01-27 &   1 & 2022-10-03 &   3 \\
    2020-03-25 &   1 & 2020-05-19 &   4 & 2022-02-23 &  27 & 2022-10-06 &   3 \\
    2020-03-26 &   1 & 2020-06-11 &  23 & 2022-02-24 &   1 & 2022-10-07 &   1 \\
    2020-03-27 &   1 & 2020-06-12 &   1 & 2022-02-28 &   4 & 2022-10-10 &   3 \\
    2020-03-30 &   1 & 2020-06-15 &   3 & 2022-03-01 &   1 & 2022-10-11 &   1 \\
    2020-03-31 &   1 & 2020-06-16 &   1 & 2022-03-02 &   1 & 2022-10-12 &   1 \\
    2020-04-01 &   1 & 2020-06-17 &   1 & 2022-03-03 &   1 & 2022-10-13 &   1 \\
    2020-04-02 &   1 & 2020-06-18 &   1 & 2022-03-04 &   1 & 2022-10-14 &   3 \\
    2020-04-03 &   1 & 2020-06-19 &   1 & 2022-03-07 &   1 & 2022-10-17 &   1 \\
    2020-04-06 &   1 & 2020-06-22 &   3 & 2022-03-08 &   1 & 2022-10-18 &   1 \\
    2020-04-07 &   1 & 2020-06-23 &   1 & 2022-03-09 &   1 & 2022-10-19 &   1 \\
    2020-04-08 &   1 & 2020-06-24 &   1 & 2022-03-10 &   1 &  &    \\
    \bottomrule
  \end{tabular}
  \label{tab:breakthrough-vix}
\end{table}

\pagebreak

\bibliographystyle{plain}
\bibliography{reference}

@article{beare2010copulas,
  title={Copulas and temporal dependence},
  author={Beare, B. K.},
  journal={Econometrica},
  volume={78},
  number={1},
  pages={395--410},
  year={2010},
  publisher={Wiley Online Library}
}

@article{jammalamadaka2018multivariate,
 	title={{Multivariate {Bayesian} Structural Time Series Model}},
 	author={Jammalamadaka, S Rao and Qiu, Jinwen and Ning, Ning},
 	journal={Journal of Machine Learning Research},
 	year={2018}
 }

@article{jammalamadaka2019predicting,
 	title={{Predicting a Stock Portfolio with the Multivariate {Bayesian} Structural Time Series Model: Do News or Emotions Matter?}},
 	author={Jammalamadaka, S Rao and Qiu, Jinwen and Ning, Ning},
 	journal={International Journal of Artificial Intelligence},
 	volume={17},
 	number={2},
 	pages={81--104},
 	year={2019}
 }

@article{qiu2020multivariate,
 	title={{Multivariate time series analysis from a {Bayesian} machine learning perspective}},
 	author={Qiu, Jinwen and Jammalamadaka, S Rao and Ning, Ning},
 	journal={Annals of Mathematics and Artificial Intelligence},
 	volume={88},
 	number={10},
 	pages={1061--1082},
 	year={2020},
 	publisher={Springer International Publishing Cham}
 }

@article{ning2023iterated,
 	title={{Iterated Block Particle Filter for High-dimensional Parameter Learning: Beating the Curse of Dimensionality}},
 	author={Ning, Ning and Ionides, Edward},
 	journal={Journal of Machine Learning Research},
 	year={2023}
 }

@article{ionides2024iterated,
 	title={{An iterated block particle filter for inference on coupled dynamic systems with shared and unit-specific parameters}},
 	author={Ionides, Edward and Ning, Ning and Wheeler, Jesse},
 	journal={Statistica Sinica},
 	volume={34},
 	pages={1--22},
 	year={2024}
 }

@article{parpoula2022distribution,
  title={A distribution-free control charting technique based on change-point analysis for detection of epidemics},
  author={Parpoula, C.},
  journal={Statistical Methods in Medical Research},
  volume={31},
  number={6},
  pages={1067--1084},
  year={2022},
  publisher={SAGE Publications Sage UK: London, England}
}

@article{timmer2003change,
  title={Change point estimates for the parameters of an {AR}(1) process},
  author={Timmer, D. H. and Pignatiello Jr, J. J.},
  journal={Quality and Reliability Engineering International},
  volume={19},
  number={4},
  pages={355--369},
  year={2003},
  publisher={Wiley Online Library}
}

@book{joe1997multivariate,
  title={Multivariate models and multivariate dependence concepts},
  author={Joe, H.},
  year={1997},
  publisher={CRC press}
}

@article{chen2006estimation,
  title={Estimation of copula-based semiparametric time series models},
  author={Chen, X. and Fan, Y.},
  journal={Journal of Econometrics},
  volume={130},
  number={2},
  pages={307--335},
  year={2006},
  publisher={Elsevier}
}

@article{emura2015algorithm,
  title={An algorithm for estimating survival under a copula-based dependent truncation model},
  author={Emura, T. and Murotani, K.},
  journal={Test},
  volume={24},
  pages={734--751},
  year={2015},
  publisher={Springer}
}

@article{sun2020bayesian,
  title={A {Bayesian} inference for time series via copula-based Markov chain models},
  author={Sun, L.-H. and Lee, C.-S. and Emura, T.},
  journal={Communications in Statistics-Simulation and Computation},
  volume={49},
  number={11},
  pages={2897--2913},
  year={2020},
  publisher={Taylor \& Francis}
}

@article{JustinKuepper,
  author    = {Kuepper, J.},
  title     = {{CBOE} volatility index (VIX): What Does It Measure in Investing?},
   journal={Investopedia},
  year      = {2024},
  url       = {https://www.investopedia.com/terms/v/vix.asp\#citation-4},
  urldate   = {2024-08-07}
}

@article{tibshirani1993introduction,
  title={An introduction to the bootstrap},
  author={Tibshirani, R. J. and Efron, B.},
  journal={Monographs on statistics and applied probability},
  volume={57},
  number={1},
  pages={1--436},
  year={1993},
  publisher={Citeseer}
}

@article{clayton1978model,
  title={A model for association in bivariate life tables and its application in epidemiological studies of familial tendency in chronic disease incidence},
  author={Clayton, D. G.},
  journal={Biometrika},
  volume={65},
  number={1},
  pages={141--151},
  year={1978},
  publisher={Oxford University Press}
}

@article{ELS2021,
  author		= "Emura, T. and Lai, C.-C and Sun, L.-H.",
  title			= "Change point estimation under a copula-based Markov chain model for binomial time series",
  journal		= "Econometrics and Statistics",
  volume		= "28",
  number		= "",
  pages			= "120–137",
   year			= "2023",
  
}

@article{huang2021model,
  title={Model diagnostic procedures for copula-based Markov chain models for statistical process control},
  author={Huang, X.W. and Emura, T.},
  journal={Communications in Statistics-Simulation and Computation},
  volume={50},
  number={8},
  pages={2345--2367},
  year={2021},
  publisher={Taylor \& Francis}
}

@article{Joe1993,
  author		= "Joe, H.",
  title			= "Parametric families of multivariate distributions with given margins.  46 (2), 262–282.",
  journal		= "Journal of Multivariate Analysis",
  volume		= "46",
   number		= "2",
  pages			= "262–282",
  year			= "1993"
}

@incollection{lavielle2007adaptive,
  title={Adaptive detection of multiple change-points in asset price volatility},
  author={Lavielle, M. and Teyssiere, G.},
  booktitle={Long Memory in Economics},
  pages={129--156},
  year={2007},
  publisher={Springer}
}

@article{macdonald2014does,
  title={Does newton--raphson really fail?},
  author={MacDonald, I. L},
  journal={Statistical Methods in Medical Research},
  volume={23},
  number={3},
  pages={308--311},
  year={2014},
  publisher={Sage Publications Sage UK: London, England}
}

@article{reeves2007review,
  title={A review and comparison of changepoint detection techniques for climate data},
  author={Reeves, J. and Chen, J. and Wang, X. L. and Lund, R. and Lu, Q. Q.},
  journal={Journal of Applied Meteorology and Climatology},
  volume={46},
  number={6},
  pages={900--915},
  year={2007}
}

@inproceedings{sklar1959fonctions,
  title={Fonctions de r{\'e}partition {\`a} n dimensions et leurs marges},
  author={Sklar, M},
  booktitle={Annales de l'ISUP},
  volume={8},
  pages={229--231},
  year={1959}
}

@article{SWLEC2025,
  author		= "Sun, L.-H. and Wang, Y.-K. and Liu, L.-H. and Emura, T and Chiu, C.-Y.",
  title			= "Change-point estimation for Gaussian time series data with copula-based Markov chain models",
  journal		= "Computational Statistics",
  volume		= "40",
  pages			= "1541–1581",
  year			= "2025"
}

@article{Copula,
  author		= "Sklar, M.",
  title			= "Fonctions de repartition an dimensions et leurs marges",
  journal		= "Publications de l’Institut
de Statistique de l’Université de Paris",
  volume		= "8",
  number		= "",
  pages			= "229—231",
  year			= "1959"
}

@article{SUN2026,
  author		= "Sun, L.-H. and Huang, Z.-Y. and Huang, Y.-L.. and Chiu, C.-Y. and Ning, N.",
  title			= "Online Supplement for ''{C}hange-point estimation for Weibull time series data with copula-based Markov chain models (https://sites.google.com/view/lhsun/publication)",
  journal		= "National Central University",
  year			= "2026"
}

\end{document}